\documentclass[
 aip,
 amsmath,amssymb,
 reprint,
]{revtex4-1}
\usepackage{orcidlink}
\usepackage{booktabs,longtable,tabularx,array,siunitx}
\usepackage{graphicx}
\usepackage{dcolumn}
\usepackage{bm}
\usepackage{multirow}
\usepackage{array}

\renewcommand{\figurename}{Fig.}
\renewcommand{\tablename}{Table}

\usepackage{cleveref}
\crefname{figure}{Fig.}{Figs.}
\Crefname{figure}{Fig.}{Figs.}
\crefname{table}{Table}{Tables}
\Crefname{table}{Table}{Tables}
\crefname{section}{Sec.}{Secs.}
\Crefname{section}{Sec.}{Secs.}
\crefname{equation}{Eq.}{Eqs.}
\Crefname{equation}{Eq.}{Eqs.}
\usepackage{listings} 
\crefname{lstlisting}{Script}{Scripts}
\Crefname{lstlisting}{Script}{Scripts}

\usepackage[scaled]{beramono} 

\usepackage[utf8]{inputenc}
\usepackage[T1]{fontenc}
\usepackage{mathptmx}
\usepackage{etoolbox}
\usepackage{braket}
\usepackage[table,xcdraw,dvipsnames]{xcolor}
\usepackage[normalem]{ulem}

\usepackage{float}
\usepackage{amsthm} 
\usepackage{soul}
\DeclareMathOperator{\tr}{tr}

\DeclareMathOperator{\rank}{rank}
\newcommand{\dyad}[1]{| #1\rangle \langle #1|}
\usepackage{mathtools}

\definecolor{lightgreen}{rgb}{0.8,1,0.8}
\definecolor{lightred}{rgb}{1,0.8,0.8}
\definecolor{lightorange}{rgb}{1,0.9,0.7}

\usepackage{tikz}
\usetikzlibrary{tikzmark,calc}

\usepackage[most]{tcolorbox}
\definecolor{B11base}{HTML}{FFA630}
\definecolor{B12base}{HTML}{EBC775}
\definecolor{B13base}{HTML}{D7E8BA}
\definecolor{B21base}{HTML}{92C5B2}
\definecolor{B22base}{HTML}{4DA1A9}
\definecolor{B23base}{HTML}{3E7990}
\definecolor{B31base}{HTML}{2E5077}
\definecolor{B32base}{HTML}{483656}
\definecolor{B33base}{HTML}{611C35}

\colorlet{B11}{B11base!30!white}  
\colorlet{B32}{B12base!35!white}
\colorlet{B13}{B13base!50!white}  
\colorlet{B21}{B21base!25!white}  
\colorlet{B12}{B22base!25!white}  
\colorlet{B23}{B23base!20!white}  
\colorlet{B31}{B31base!20!white}  
\colorlet{B22}{B32base!15!white} 
\colorlet{B33}{B33base!18!white} 

\colorlet{diagcol}{B11}  
\colorlet{upcol}{B12}    
\colorlet{lowcol}{B21}   

\definecolor{contboxframebase}{HTML}{88D498}
\definecolor{theoframebase}{HTML}{114B5F}
\definecolor{lemmaframebase}{HTML}{1A936F}
\definecolor{propframebase}{HTML}{166F67}

\colorlet{contboxframe}{contboxframebase!100!white}  
\colorlet{theoframe}{theoframebase!70!white} 
\colorlet{lemmaframe}{lemmaframebase!70!white}  
\colorlet{propframe}{propframebase!70!white}  

\colorlet{contboxback}{contboxframe!7!white}  
\colorlet{theoback}{theoframe!8!white} 
\colorlet{lemmaback}{lemmaframe!8!white}  
\colorlet{propback}{propframe!8!white}

\newtcolorbox{recapbox}{
  colback=contboxback,      
  colframe=contboxframe,    
  fontupper=\normalsize, 
  title=Context,         
  boxrule=0.8pt,
  arc=2pt,             
  left=1em, right=1em, top=0.5em, bottom=0.5em
}
\newtcolorbox{theorem}{
  colback=theoback,    
  colframe=theoframe,  
  fontupper=\normalsize,
  title=Theorem,
  boxrule=0.8pt,
  arc=2pt,
  left=1em, right=1em, top=0.5em, bottom=0.5em
}

\newcounter{lemcounter}
\newtcolorbox{lemma}{
  colback=lemmaback,    
  colframe=lemmaframe,  
  fontupper=\small,
  boxrule=0.8pt,
  arc=2pt,
  left=1em, right=1em, top=0.5em, bottom=0.5em,
  before upper={\refstepcounter{lemcounter}}, 
  title=Lemma~\arabic{lemcounter},
}

\newcounter{propcounter}
\newtcolorbox{proposition}{
  colback=propback,    
  colframe=propframe, 
  fontupper=\normalsize,
  boxrule=0.8pt,
  arc=2pt,
  left=1em, right=1em, top=0.5em, bottom=0.5em,
  before upper={\refstepcounter{propcounter}},
  title=Proposition~\arabic{propcounter},
}

\newcommand{\ThreePmatrixMainFigure}[9]{%
  \begin{figure}[H]
  \centering
  \setlength{\tabcolsep}{0pt}%
  \renewcommand{\arraystretch}{1}%

  \begin{tabular}{@{}c c c@{}}  
      & \textbf{Theoretical $P$-matrix} 
      & \textbf{Experimental reconstruction #9} \\[0.15cm]

    \rotatebox[origin=c]{90}{\textbf{$d=3$}} &
    \multicolumn{1}{@{\hspace{0.12cm}}c}{%
        \begin{minipage}[c]{0.46\linewidth}
            \centering
            \includegraphics[width=\linewidth]{#1}
        \end{minipage}
    } &
    \multicolumn{1}{@{\hspace{0.12cm}}c}{%
        \begin{minipage}[c]{0.46\linewidth}
            \centering
            \includegraphics[width=\linewidth]{#2}
        \end{minipage}
    } \\[-0.05cm]

      \rotatebox[origin=c]{90}{\textbf{$d=5$}} &
      \begin{minipage}[c]{0.46\linewidth}
          \centering
          \includegraphics[width=\linewidth]{#3}
      \end{minipage} &
      \begin{minipage}[c]{0.46\linewidth}
          \centering
          \includegraphics[width=\linewidth]{#4}
      \end{minipage} \\[-0.05cm]

      \rotatebox[origin=c]{90}{\textbf{$d=7$}} &
      \begin{minipage}[c]{0.46\linewidth}
          \centering
          \includegraphics[width=\linewidth]{#5}
      \end{minipage} &
      \begin{minipage}[c]{0.46\linewidth}
          \centering
          \includegraphics[width=\linewidth]{#6}
      \end{minipage} \\
  \end{tabular}

  \caption{#7}
  \label{#8}
  \end{figure}%
}

\renewcommand{\thefootnote}{\fnsymbol{footnote}}

\makeatletter
\def\@email#1#2{%
 \endgroup
 \patchcmd{\titleblock@produce}
  {\frontmatter@RRAPformat}
  {\frontmatter@RRAPformat{\produce@RRAP{*#1\href{mailto:#2}{#2}}}\frontmatter@RRAPformat}
  {}{}
}%
\makeatother

\renewcommand{\thefootnote}{\fnsymbol{footnote}}

\makeatletter
\def\blfootnote{\gdef\@thefnmark{}\@footnotetext}
\makeatother

\usepackage{etoolbox}
\makeatletter
\patchcmd{\@maketitle}{\@author}{\@author\show\@thanks}{}{}
\makeatother

\begin{document}

\preprint{AIP/123-QED}

\title{
Robust certification of high-dimensional quantum devices}

\author{Javier~Fernández$^{\ast~\orcidlink{0009-0006-5765-0129}}$}

\affiliation{Departament de F\'isica, Universitat Aut\`onoma de Barcelona, E-08193 Bellaterra, Spain.}

\author{Albert~Rico$^{\dagger~\orcidlink{0000-0001-8211-499X}}$}

\affiliation{Naturwissenschaftlich-Technische Fakult\"{a}t, Universit\"{a}t Siegen, Walter-Flex-Stra\ss e 3, 57068 Siegen, Germany}

\author{David~Viedma$^{~\orcidlink{0000-0002-1675-4523}}$}
\affiliation{Departament de F\'isica, Universitat Aut\`onoma de Barcelona, E-08193 Bellaterra, Spain.}

\author{Evelyn~A.~Ortega$^{~\orcidlink{0000-0002-4347-7971}}$}
\affiliation{ICFO – Institut de Ciencies Fotoniques, The Barcelona Institute of Science and Technology, 08860 Castelldefels, Spain}

\author{Valerio~Pruneri$^{~\orcidlink{0000-0002-6425-9332}}$}
\affiliation{ICFO – Institut de Ciencies Fotoniques, The Barcelona Institute of Science and Technology, 08860 Castelldefels, Spain}
\affiliation{ICREA. Lluis Companys 23, 08010 Barcelona, Spain.}

\author{Adam~Vallés$^{~\orcidlink{0000-0002-5200-4914}}$}
\affiliation{Departament de F\'isica, Universitat Aut\`onoma de Barcelona, E-08193 Bellaterra, Spain.}

\author{Verònica~Ahufinger$^{~\orcidlink{0000-0002-6628-9930}}$}
\affiliation{Departament de F\'isica, Universitat Aut\`onoma de Barcelona, E-08193 Bellaterra, Spain.}
\affiliation{ICFO – Institut de Ciencies Fotoniques, The Barcelona Institute of Science and Technology, 08860 Castelldefels, Spain}

\author{Anna~Sanpera$^{\ddagger~\orcidlink{0000-0002-8970-6127}}$}
\affiliation{Departament de F\'isica, Universitat Aut\`onoma de Barcelona, E-08193 Bellaterra, Spain.}
\affiliation{ICREA. Lluis Companys 23, 08010 Barcelona, Spain.}

\author{Some~S.~Bhattacharya$^{~\orcidlink{0000-0001-6464-5068}}$}
\affiliation{Departament de F\'isica, Universitat Aut\`onoma de Barcelona, E-08193 Bellaterra, Spain.}

\date{\today}

\begin{abstract}
Certifying quantum behavior from classically accessible data is essential for secure communication and scalable quantum technologies. While powerful certification methods such as Bell nonlocality and quantum steering exist, their implementation typically requires entanglement or additional assumptions, and experimental demonstrations mainly focus on low-dimensional systems. In minimal prepare-and-measure scenarios, where a sender encodes information into quantum states and a receiver performs a single measurement, robust certification becomes particularly challenging, especially in the presence of noise and in higher-dimensional Hilbert spaces. Here, we propose, design, and experimentally implement a protocol that certifies quantumness between two distant parties without the need for preshared resources or measurement incompatibility. The experiments are carried out using the orbital angular momentum degrees of freedom of single photons, chosen for providing increased dimensionality that is scalable. We demonstrate the robustness of the protocol through rank‑stability analysis of the observed correlations, which enables the certification of non‑classicality even in the presence of noise. Our results provide a practical route to validate high-dimensional quantum communication systems and open new possibilities for secure and dimension-efficient quantum information processing.
\end{abstract}

\maketitle

\blfootnote{$^\ast$
Javier.Fernandez.Cano@uab.cat\\
$^\dagger$Albert.RicoAndres@uni-siegen.de\\
$^\ddagger$Anna.Sanpera@uab.cat}
Quantum mechanics enables correlations and information-processing capabilities that cannot be explained within a classical framework, forming the foundation of emerging quantum technologies such as secure communication, quantum simulation, and quantum computation~\cite{bell1964einstein,bennett1993Teleport,BB84,bennett1996concentrating,Teleport2001polarized}. A central challenge in quantum information science is therefore to identify and certify quantum behaviour using only classically accessible data, without relying on detailed assumptions about the underlying devices~\cite{brunner2014bell,Wiseman2007SteeringEntNonloc,kochenSpeckerContext2011}. This task, known as \emph{quantum certification}, exploits the fact that classical theories cannot reproduce the full range of correlations achievable in quantum systems. Prominent manifestations include Bell nonlocality~\cite{bell1964einstein,brunner2014bell}, quantum steering~\cite{Wiseman2007SteeringEntNonloc,Uola2020SteeringReview}, and contextuality~\cite{kochenSpeckerContext2011,Budroni2022KScontextualityReview}, each revealing fundamental departures from classical hidden-variable descriptions. More recently, related approaches have explored quantum signatures encoded in correlations between state preparations and measurement outcomes, where incompatibility of quantum measurements provides an operational resource~\cite{Gallego2010DVIwitCQdims,Pawlowski2011SDVIsecureOne-Way,tavakoli2020self,puliyil2025SDVIchannnelCommMats,Halder2025QAUnified}.

Certifying quantum properties is particularly important for reliable quantum communication protocols, which in thier simplest version involve only three ingredients: Alice encodes an input, transmits it to Bob, and Bob performs a decoding measurement without an external input~\cite{RFreeNClassPCOncept_Ma2023,Ding2024QadvQubitPM,diebra2026StateExclPM,Chakraborty2025AdvQubitComPM,doolittle2025OpFrQComNettsPM}. Here Alice and Bob do not share any correlations, and Alice's input is identically independently distributed at each round. The behaviors obtained under these assumptions are known as {\em communication matrices (CM)}~\cite{Heino24SimpleUnboundQAdvtg,Heinosaari2024MaxElsQComPM}. 
These are based on different assumptions from standard approaches, which require additional resources, for example pre-shared entanglement or measurements with external inputs~\cite{tavakoli2020self,Piveteau2022EAQCommSimpMeas,JulioVicente2017Srand}. In the absence of these resources, quantum behavior reduces to the ability to prepare and measure sets of states with superposition, namely that do not have a simplicial classical description~\cite{Streltsov_2017CoherenceReview}. Then, quantum behavior can be certified in a set of quantum states that do not admit such classical description under a fixed basis~\cite{Designolle2021CoherenceSet}.

Despite recent progress in protocols to certify quantumness via CM, several key limitations remain. First, existing experimental demonstrations are largely restricted to qubit (two-dimensional) systems~\cite{RFreeNClassPCOncept_Ma2023,Ding2024QadvQubitPM}, even though higher-dimensional systems are prevalent in nature and offer increased versatility across several fields of quantum technology \cite{forbes2025progress}. Second, while existing certification criteria rely on reproducing exact correlations displaying large quantum advantage~\cite{Heino24SimpleUnboundQAdvtg}, addressing robustness to experimental imperfections is needed for implementations. Third, current theoretical frameworks frequently depend on generalized measurements whose physical implementation requires dimensional extension of the system~\cite{Patra2024classicalanalogueofCM}, which limit practical applicability.

\begin{figure*}[htbp]
\centering
\includegraphics[width=0.8\linewidth]{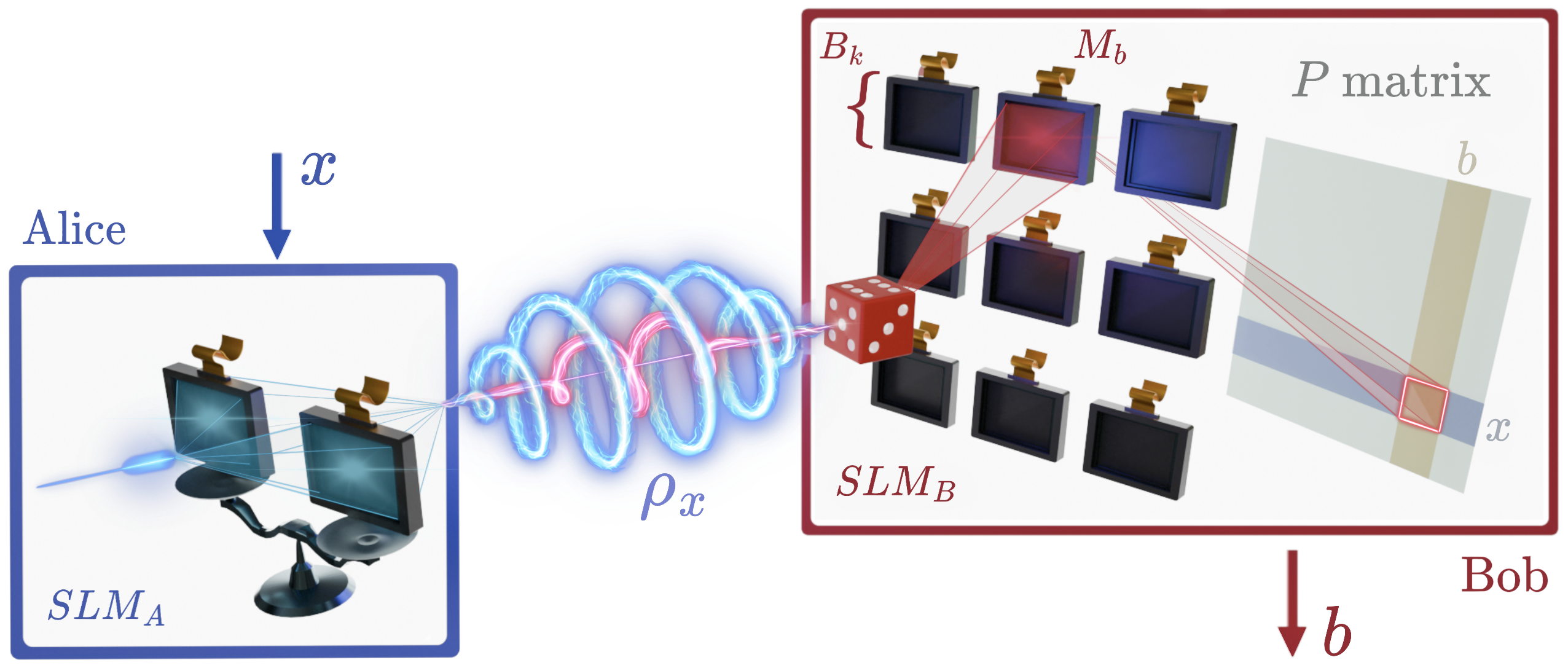}
\caption{\textbf{Conceptual sketch of the communication protocol.} 
The implemented setup decomposes the POVM into a set of weighted projectors. In the conceptual protocol experiment \textit{illustrated here for $d_Q=3$}, measuring the full probability matrix $P$ requires many iterations. In each run, given an input $x$, Alice prepares the state $\rho_x$ and sends it to Bob. Bob then randomly selects a basis ($B_k$) and a projector ($M_b$) to measure the state. Both parties use a spatial light modulator to encode or decode $\rho_x$.}
\label{fig:img-protocol}
\end{figure*}

In this work we design and experimentally implement a robust
high-dimensional 
prepare-measurement protocol to certify the ability of remote parties to prepare superpositions. 
Our setup does not make use of additional resources like ancillas for implementation of generalized measurements~\cite{RFreeNClassPCOncept_Ma2023}, shared randomness~\cite{Frenkel_2015ComMats}, entanglement~\cite{Piveteau2022EAQCommSimpMeas} or contextuality~\cite{Amselem2009SIContextSingPhot}. We use orbital angular momentum (OAM) degree of freedom of light, which offers access to a discrete and in principle unbounded Hilbert space, to demonstrate robust certification of quantum devices of dimension up to seven. We emphasize this dimension strictly as a proof of principle, as neither the protocol nor the experimental setup restricts scaling to higher dimensions, other than the limits imposed by the physical apertures of the optical channel. The proposed protocol with OAM using single photons enables compact certification of high-dimensional quantum carriers~\cite{malik2026HDopticReview} in quantum computation~\cite{imany2019HDimLogicPhotons} and communication with increased channel capacity \cite{barreiro2008beating}, security \cite{bouchard2017high}, and robustness to noise \cite{ecker2019overcoming}. This proposal is particularly applicable to quantum communication setups with limited carrier system size, in free space links and few-mode optical fibers for future long-distance communication networks.

\section*{\label{sec:Protocol} Protocol and robustness}

We consider a prepare-and-measure communication task between two parties without any shared resources, Alice and Bob~\cite{Patra2024classicalanalogueofCM,Heino24SimpleUnboundQAdvtg}. In each round, Alice encodes a classical input $x\in X$ into a physical system and sends it to Bob. Bob performs a measurement without external inputs, and obtains an output $b$. In the ideal (quantum) realization, Alice prepares quantum states $\varrho_x$ and Bob measures them with quantum measurement $\{M_b\}_b$, yielding the conditional distribution $p(b|x)=\tr(\varrho_xM_b)\,$. The communication between Alice and Bob is assumed effectively noiseless, but constrained in its bandwidth $d$. Here $d$ denotes the \emph{operational dimension}, namely the maximum number of states (in an arbitrary theory, e.g. classical or quantum) that can be perfectly distinguished by a single-shot measurement on the transmitted system. Since quantum strategies are more general than classical strategies of the same operational dimension, we will consider $d$ to be the dimension of the underlying Hilbert space.

Our protocol 
uses the \emph{weak exclusion} condition~\cite{Brun2002QSExcl,Brand2014ConclQSExcl}
\begin{equation}
    p(b=x|x)=\tr(\varrho_xM_x)=0 \quad\forall x\in X\,,
\end{equation}
i.e., for each preparation $x$ there exists a corresponding outcome $b=x$ that never occurs (see \cref{fig:img-protocol}); and the additional condition
$p(b\neq x|x)=\tr(\varrho_xM_b)\geq t>0~\forall b\neq x$, where we maximize $t$ over density matrix states with fixed measurements with semidefinite programming. 

For the target correlation $P$, we will compare the Hilbert space dimension $d_Q$ used in our protocol with the operational dimension $d_C$ that would be needed in order to reproduce $P$ by classical preparations and measurements. In the prepare-and-measure setting considered, $d_C$ coincides with the so-called \emph{non-negative rank} of $P$~\cite{fiorini2012NNrankdC,Heino24SimpleUnboundQAdvtg}, i.e.\ $d_C=\rank_+(P)$, which is the number of rank-1 entrywise nonnegative matrices $R_i$ needed to decompose $P$, as $P=\sum_{i=1}^{\rank_+}R_i$~\cite{cohen1993NNrank,Fawzi_2014LBNNrank,gillis2012geometricNNrank,shitov2014upperBoundNNrank}. Since $\rank_+(P)$ is NP-hard to compute in general~\cite{shitov2017complexityNNRank,vavasis2010complexityNNrank,Fawzi_2014LBNNrank}, we will use the lower bound
\begin{equation}
    \rank(P)\le \rank_+(P)=d_C\,,
\end{equation}
where $\rank(P)$ is the standard rank of $P$ (i.e. the number of linearly independent rows). Note that this lower bound was found earlier in~\cite{JulioVicente2017Srand} for more general setups and applies in particular to the case considered here. A \emph{quantum advantage} is then witnessed by correlations $P$ satisfying $d_C>d_Q$~\cite{Heino24SimpleUnboundQAdvtg,Heinosaari2024MaxElsQComPM}, meaning that reproducing $P$ requires a strictly larger operational dimension classically than with quantum strategies. A sufficient condition for quantum certification is thus
\begin{equation}\label{eq:rank>dQ}
    \rank(P)>d_Q\,,
\end{equation}
assuming that no more than $\rank(P)$ operational dimensions are allowed by either of the systems nor the carrier channel. 

Although most existing works focus on targeting specific distributions $P$~\cite{Ding2024QadvQubitPM,Heinosaari2024MaxElsQComPM,Chakraborty2025AdvQubitComPM,doolittle2025OpFrQComNettsPM,Heino24SimpleUnboundQAdvtg}, 
for the purpose of experimental implementation we need to consider that the state correlations are subject to noise. As such, we require to ensure that Eq.~\eqref{eq:rank>dQ} is satisfied for the noisy distribution $P'$ obtained in the experiment. 
For that it is convenient to consider the error matrix $E := P' - P$.
As detailed in \cref{sec:Suppl-Noise} of the Supplementary Information, a sufficient condition to certify stability of $\rank(P')$ is
\begin{equation}\label{eq:rank_condition}
    \|E\|_2<\sigma_r(P)\,,
\end{equation}
where $\|\,\cdot\,\|_2$ is the spectral norm and $\sigma_r(P)$ is the $r$-th singular value of $P$ (ordered non-increasingly). Conceptually, our strategy is thus to reproduce target correlations whose $r$-th singular value is larger than the potential deviations due to the noise effects, so that the rank of the matrix $P$ is assured to be at least the target value $r$. We thus consider the experimental figure of merit
\begin{equation}
  d^{\mathrm{exp}}_C:=\max\big\{\, r : \|E\|_2<\sigma_r(P)\,\big\},
\end{equation}
so that $d_C\ge d^{\mathrm{exp}}_C$ even when experimental noise is considered. 
We refer to $d^{\mathrm{exp}}_C$ as the \emph{dimension certification bound}.

\section*{State preparations and measurements}

\begin{figure*}[tbp]
    \vspace{0.1cm}
    \centering
    \includegraphics[width=0.88\linewidth]{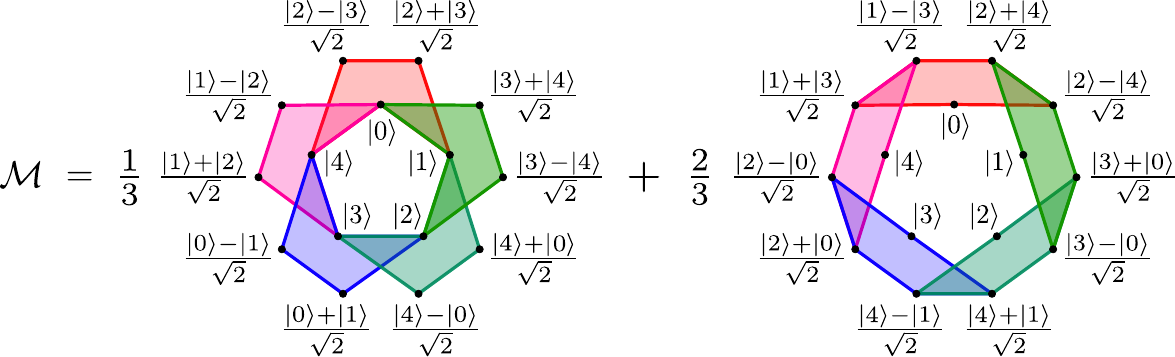}
    \caption{{\bf Decomposition of the optimal measurement} $\mathcal{M}$ in Eqs.~\eqref{eq:measdiag},~\eqref{eq:meas+} and~\eqref{eq:meas-} into projective measurements, for dimension $d=5$.
    When Bob receives Alice's message, he performs a projective measurement (Strategy A) on a randomly chosen basis among five possibilities $B_1,\dots,B_5$. Each basis is represented by a colored polygon whose vertices correspond to the basis projectors. A second projective measurement family (Strategy B) is defined by choosing randomly among five bases containing the same projectors but with different repetitions. The statistics of the target POVM $\mathcal{M}$ are obtained as a convex combination of the statistics in both strategies, so that the protocol yields the probability matrix $P$ using projective measurements only (see Appendix~\ref{appSubsec:DecPOVMs} for further details). In this way, Alice and Bob can perform the protocol using a five-dimensional quantum system via projective measurements, while requiring a 10-dimensional system to simulate it classically.}
    \label{fig:PVMd=5}
    \vspace{0.2cm}
\end{figure*}

In our protocol, depicted in~\cref{fig:PVMd=5} for dimension $d_Q=5$, Bob's measurement is a $3d_Q$-outcome POVM $\mathcal{M}=\{M_b\}_{b=0}^{3d-1}$ on $\mathbb{C}^{d}$ defined, for $k=0,\dots,d-1$, by
\begin{align}
M_k &= p\,\dyad{k}\label{eq:measdiag}\\
M_{d_Q+k} &= \frac{1-p}{2}\,\frac{\ket{k}+\ket{k\oplus 1}}{\sqrt{2}}\frac{\bra{k}+\bra{k\oplus 1}}{\sqrt{2}},\label{eq:meas+}\\
M_{2d_Q+k} &= \frac{1-p}{2}\,\frac{\ket{k}-\ket{k\oplus 1}}{\sqrt{2}}\frac{\bra{k}-\bra{k\oplus 1}}{\sqrt{2}}
\label{eq:meas-}\,,
\end{align}
where $0\leq p\leq 1$ and $\oplus$ denotes addition modulo $d_Q$. The parameter $p$ controls the balance between incoherent (computational) and coherent (neighbor superposition) projections. Each state preparation $\rho_x$ on Alice's system is a probabilistic mixture $\rho_x=\sum_{M_b\perp M_x}\lambda_{b,x}M_b$ of measurements $M_b$ orthogonal to $M_x$, $\tr(M_xM_b)=0$, where the probabilities $\lambda_{b,x}$ are optimized with semidefinite programming to maximize the smallest off-diagonal probabilities $P_{x,b\neq x}$. 

Experimentally, we implement $\mathcal{M}$ by decomposing it into a convex combination of projective measurements on randomly selected orthonormal bases. Concretely, Bob draws a basis label and performs a projective measurement (PVM); the resulting statistics reproduce the POVM outcome probabilities up to known rescaling factors (the prefactors in \cref{eq:measdiag} to \cref{eq:meas-}).
For $d=3$, the choice $p=1/3$ is implemented directly. For $d=5$ and $d=7$, we realize two decompositions (see Fig. \ref{fig:PVMd=5}):
\begin{itemize}
    \item[\textit{i)}] \emph{Strategy A:} incoherent projections appear $d-2$ times and coherent projections appear once,
    giving $p=3/5$ for $d=5$ and $p=5/7$ for $d=7$;
    \item[\textit{ii)}] \emph{Strategy B:} incoherent projections appear once and coherent projections appear $(d-1)/2$ times,
    giving $p=1/5$ for $d=5$ and $p=3/7$ for $d=7$.
\end{itemize}
The ideal $p=1/3$ is then achieved by a convex combination of these two strategies. For both $d=5$ and $d=7$, this is achieved by mixing $1/3$ of Strategy~A with $2/3$ of Strategy~B. Given the fixed measurement $\mathcal{M}$, Alice's mixed preparations $\{\varrho_x\}$ are obtained via semidefinite programming so to (i) minimize the probability of correlated events, $\tr(\varrho_x M_x)=0$, and (ii) maximize the smallest probability of non-correlated events, namely maximize $t$ such that $\tr(\varrho_x M_b)\geq t$ for all $x\neq b$.
For later reconstruction, we express all vectors in the computational basis $\{\ket{0},\dots,\ket{d-1}\}$. 

Note that the overall implementation of $\mathcal{M}$ requires only a classical \emph{flag state} (corresponding to random selection of orthonormal bases). This reduces the experimental overhead compared to the implementation of Trine \cite{RFreeNClassPCOncept_Ma2023} or SIC-POVMs \cite{Bent2015}, which require a quantum ancilla {\it i.e.} an extended Hilbert space, for practical implementation. 
We highlight that, although incompatible measurements are used to decompose the target POVM into projective measurements, the statistics are obtained by averaging over different choices. This implies that no input for Bob is required, in contrast to most existing protocols employing multiple measurements~\cite{Gallego2010DVIwitCQdims,Pawlowski2011SDVIsecureOne-Way,tavakoli2020self,puliyil2025SDVIchannnelCommMats,Halder2025QAUnified}. Our protocol does indeed not use contextuality, since one can assign deterministic labels to all projectors appearing in the measurements consistently (as in Fig.~\ref{fig:PVMd=5}).

\section*{\label{sec:Results} Implementation \& Experimental Setup}

Alice prepares and sends states $\rho_x$ to Bob. For each received state, Bob (conceptually) selects a measurement basis $B_k$, each containing a set of projectors $\{M_b\}$ used to measure the state (see \cref{fig:img-protocol}). 
In the experiment, the POVM elements are realized by a numerically specified, weighted set of projectors. Each target density matrix state $\rho_x$ describes a  probabilistic ensemble of two-mode OAM superposition states, and is given by their weighted convex combination. Operationally, Alice transmits the two constituent OAM superpositions sequentially (rather than transmitting their weighted sum in a single shot). Bob measures each incoming component by projecting onto the desired OAM superposition $M_b$. There is no physical random basis choice during acquisition: all relevant overlaps are measured, and the required classical randomness and weights (Alice's preparation weights and Bob's intended basis-selection probabilities) are applied \textit{a posteriori}. This is operationally equivalent to Bob performing the intended POVM $\{M_b\}$ on the original state $\rho_x$. 
From the raw data in the $49\times 49$ matrix shown in \cref{sec:data-49}, of measured overlaps between the states $\ket{M_b}$ from \cref{eq:measdiag} to \cref{eq:meas-}, we first select the entries corresponding to the logical $d$-dimensional qudit subspace used by the protocol. These selected overlaps, together with the numerical weights defining the superpositions $\ket{M_b}$, are then combined to reconstruct the conditional probabilities and hence the matrix $P$.

The experimental results are obtained by post-processing projective-measurement data to reconstruct the protocol probability matrices $P'$. 
Although we emphasize $d=5$ in the main text, we also certify quantumness for $d=3$ and $d=7$ (see \cref{sec:Ap0}). In all cases, the required state preparations and projections are drawn from a sample of a global tomographically complete set of $49\times49$ overlap measurements spanning pairwise superpositions of two OAM modes within the seven charges $\ell\in\{-3,-2,\ldots,3\}$. After reconstructing $P'$, we quantify the agreement with the ideal prediction $P$ through the spectral-norm deviation. The protocol is designed to enhance diagonal contrast and increase the certified classical dimension, without increasing the number of physical states. Concretely, it implements a convex combination of two projective-measurement strategies (PVM$_1$ and PVM$_2$), each using the minimal set of $n=3d_Q$ states (\cref{fig:PVMd=5}). Operationally, this corresponds to a systematic redistribution of repetitions within the projective bases (Bob effectively samples from a ``richer die''), yielding a non-projective POVM while keeping the same underlying projectors. 

The dimension certification is obtained from the rank-stability criterion, by assuring that all experimental matrices $P'$ obtained by the error propagation describing our setup whose error deviation from $P$ satisfies $\|E\|_2<\sigma_r(P)$, will be subject to the lower bound $\mathrm{rank}(P')\ge r$, and therefore must satisfy $d_C\ge r$. We summarize the robust certification results in \cref{tab:robust-cert} by reporting the experimentally certified dimension $d^{\mathrm{exp}}_C$ and the corresponding bound $\sigma_{d^{\mathrm{exp}}_C}(P)$. We also consider an alternative maximal POVM that uses additional experimental resources---increasing the total number of states ($n$) from $3d_Q$ to $d_Q^2$---to therefore achieve certification for higher dimensions (boosting from $2d_Q$ to $d_Q(d_Q+1)/2$). Other protocols that prioritize coherent over incoherent measures, or vice versa, are detailed in Supplementary Information \cref{sec:Approaches}.

\begin{table}[tbp]
  \centering
\caption{\textbf{Robust quantum certification.}
For each protocol, the ideal correlation matrix of size $n\times n$ is constructed with a device of quantum dimension $d_Q$. We compute the spectral-norm deviation $\|E\|_2$ and define
$
d^{\mathrm{exp}}_C = \max\{ r : \|E\|_2 < \sigma_r(P) \}
$ as an experimental figure of merit: by Weyl's inequality, we have $\mathrm{rank}(P') \ge d^{\mathrm{exp}}_C$ and thus any classical distribution requires $d_C \ge d^{\mathrm{exp}}_C$.
We report the certification bound $\sigma_{d^{\mathrm{exp}}_C}(P)$ and the next singular value $\sigma_{d^{\mathrm{exp}}_C+1}(P)$, which lies below $\|E\|_2$ and marks the point where certification stops. The red-boxed results represent a fundamental improvement over existing protocols, since no alternative approach (e.g., contextuality-based methods) achieves linear scaling. Theoretical certification is achieved for dimension up to $d_C=2d_Q$, and the certified value  $d^{\mathrm{exp}}_C$ is smaller due to experimental noise. 
In the last three rows we achieve certification for higher classical dimension $d^{\mathrm{exp}}_C=d_C=d_Q(d_Q+1)/2$ at the expense of using a larger number $n$ of preparations and measurements.
Note that in the $d^{\mathrm{exp}}_C=12$ case, the certification is upheld with a statistical significance of $14\sigma$.}
  \label{tab:robust-cert}

  \setlength{\tabcolsep}{3pt}
  \renewcommand{\arraystretch}{1.4}

\resizebox{0.485\textwidth}{!}{%
\begin{tabular}{|c c c|c c c|}
  \hline
   $n$ & $d_Q$ & $d^{\mathrm{exp}}_C$ & $\sigma_{d^{\mathrm{exp}}_C}(P)$ & $\|E\|_2$ & $\sigma_{d^{\mathrm{exp}}_C+1}(P)$ \\
  \hline
  \tikzmarknode{TL}{\strut 9} & 3 &  6 &
  $2.00\,\text{\footnotesize$\times 10^{-1}$}$ &
  $(6.57\pm0.29)\,\text{\footnotesize$\times 10^{-2}$}$ &
  $9\,\text{\footnotesize$\times 10^{-17}$}$ \\
     15 & 5 &  8 &
  $1.24\,\text{\footnotesize$\times 10^{-1}$}$ &
  $(5.87\pm0.13)\,\text{\footnotesize$\times 10^{-2}$}$ &
  $4.00\,\text{\footnotesize$\times 10^{-2}$}$ \\
     21 & 7 & 12 &
  $8.10\,\text{\footnotesize$\times 10^{-2}$}$ &
  $(6.41\pm0.12)\,\text{\footnotesize$\times 10^{-2}$}$ &
  \tikzmarknode{BR}{\strut $2.00\,\text{\footnotesize$\times 10^{-2}$}$} \\
  \hline
      9 & 3 &  6 &
  $6.67\,\text{\footnotesize$\times 10^{-2}$}$ &
  $(2.21\pm0.09)\,\text{\footnotesize$\times 10^{-2}$}$ &
  $2\,\text{\footnotesize$\times 10^{-17}$}$ \\
     25 & 5 & 15 &
  $1.44\,\text{\footnotesize$\times 10^{-2}$}$ &
  $(1.22\pm0.02)\,\text{\footnotesize$\times 10^{-2}$}$ &
  $2\,\text{\footnotesize$\times 10^{-17}$}$ \\
     49 & 7 & 28 &
  $1.061\,\text{\footnotesize$\times 10^{-2}$}$ &
  $(1.045\pm0.007)\,\text{\footnotesize$\times 10^{-2}$}$ &
  $2\,\text{\footnotesize$\times 10^{-17}$}$ \\
  \hline
\end{tabular}
}
\begin{tikzpicture}[overlay,remember picture]
  \draw[red,line width=1pt]
    ($(TL.north west)+(-\tabcolsep-\arrayrulewidth/2-1.7pt, \arrayrulewidth/2+4.1pt)$)
    rectangle
    ($(BR.south east)+(\tabcolsep+\arrayrulewidth/2+8.8px,-\arrayrulewidth/2pt-1.1pt)$);
\end{tikzpicture}
\end{table}

\begin{figure*}[tbp]
\centering
\includegraphics[width=0.95\linewidth]{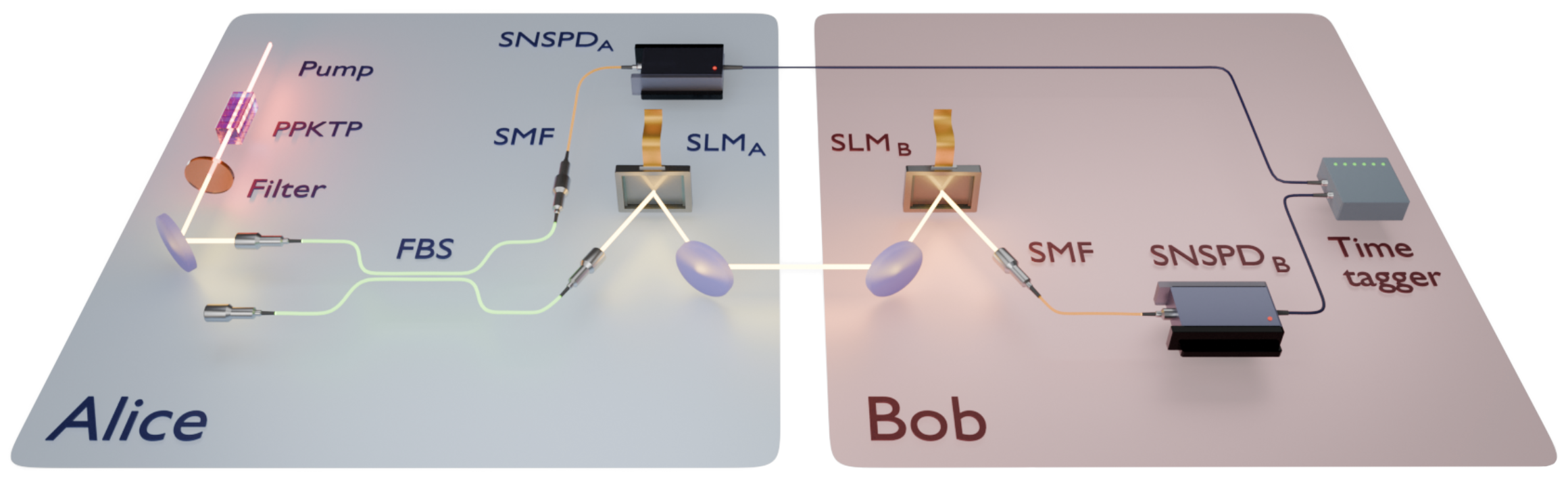}
\caption{\textbf{Simplified scheme of the setup.}  PPKTP: periodically poled potassium titanyl phosphate (non-linear crystal); FBS: fibre beam splitter; SLM: spatial light modulator; SMF: single-mode fibre; SNSPD: superconducting nanowire single-photon detector. The subscripts A and B stand for Alice and Bob.}
\label{fig:setup-simple}
\end{figure*}

The prepare-and-measure implementation is depicted in \cref{fig:setup-simple}. A photon-pair source is realized via type-0 spontaneous parametric down-conversion (SPDC) in a 5\,mm-long periodically poled potassium titanyl phosphate (PPKTP) crystal. After spectral filtering to reject residual pump light, the down-converted photons at 1550\,nm are coupled into a single-mode fibre beam splitter (FBS), yielding temporally correlated signal and idler photons in the fundamental spatial mode ($\ell=0$). The idler photon serves as a trigger and is detected directly on a superconducting nanowire single-photon detector (SNSPD). The signal photon is directed to Alice’s spatial light modulator (SLM$\text{A}$) for state preparation, where a hologram encodes either a single OAM mode or a superposition of OAM modes. It then propagates to Bob’s spatial light modulator (SLM$\text{B}$), which performs the state projection. Finally, the photon is coupled into a single-mode fibre (SMF) and detected by a second SNSPD, while coincidences between the signal and idler detections are recorded by a time-tagging unit.
A projection onto the exact phase-conjugate of the prepared mode (e.g.\ $\ell_{\rm prep}=+1$ and $\ell_{\rm meas}=-1$) cancels the OAM phase structure and yields efficient coupling into the SMF (which supports only the fundamental mode), producing maximal coincidence counts. Partial overlap between preparation and projection reduces the coupling efficiency and hence the coincidence rate, providing direct access to the desired overlap probabilities. Additional implementation details are given in \cref{sec:s-details}, and alignment procedures in \cref{sec:s-align} of the Supplementary Information.

\bigskip

\section*{Discussion}

The protocol is intrinsically dimension-scalable: for any target qudit dimension $d_Q$, one constructs an anticorrelated witness matrix $P$ from a $3d_Q$-outcome POVM implemented as a convex combination of projective measurements. Experimentally, the same post-processing principle applies across dimensions, and in this work we demonstrate certification for $d=3,5,7$ (see Supplementary Information \cref{sec:Ap0}). Extending to larger $d_Q$ is limited primarily by mode quality and stability of the high-dimensional OAM superpositions, rather than by the structure of the certification criterion. 
In our protocol, quantumness is embedded in the discrete OAM Hilbert space of single photons, where the information-processing capabilities are subject to superposition, measurement back-action, and the non-cloning ~\cite{Wootters1982}. These features cannot be replicated within a classical-wave description and are essential for the security guarantees of the scheme. Operating in the single-photon regime, heralded by coincidence detection rather than single-detector counts, provides an additional layer of security.  This guarantees that the certified behavior genuinely relies on irreducible quantum resources rather than classical optical interference.

Direct single-shot implementation of our three-outcome POVM (rather than reconstructing it from projective measurements) would require an interferometric decomposition into partial projections and path-dependent unitaries, analogous to the variational triangular polarimeter demonstrated for polarization in~\cite{Ding2024QadvQubitPM}, and in principle translatable to OAM qudits (see Supplementary Information \cref{sec:s-future}).

To certify quantumness in larger dimensional or more noisy systems, we optionally suggest a final symmetry-based post-processing to reduce the overall error $\|E\|_2$. Since Alice and Bob may freely identify the logical labels $(q)$ with the OAM labels $(\ell)$, the reconstructed matrix $P'$ can be averaged over an appropriate family of relabelings to dilute configuration-dependent systematic effects (e.g.,\ SLM pixel discretization, fixed small misalignments, or fixed beam-size differences). Details are provided in Supplementary Information \cref{sec:s-perms}.

In the present implementation, the POVM is realized through acquisition of overlaps followed by \textit{a posteriori} weighting, which is operationally equivalent to randomized basis choice. A natural next step is to investigate real-time basis sampling and faster acquisition strategies, as well as the impact of finite coincidence windows and temporal-mode effects on the reconstructed anticorrelations.

\section*{Conclusions}
We have introduced and experimentally demonstrated a robust method to certify high-dimensional quantum behaviour in a prepare-and-measure scenario under an operational-dimension constraint. The certification uses exclusion correlations that generate communication matrices with large rank, and it is made resilient to experimental imperfections by a rank-stability criterion that certify a lower bound for the classical communication dimension $d_C\ge d^{\mathrm{exp}}_C$. 

Experimentally, we implement the protocol with single photons encoded in OAM and reconstruct the communication matrices from a tomographically complete overlap dataset. For the first protocol employing $n=3d_Q$ outcomes, we certify $d_C$ beyond $d_Q$ up to $d_Q=7$; for the second protocol employing $n=d_Q^2$ outcomes, we certify substantially larger classical dimensions (see \cref{tab:robust-cert}). These results establish a practical route to certify non-classical prepare-and-measure devices in a high-dimensional communication framework without requiring entanglement distribution or full device tomography.


\begin{acknowledgments}
We are thankful to Riccardo Castellano, Carlos de Gois, Marc Olivier Renou, Jef Pauwels, Martin Plávala, Alejandro Pozas-Kerstjens, Armin Tavakoli, Lucas Tendick and Marco Túlio Quintino for comments and discussions. 
This work was partially funded by CEX2024-001490-S [MICIU/AEI/10.13039/501100011033]. AR acknowledges financial support
by the Deutsche Forschungsgemeinschaft (DFG, German
Research Foundation), project number 563437167 and Project BeRyQC, Grant No.
13N17292). We acknowledge support from  Spanish MICIN (project: PID2022:139099NBI00) with the support of FEDER funds, the Spanish Goverment with funding from European Union NextGenerationEU (PRTR-C17.I1), the Generalitat de Catalunya, the Ministry for Digital Transformation and of Civil Service of the Spanish Government through the QUANTUM ENIA project -Quantum Spain Project- through the Recovery, Transformation and Resilience Plan NextGeneration EU within the framework of the Digital Spain 2026 Agenda. DV acknowledges funding from MCIN (MCIN/AEI/10.13039/501100011033, projects~PID2023-149988NB-C21;PID2024-160393NB-I00) and funding from the European Union NextGenerationEU (PRTR-C17.I1). AV acknowledges financial support from the Ram\'on y Cajal Fellowship RYC2023-043066-I, and Grant Nos. PID2024:160393NB-I00;156240OB-C22 funded by MICIU/AEI/10.13039/501100011033 and FSE+. SSB acknowledges financial support from the Beatriu de Pinós programme of the Ministry of Research and Universities of the Government of Catalonia (grant number 2024 BP 00267).

\end{acknowledgments}

\section*{Data Availability Statement}
The data that support the findings of this study are available from the corresponding authors upon reasonable request.

\section*{References}
\bibliography{mybibfile}

\onecolumngrid

\newpage
\section*{\label{sec:SI} Supplementary Information}

\subsection{Decomposition of the POVMs}\label{appSubsec:DecPOVMs}
\begin{figure}[H]
    \vspace{0.3cm}
    \centering
    \includegraphics[width=0.9\linewidth]{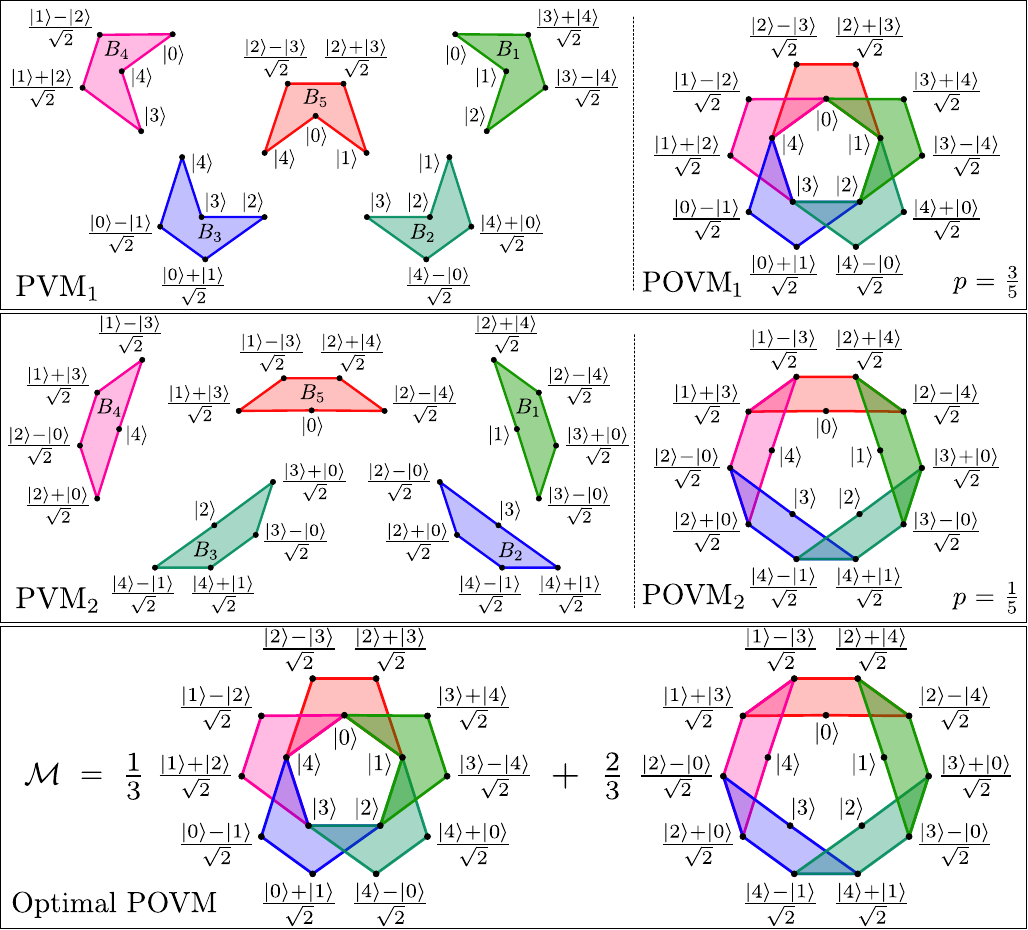}
    \caption{{\bf Decomposition of the optimal measurement $\mathcal{M}$ into projective measurements, for dimension $d=5$.} In the first rectangle (left) we depict the five possible bases that Bob can choose where to perform a projective measurement in strategy A (POVM$_1$). Each colored polygon describes a different basis. These overlap at points where different bases share the same basis vectors, which leads to their combination depicted in the right side of the top rectangle. The statistics considering repetition of overlapping events give rise to the POVM of Eqs~\eqref{eq:measdiag},~\eqref{eq:meas+} and~\eqref{eq:meas-} with $p=3/5$. The middle rectangle represents a similar procedure to reproduce strategy B: projective measurements represented in colored polygons, with overlapping outcomes, combine together into the measurement in Eqs~\eqref{eq:measdiag},~\eqref{eq:meas+} and~\eqref{eq:meas-} with $p=1/5$, leading to strategy B (POVM$_2$). At the end, the statistics are obtained by a convex combination of the statistics resulting from each strategy (bottom rectangle).}
    \label{fig:PVMd=5Appendix}
    \vspace{0.2cm}
\end{figure}

The procedure to obtain the probability matrix $P$ consists of decomposing the POVM quantum measurement in Eqs~\eqref{eq:measdiag},~\eqref{eq:meas+} and~\eqref{eq:meas-} into two different POVMs (POVM$_1$ and POVM$_2$), each of which admits a decomposition into projective measurements. The procedure is described graphically at Fig.~\ref{fig:PVMd=5Appendix}.

\subsection{Noise analysis}\label{sec:Suppl-Noise}
We quantify the deviation between the ideal witness matrix $P$ and its experimental reconstruction $P'$ through the (spectral) $2$-norm of the error:
\begin{equation}
E:=P'-P,
\qquad
\|E\|_2=\sigma_1(E).
\end{equation}
This section provides (a) a perturbation bound connecting $\|E\|_2$ to the singular values of $P$ and $P'$, and (b) a first-order uncertainty propagation rule to attach a standard uncertainty to $\|E\|_2$ from entrywise uncertainties of $E$.

\subsubsection*{Perturbation bound}

Let $\sigma_1(\cdot)\ge\sigma_2(\cdot)\ge\cdots\ge0$ denote singular values in non-increasing order. Weyl's inequality yields, for every index $r$:
$\big|\sigma_r(P')-\sigma_r(P)\big|
\le \sigma_1(P'-P)=\|E\|_2.$

If $\exists~M\in\mathbb{R}^+$ such that $\big|\sigma_r(P')-\sigma_r(P)\big|\leq M<\sigma_r(P)$ then $\mathrm{rank}(P')\ge r$.
\begin{proof}
If $\sigma_r(P')<\sigma_r(P)$, then we have:
$-\sigma_r(P')+\sigma_r(P)\leq M
\implies    
-\sigma_r(P')\leq M - \sigma_r(P)
\implies    
\sigma_r(P) - M \leq \sigma_r(P')
$.\\

But since $ M<\sigma_r(P)\implies 0<\sigma_r(P)-M$, and we can conclude that:
$
0 < \sigma_r(P')
\implies
\mathrm{rank}(P')\ge r
$.\\

The case $\sigma_r(P)\leq\sigma_r(P')$ is direct, since $0<M<\sigma_r(P)\implies0<\sigma_r(P')\implies
\mathrm{rank}(P')\ge r$.
\end{proof}

Applying Weyl with $M=\|E\|_2$, we obtain the practical certification condition:
\begin{equation}\label{eq:cert-cond-sigr}
\|E\|_2 < \sigma_r(P)
\implies
\mathrm{rank}(P')\ge r.
\end{equation}
Finally, since $\mathrm{rank}_+(P')\ge \mathrm{rank}(P')$, the above implies the classical dimension lower bound $d_C=\mathrm{rank}_+(P')\ge r$.

\subsubsection*{Uncertainty propagation}

In practice, $E$ is obtained experimentally, and each entry $E_{ij}$ comes with a standard uncertainty $(s_E)_{ij}$. Our goal is to propagate these entry-wise uncertainties to an uncertainty on $y=\|E\|_2$,
i.e. on the largest singular value of $E$.\\

Let $E=U\Sigma V^{\mathsf T}$ be an SVD, with singular values $\sigma_1(E)\ge \sigma_2(E)\ge \cdots \ge 0$ in $\Sigma$ and corresponding left/right singular vectors $u_k$ and $v_k$ (columns of $U$ and columns of $V$). We assume the dominant singular value is simple $\sigma_1(E)>\sigma_2(E)$, which implies that the map $E\mapsto \sigma_1(E)=\|E\|_2$ is differentiable at $E$ and has a unique (Fréchet) derivative and gradient at that point.\\

Stewart and Sun show that, for a simple singular value $\sigma_i$ of $A$ and a perturbation $\tilde A = A+A_1$, so that: 
$\tilde\sigma_i~=~\sigma_i~+~u_i^{\mathsf T}A_1v_i~+~O(\|A_1\|^2),
$\cite[p.~265, after eq.~(4.13)]{stewart1990mpt} ~i.e.\ the linear term is $u_i^{\mathsf T}A_1v_i$. Equivalently, Horn and Johnson show that, if $\sigma_1$ is a simple nonzero singular value of $A$ with corresponding left/right singular vectors $u_1$ and $v_1$ (in general $Au_k=\sigma_k v_k$), then \cite[p.~453, eq.~7.3.12]{horn2012matrix}:
\[
\left.\frac{\mathrm d}{\mathrm dt}\sigma_1(A+tA_1)\right|_{t=0}=\Re\!\left(u_1^{\ast}A_1v_1\right),
\quad\text{i.e.}\quad
\mathrm d\sigma_1=\Re\!\left(u_1^{\ast}(\mathrm dA)v_1\right)
\ \, \therefore \ \ \text{for $A$}\in\mathbb{R}^{n\times m} \ \ \mathrm d\sigma_1=u_1^{\mathsf T}(\mathrm dA)v_1.
\]
Where $A+tA_1$ is the 1-parameter perturbation of $A$, since $t\in\mathbb{R}$.\\

To match the notation, we identify
$A \longleftrightarrow E$ and
$A_1 \longleftrightarrow \Delta E$, to consider the 1-parameter perturbation $E+t\,\Delta E$. Under the simplicity assumption, Horn--Johnson's formula gives the directional derivative at $t=0$ in the direction $\Delta E$.\\

In the real case, we obtain the Fréchet derivative:
\begin{equation}
Dy(E)[\Delta E]=u_1^{\mathsf T}(\Delta E)v_1
=\langle u_1v_1^{\mathsf T},\,\Delta E\rangle_F=\sum_{i,j}(u_1v_1^{\mathsf T})_{ij}\,(\Delta E)_{ij}
=
\sum_{i,j}(u_1)_i\,(v_1)_j\,(\Delta E)_{ij},,
\end{equation}
where $\langle A,B\rangle_F:=\sum_{i,j}A_{ij}B_{ij}=\mathrm{Tr}(A^{\mathsf T}B)$ is the Frobenius inner product. \\

We identify the partial derivatives with respect to each entry $E_{ij}$:
\begin{equation}\label{eq:grad-norm2}
\frac{\partial \|E\|_2}{\partial E_{ij}}=(u_1)_i\,(v_1)_j,
\qquad\text{i.e. }
\nabla_E\|E\|_2 = u_1v_1^{\mathsf T}.
\end{equation}
Intuitively: entries of $E$ aligned with large components of $u_1$ and $v_1$ influence $\|E\|_2$ the most (to first order).\\

The singular vectors in \eqref{eq:grad-norm2} are exactly the leading vectors returned by an SVD of $E$:
\begin{verbatim}
U, S, VT = np.linalg.svd(E, full_matrices=False); u1 = U[:, 0]; v1 = VT[0, :]
\end{verbatim}
Here \texttt{S[0]} is $\sigma_1(E)$, \texttt{U[:,0]} is $u_1$, and \texttt{VT[0,:]} is $v_1^{\mathsf T}$ (so \texttt{v1} stores the components of $v_1$ as a 1D array). With these, the gradient matrix is implemented by \texttt{np.outer(u1, v1)}, which forms $u_1v_1^{\mathsf T}$.

We now treat the entries of $E$ as the inputs: $x_k \ \longleftrightarrow \ E_{ij}$,
$u(x_k)\ \longleftrightarrow\ (s_E)_{ij}$ and $
y=f(E)=\|E\|_2$. 
The NIST first-order law of propagation of uncertainty\cite{taylor2009guidelines} states that for a scalar output $y=f(x_1,\dots,x_p)$,
\begin{equation}\label{eq:nist-A3}
u_c^2(y)=\sum_k\Big(\frac{\partial f}{\partial x_k}\Big)^2u^2(x_k)
+2\sum_{k<\ell}\frac{\partial f}{\partial x_k}\frac{\partial f}{\partial x_\ell}\,u(x_k,x_\ell).
\end{equation}
In our case, the index $k$ is simply a relabeling of the pair $(i,j)$, so the first sum runs over all matrix entries.\\

Assuming entrywise independence, i.e. $u(E_{ij},E_{kl})=0\quad \text{for }(i,j)\neq(k,l)$,
the covariance term (the double sum in \eqref{eq:nist-A3}) vanishes. Using the sensitivities from \eqref{eq:grad-norm2} then yields the first-order approximation
\begin{equation}\label{eq:var-norm2E}
\mathrm{Var}(\|E\|_2)\approx
\sum_{i,j}\Big((u_1)_i (v_1)_j\Big)^2\,(s_E)_{ij}^2,
\qquad
\mathrm{Std}(\|E\|_2)=\sqrt{\mathrm{Var}(\|E\|_2)}.
\end{equation}
This is exactly the usual propagation of uncertainty rule, for the case of $f(E)=\|E\|_2$. And can be implemented by forming the gradient matrix
$\nabla_E\|E\|_2=u_1v_1^{\mathsf T}$ and performing an elementwise weighted sum. Finally, using \eqref{eq:cert-cond-sigr} we get a statistically robust certification threshold.

\subsection{\label{sec:data-49} Unprocessed experimental data}
\begin{figure}[H]
\centering
\includegraphics[width=0.48\linewidth]{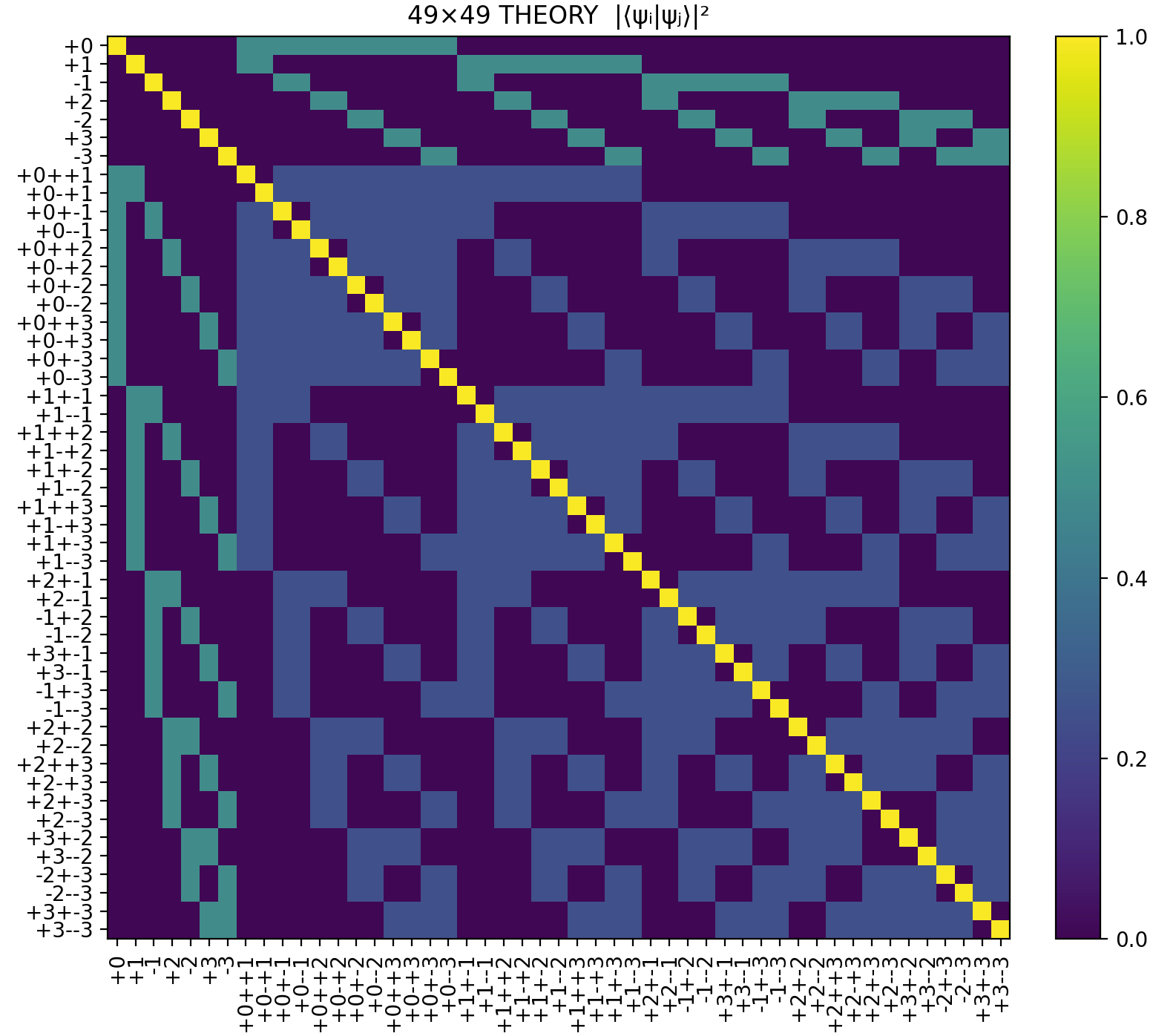}
\hspace{0.3cm}
\includegraphics[width=0.48\linewidth]{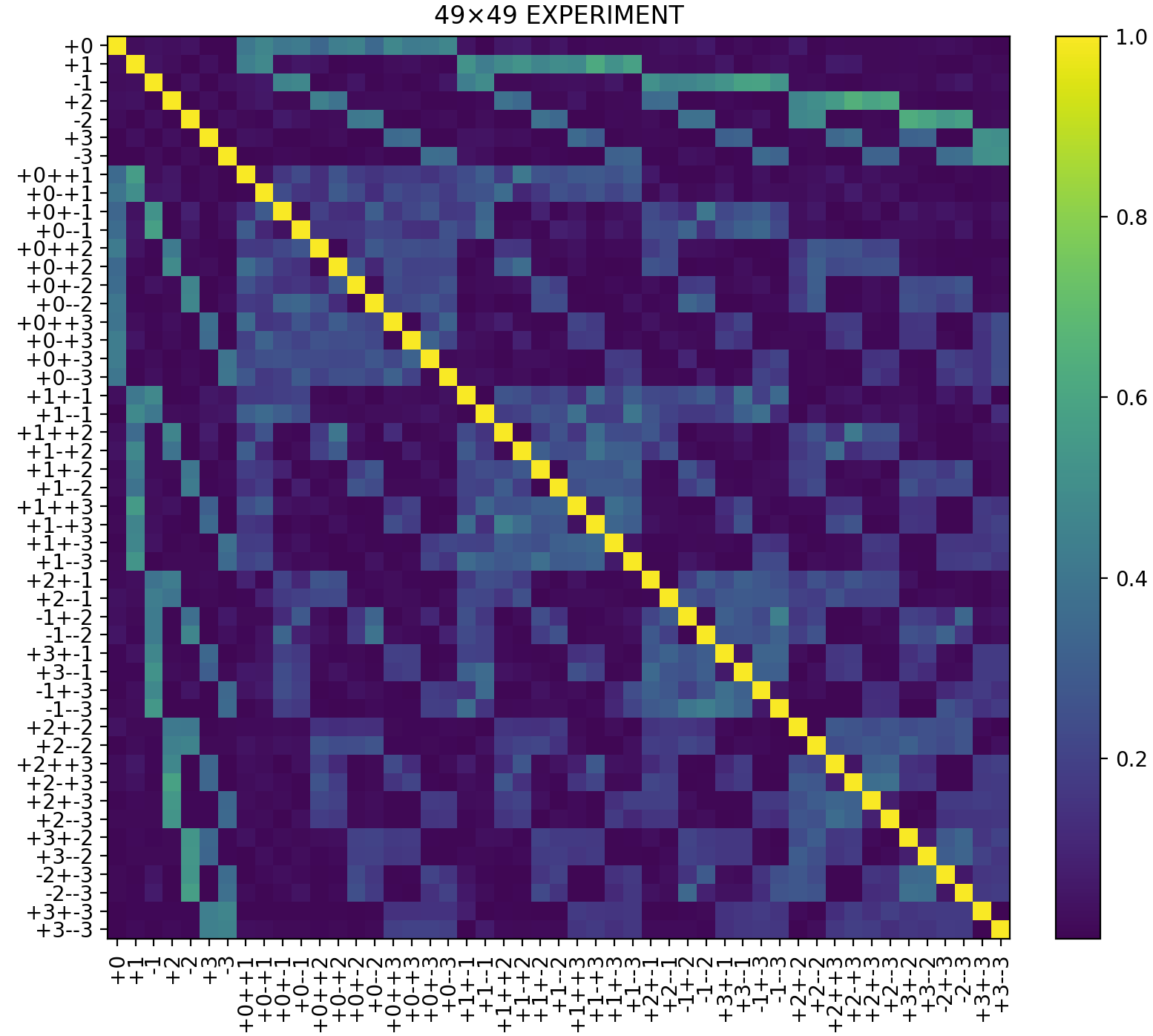}
\caption{\textbf{Raw data for the protocol.} Theoretical (left) and experimental (right) overlap matrices for the states $\ket{M_j}$ defined in Eqs. 5-7, with prepared OAM superpositions as columns and measurement superpositions as rows. Labels denote equal‑weight superpositions of two OAM eigenstates: for example, \texttt{+1+-2} represents $(\ket{\ell=+1}+\ket{\ell=-2})/\sqrt{2}$, here the leading signs indicate the $\ell$ indices (not relative phases). 
}
\label{fig:raw}
\end{figure}

From the $49\times49$ coincidence‑count matrix (\cref{fig:raw}), we extracted the protocols for each Hilbert‑space dimension $d$. For each $d$, we selected the OAM labeling that minimizes the parameter $\|E\|_2$ between theory and experiment, defined as the spectral norm of the error matrix, $\|E\|_2=\|P'-P\|_2$ (see \cref{sec:Suppl-Noise}). Consequently, the theoretical qudit labels are defined only up to a permutation of the OAM indices—i.e., any theoretical qudit basis state may correspond to any OAM mode under the optimal labeling.

\subsection{\label{sec:Approaches} Different protocol approaches}
To contextualize our design choices, \cref{tab:protocol-variants} summarizes the four protocol variants considered.

\begin{table}[!h]
\centering
\small
\renewcommand{\arraystretch}{1.2}
\setlength{\tabcolsep}{2pt}
\begin{tabular}{@{}llll@{}}
\hline
\parbox[t]{0.14\linewidth}{\raggedright \textbf{Variant}} &
\parbox[t]{0.25\linewidth}{\raggedright \textbf{Use when}} &
\parbox[t]{0.26\linewidth}{\raggedright \textbf{Pros}} &
\parbox[t]{0.22\linewidth}{\raggedright \textbf{Cons}} \\
\hline
\parbox[t]{0.14\linewidth}{\raggedright \textbf{``Convex''}} 
&
\parbox[t]{0.25\linewidth}{\raggedright No preference between two–mode superpositions and single-mode states, and you want to keep the basis size minimal.}
&
\parbox[t]{0.26\linewidth}{\raggedright Clearer zero-diagonal than Approaches~1 and~2 overall; retains the smallest basis size; operationally simple (reordering within bases).}
&
\parbox[t]{0.22\linewidth}{\raggedright Requires acquiring and mixing two PVMs to a single POVM, which translates as the implementation of a richer die. Zero-diagonal not as pronounced as Approach~3.} \\
\hline
\parbox[t]{0.14\linewidth}{\raggedright \textbf{A1: ``Coherent''}} 
&
\parbox[t]{0.25\linewidth}{\raggedright Single-mode states are costly and you still wish to minimize the basis size.}
&
\parbox[t]{0.26\linewidth}{\raggedright Zero-diagonal clearer than Approach~2 for two–mode superpositions (i.e. coherent states); smallest basis size.}
&
\parbox[t]{0.22\linewidth}{\raggedright Zero-diagonal less pronounced for single-mode elements.} \\
\hline
\parbox[t]{0.14\linewidth}{\raggedright \textbf{A2: ``Incoherent''}} 
&
\parbox[t]{0.25\linewidth}{\raggedright Two–mode superposition states are costly and you still wish to minimize the basis size.}
&
\parbox[t]{0.26\linewidth}{\raggedright Zero-diagonal clearer than Approach~1 for single-mode  (i.e. incoherent) states; smallest basis size.}
&
\parbox[t]{0.22\linewidth}{\raggedright Overall least pronounced zero-diagonal in general (as there are more coherent than incoherent states).} \\
\hline
\parbox[t]{0.14\linewidth}{\raggedright \textbf{A3: ``Maximal''}} 
&
\parbox[t]{0.25\linewidth}{\raggedright No preference between state types and no need to optimize basis size.}
&
\parbox[t]{0.26\linewidth}{\raggedright Clearest and most visually distinct zero-diagonal in $P$; the die implemented is the simplest.}
&
\parbox[t]{0.22\linewidth}{\raggedright Highest resource cost (maximal number of states).} \\
\hline
\end{tabular}
\caption{\textbf{Summary of protocol variants.} The Main approach forms a convex combination of two PVMs (cf. Approaches~1 and~2) while keeping the minimal-state budget (e.g. $d=5\implies$ 15 states). Approach~3 achieves better contrast by using a basis with all available states (e.g. $d=5\implies$ 25 states).}
\label{tab:protocol-variants}
\end{table}

\newpage

\subsubsection{\label{sec:Ap0} Combined data \vspace{-1cm}}
\ThreePmatrixMainFigure
  {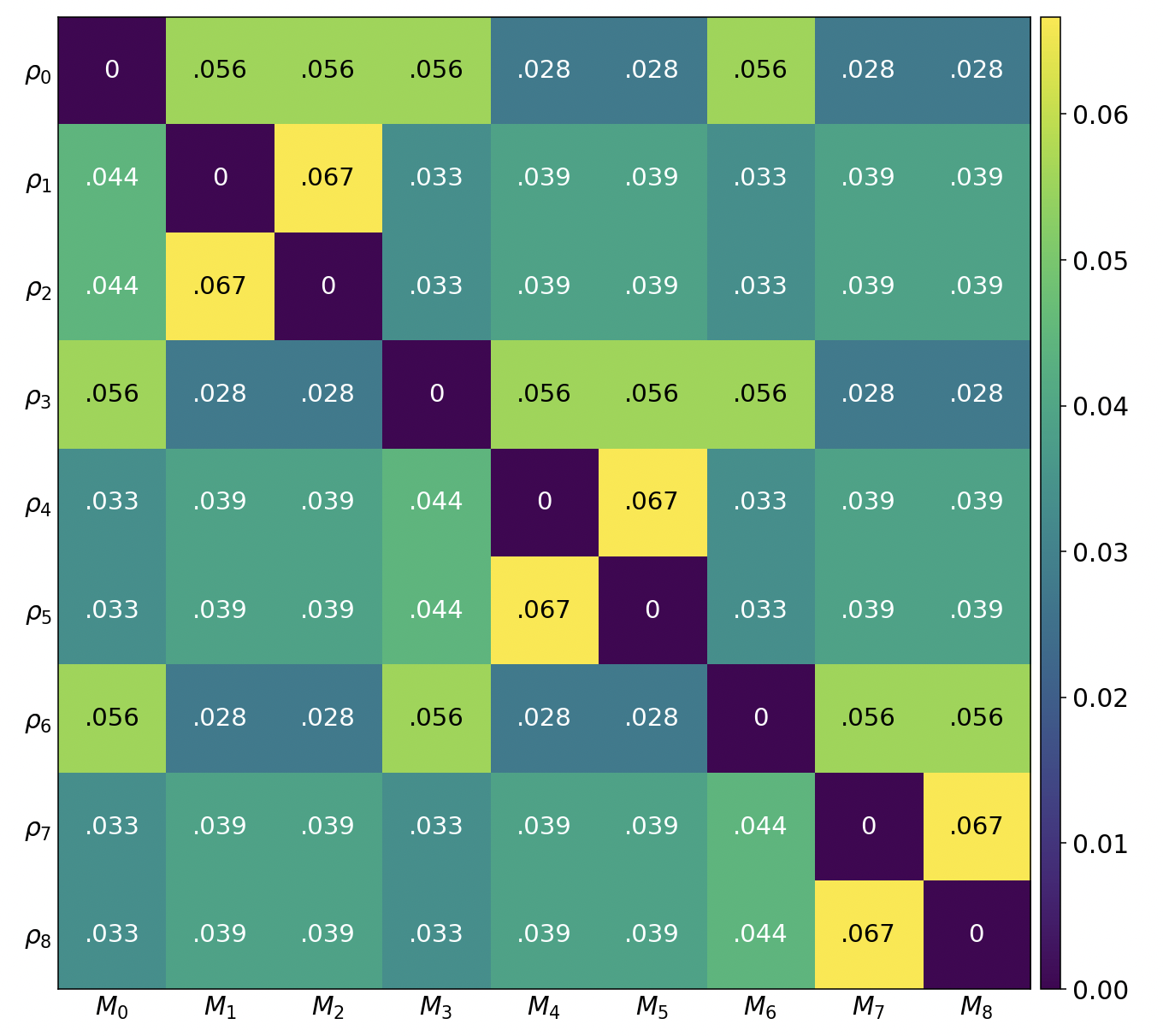}
  {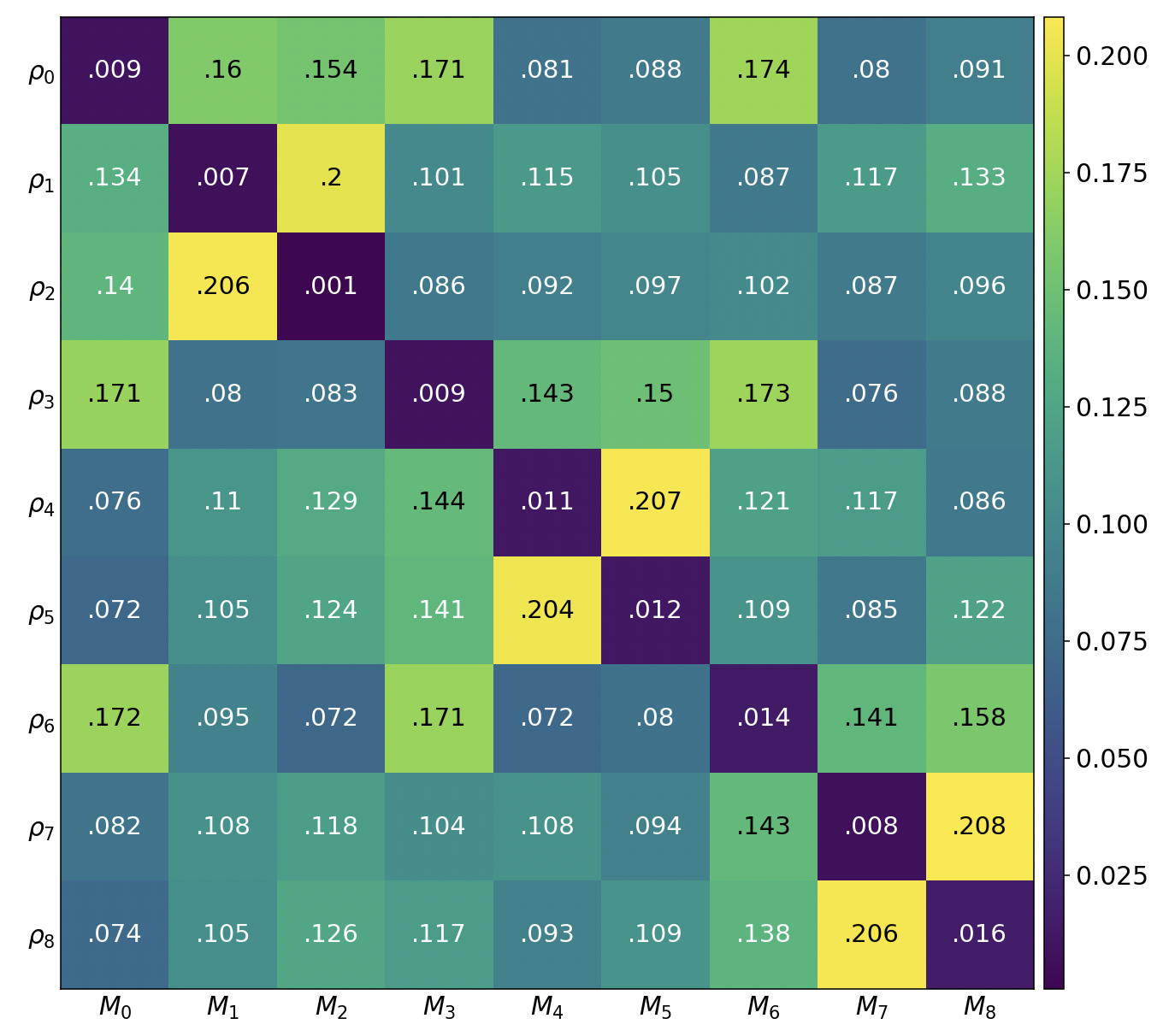}
  {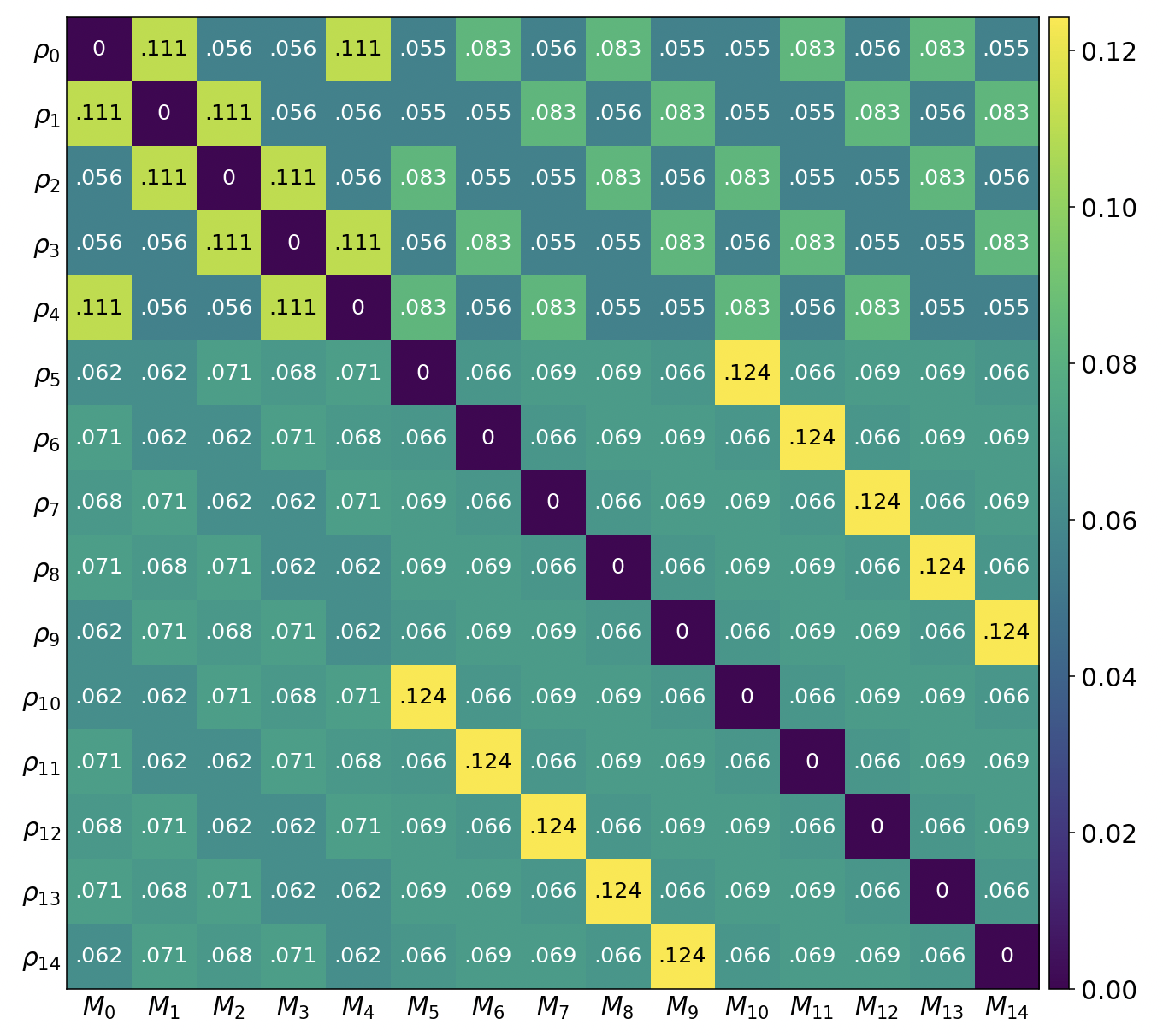}
  {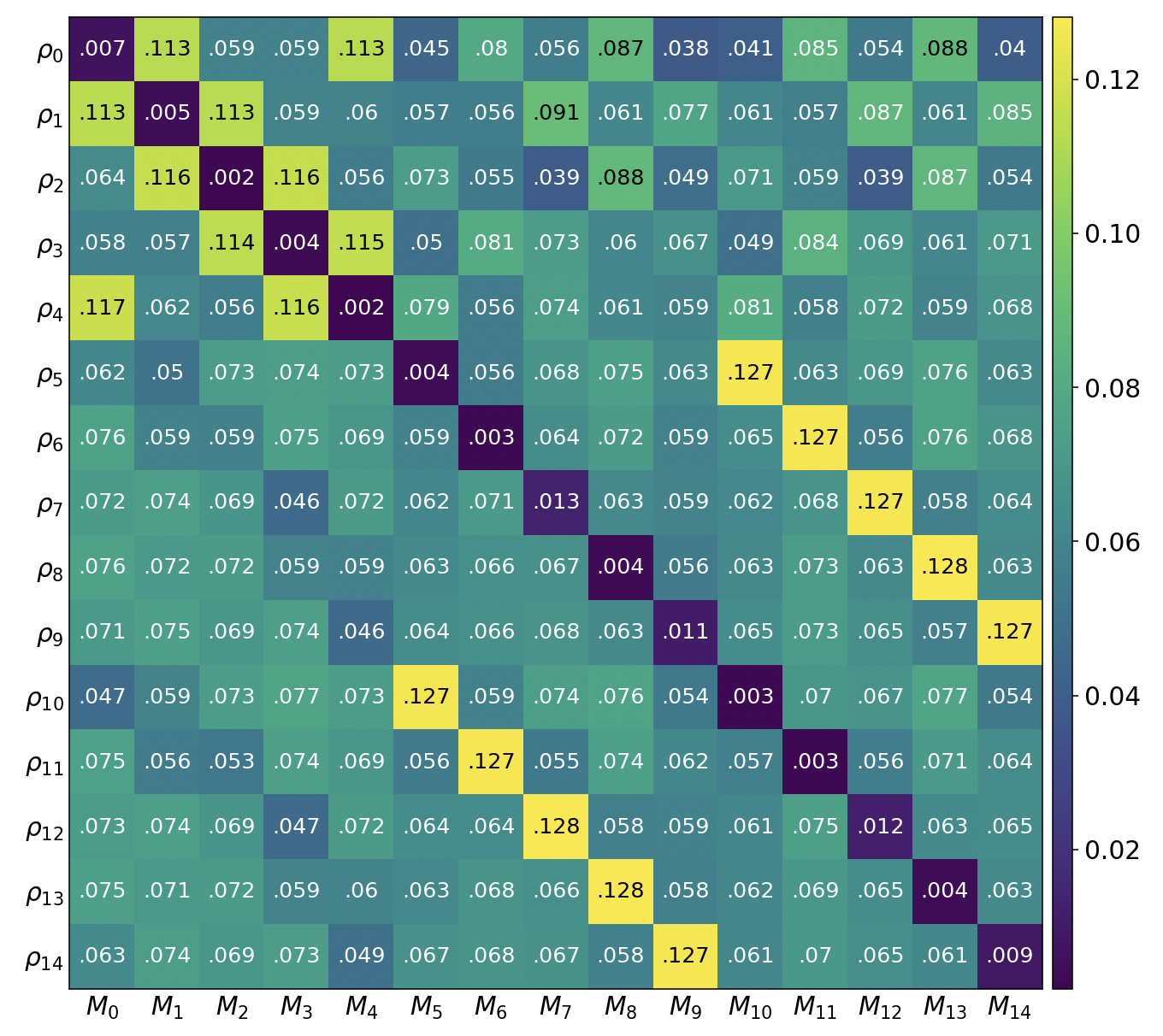}
  {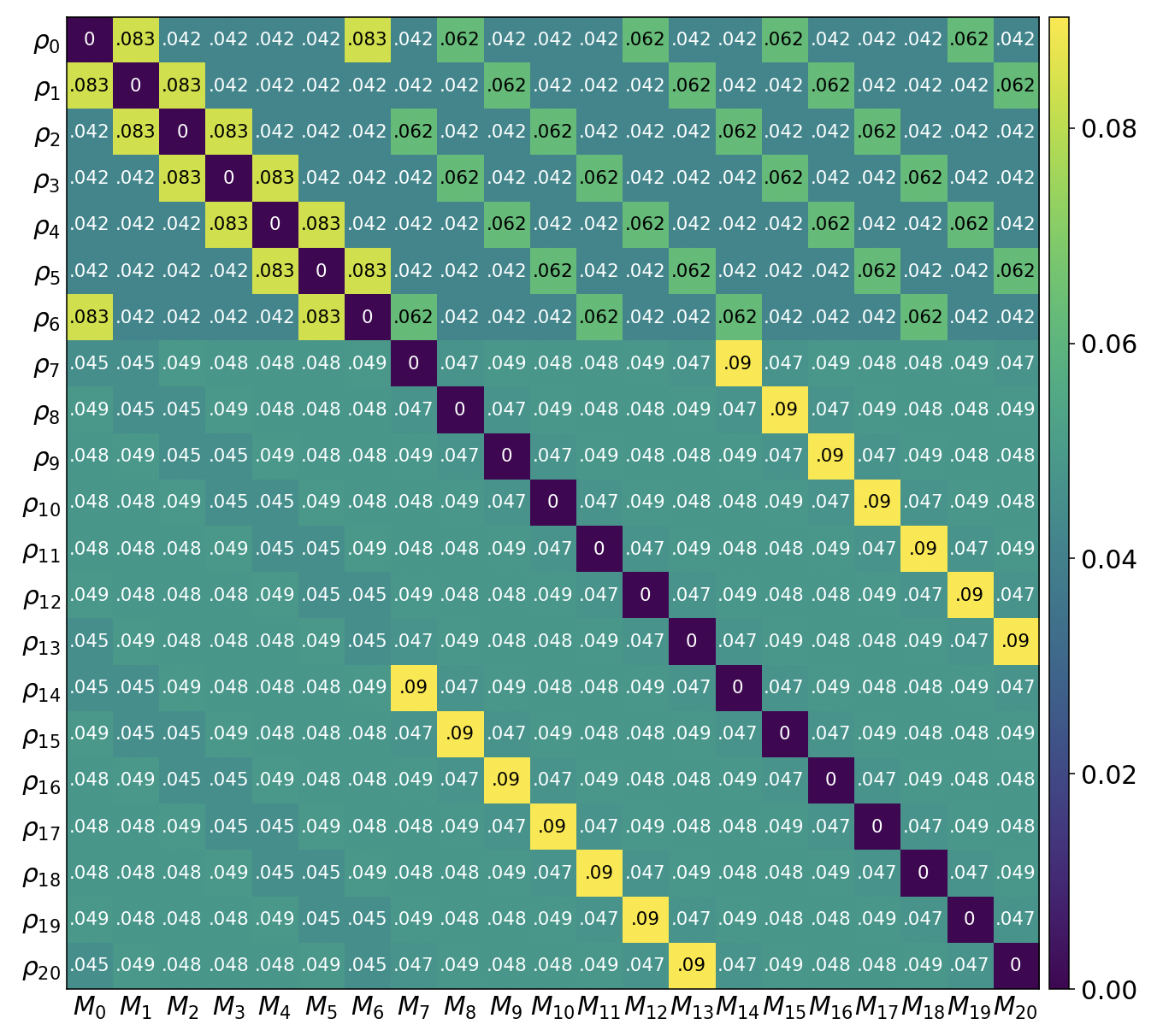}
  {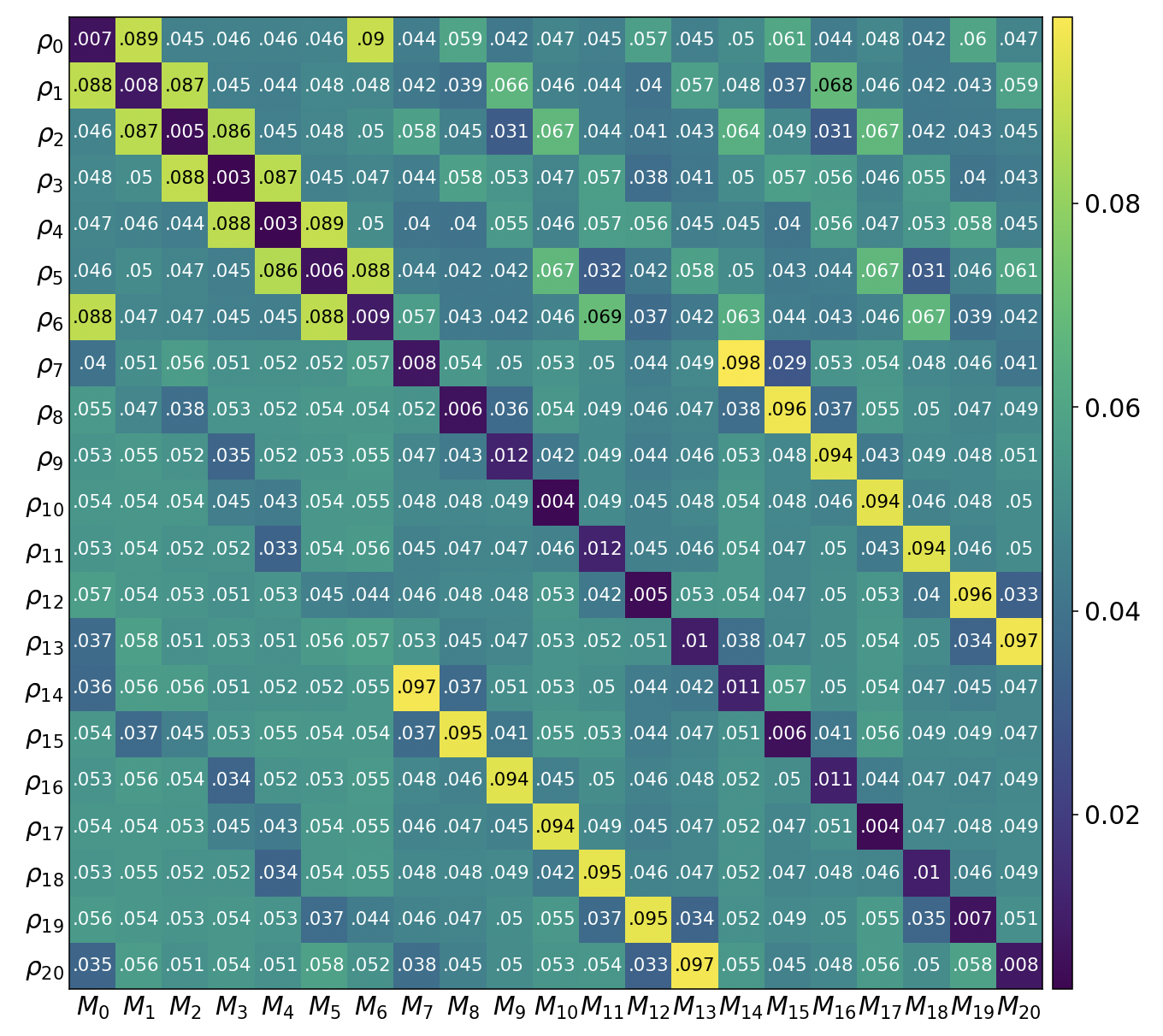}
  {\textbf{Experimental results taken for the convex protocol.} 
   Theoretical $P$-matrices with zeros on the diagonal (left) and their experimentally reconstructed counterparts (right) for $d=3,5,7$ (top–bottom).
   The OAM ($\ell$) $\leftrightarrow$ qudit mappings (order of $\ket{0},\ket{1},\ldots,\ket{6}$) chosen to minimize $\|E\|_2$ are: 
   $P_{3}\!:\!(\ell=+0,+1,-1)$; 
   $P_{5}\!:\!(\ell=+1,+2,-2,+3,-3)$; 
   $P_{7}\!:\!(\ell=+0,+1,+2,-3,+3,-2,-1)$.}%
  {fig:convex-noperms}
  {{}}

\subsubsection{\label{sec:Ap1} Coherent \vspace{-1cm}}

\ThreePmatrixMainFigure
  {qcom-outputs-M3/Theory/d3}
  {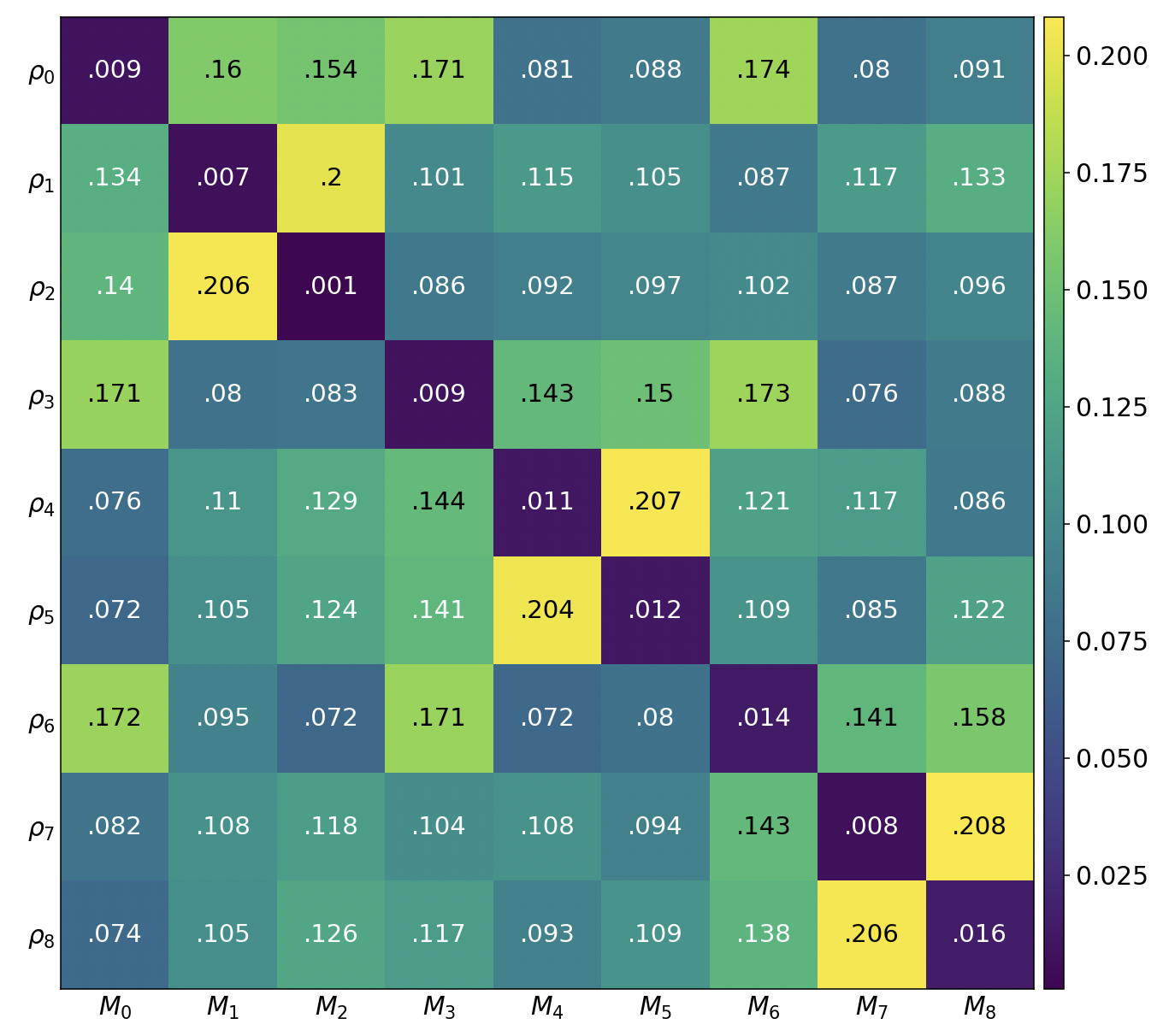}
  {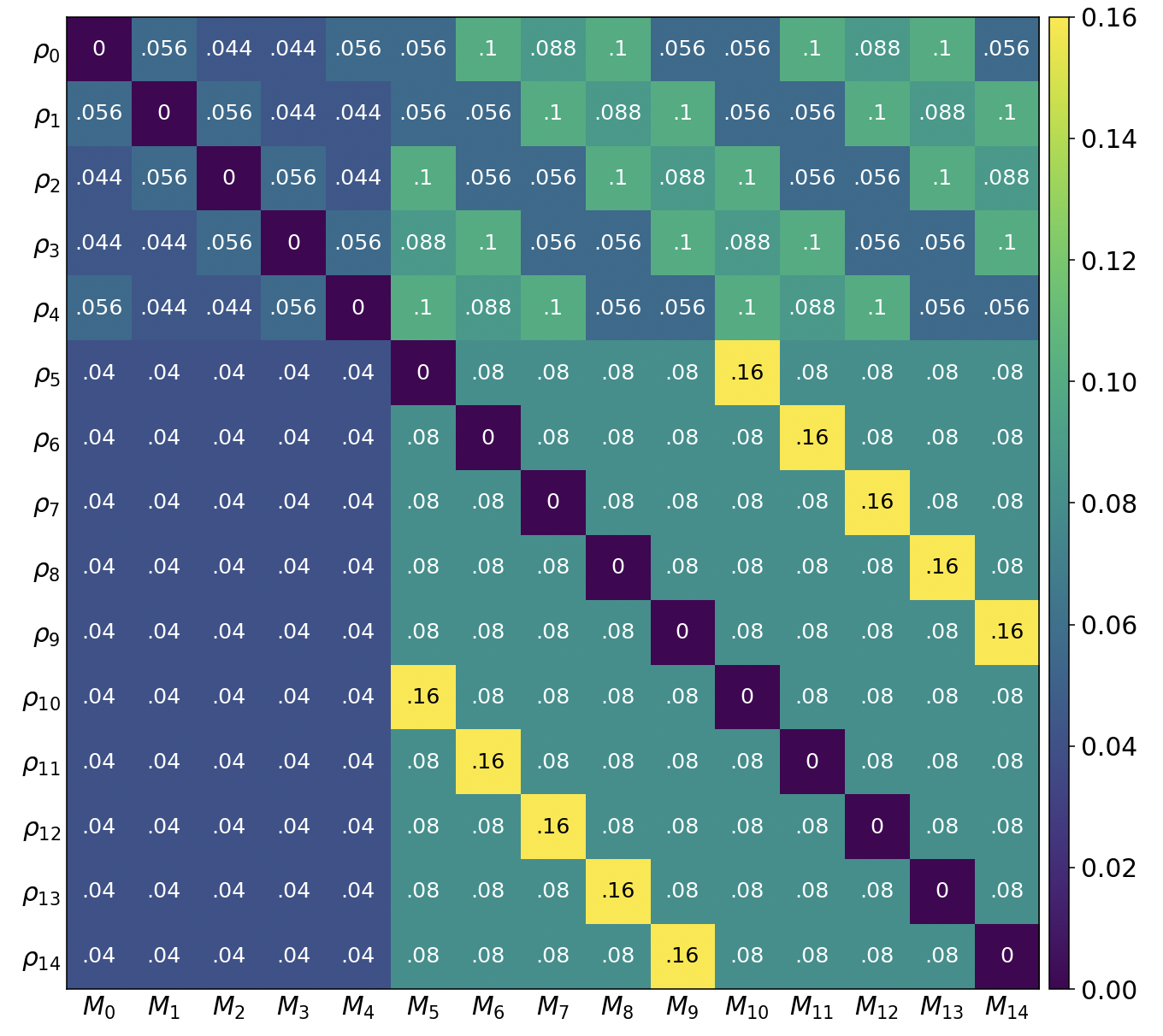}
  {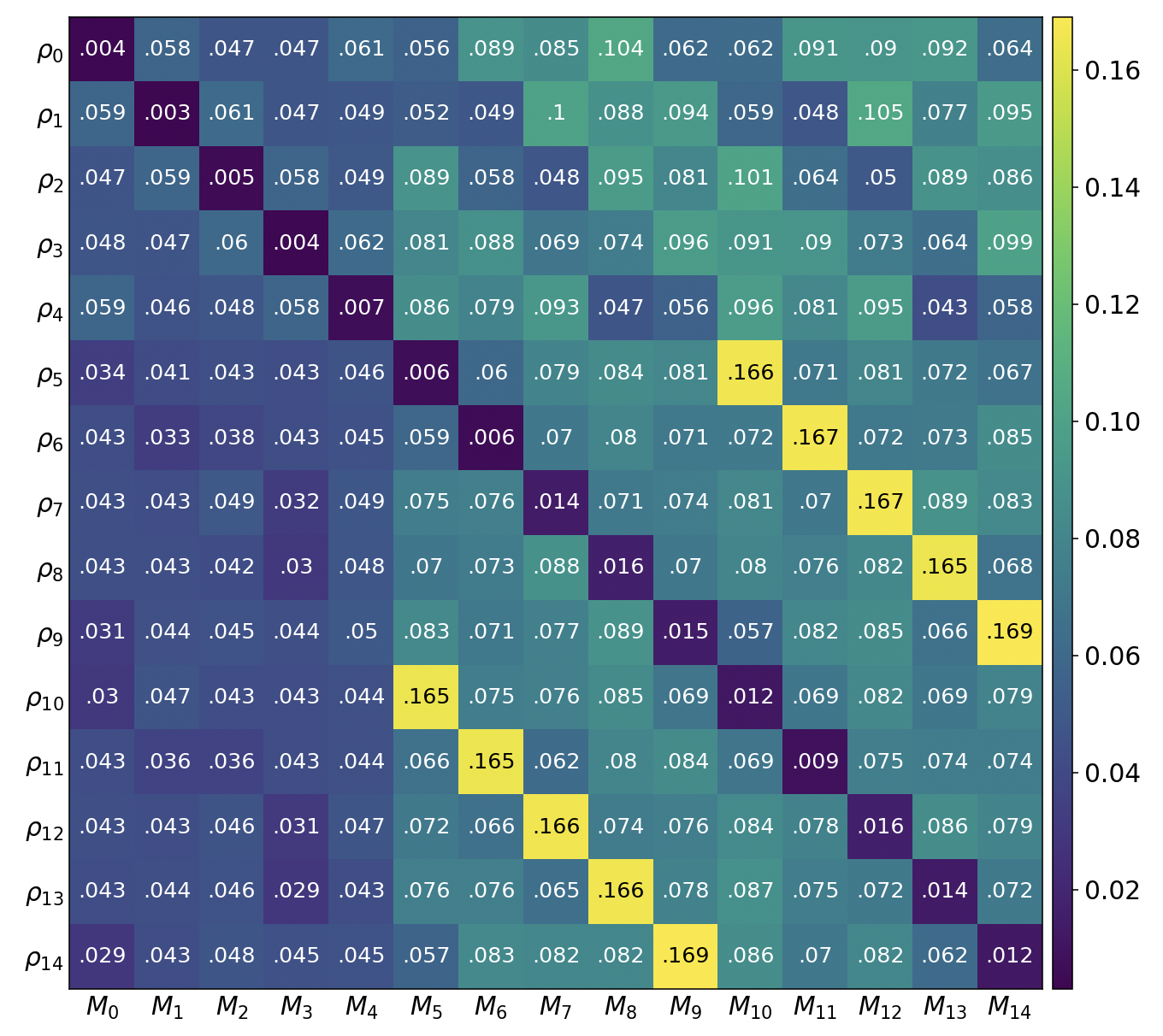}
  {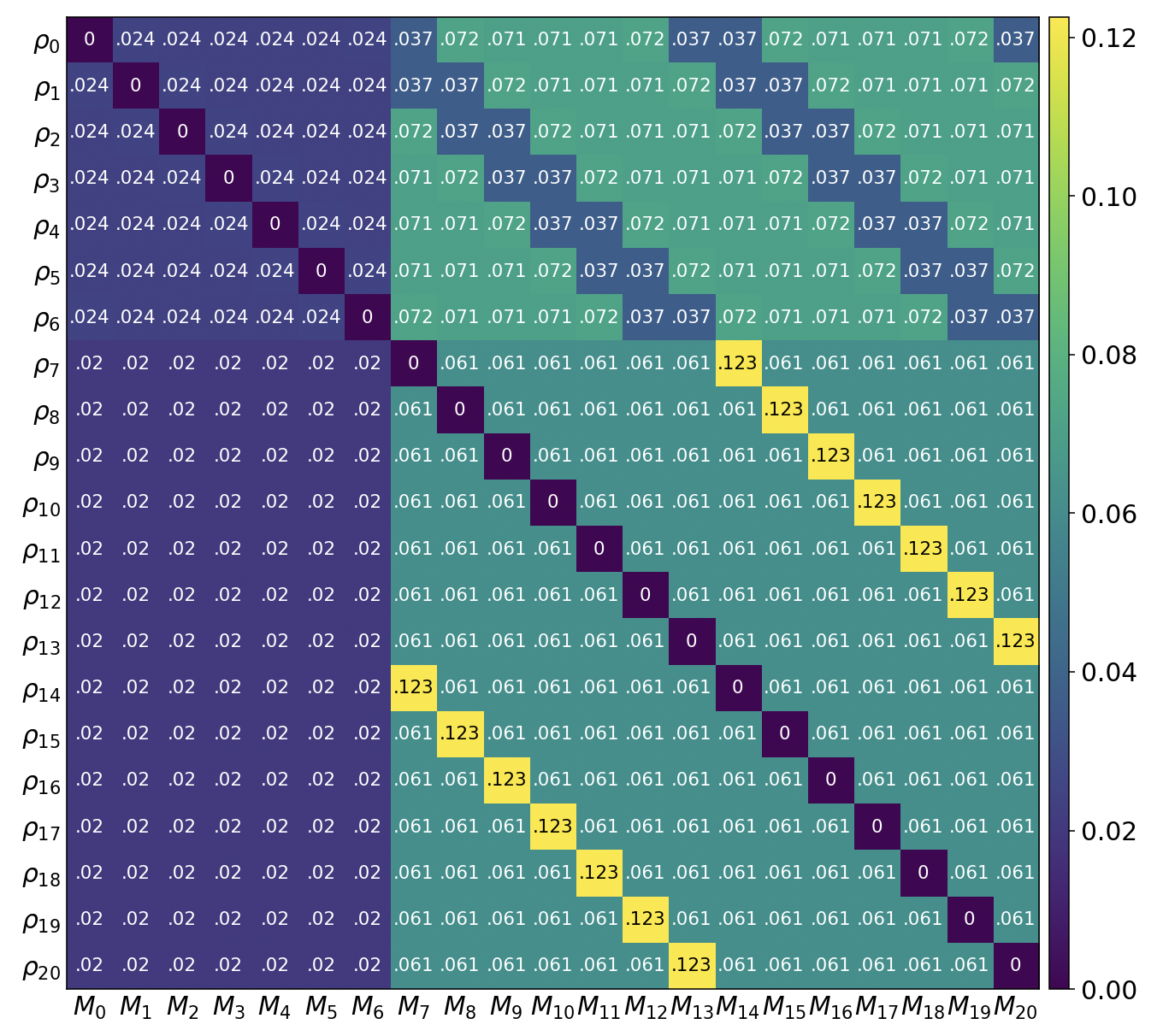}
  {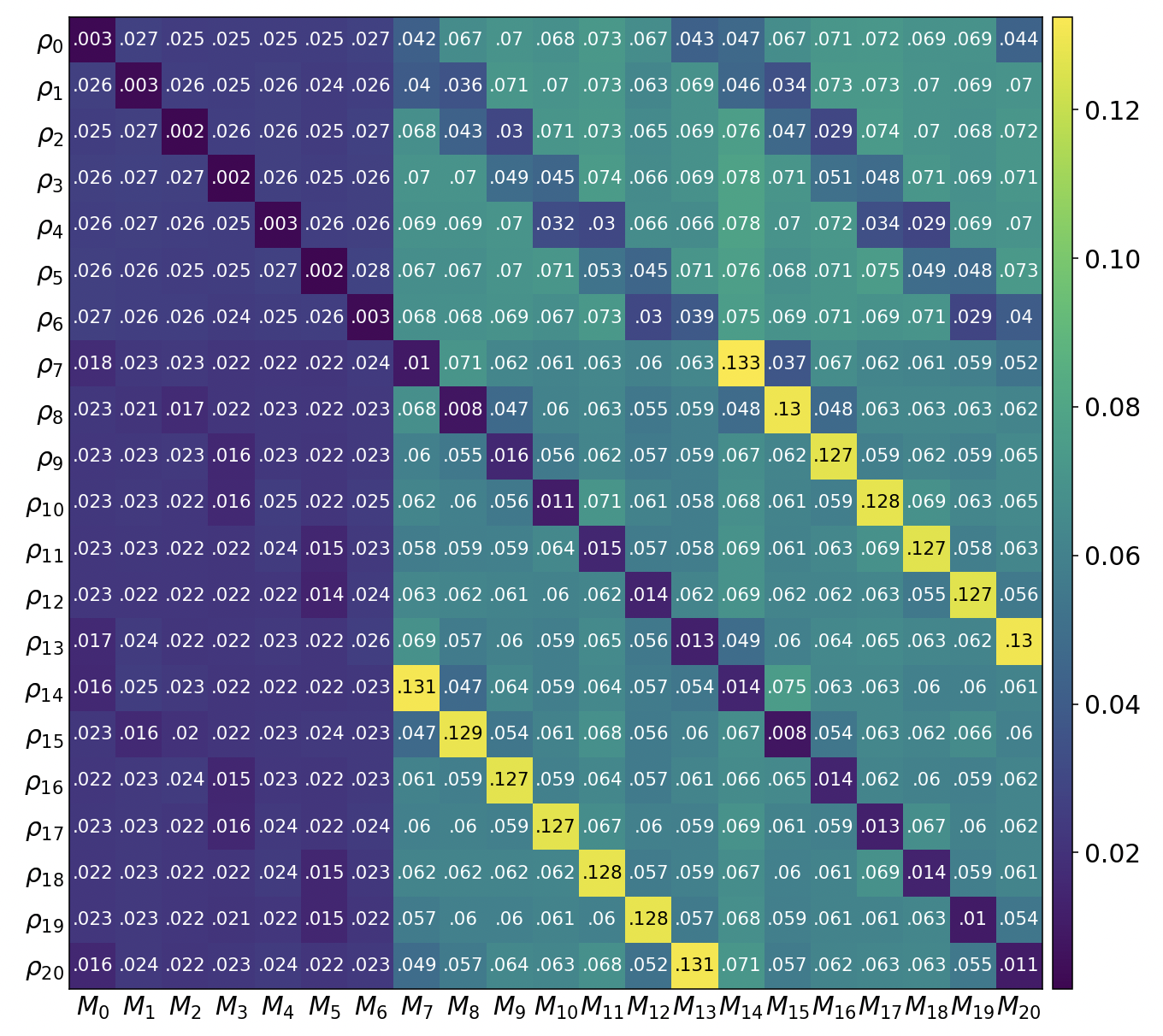}
  {\textbf{Experimental results taken for the coherent protocol.} 
   Theoretical $P$-matrices with zeros on the diagonal (left) and their experimentally reconstructed counterparts (right) for $d=3,5,7$ (top–bottom).
   The OAM ($\ell$) $\leftrightarrow$ qudit mappings (order of $\ket{0},\ket{1},\ldots,\ket{6}$) chosen to minimize $\|E\|_2$ are: 
   $P_{3}\!:\!(\ell=+0,+1,-1)$; 
   $P_{5}\!:\!(\ell=+0,+1,-2,-3,-1)$; 
   $P_{7}\!:\!(\ell=+0,+1,+2,-3,-2,+3,-1)$.}%
  {fig:coherent-noperms}
  {{}}

\subsubsection{\label{sec:Ap2} Incoherent heavy\vspace{-1cm}}

\ThreePmatrixMainFigure
  {qcom-outputs-M3/Theory/d3}
  {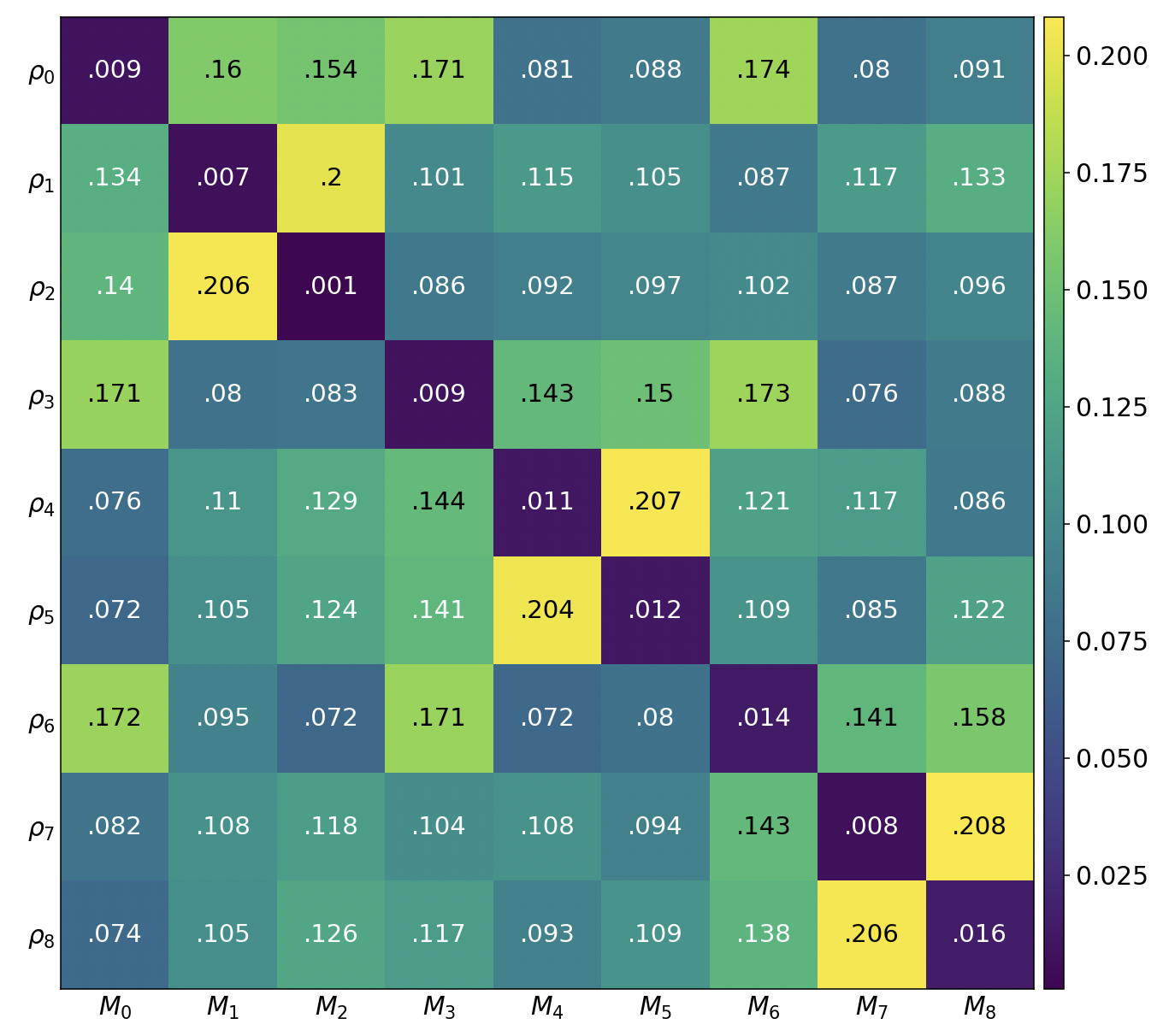}
  {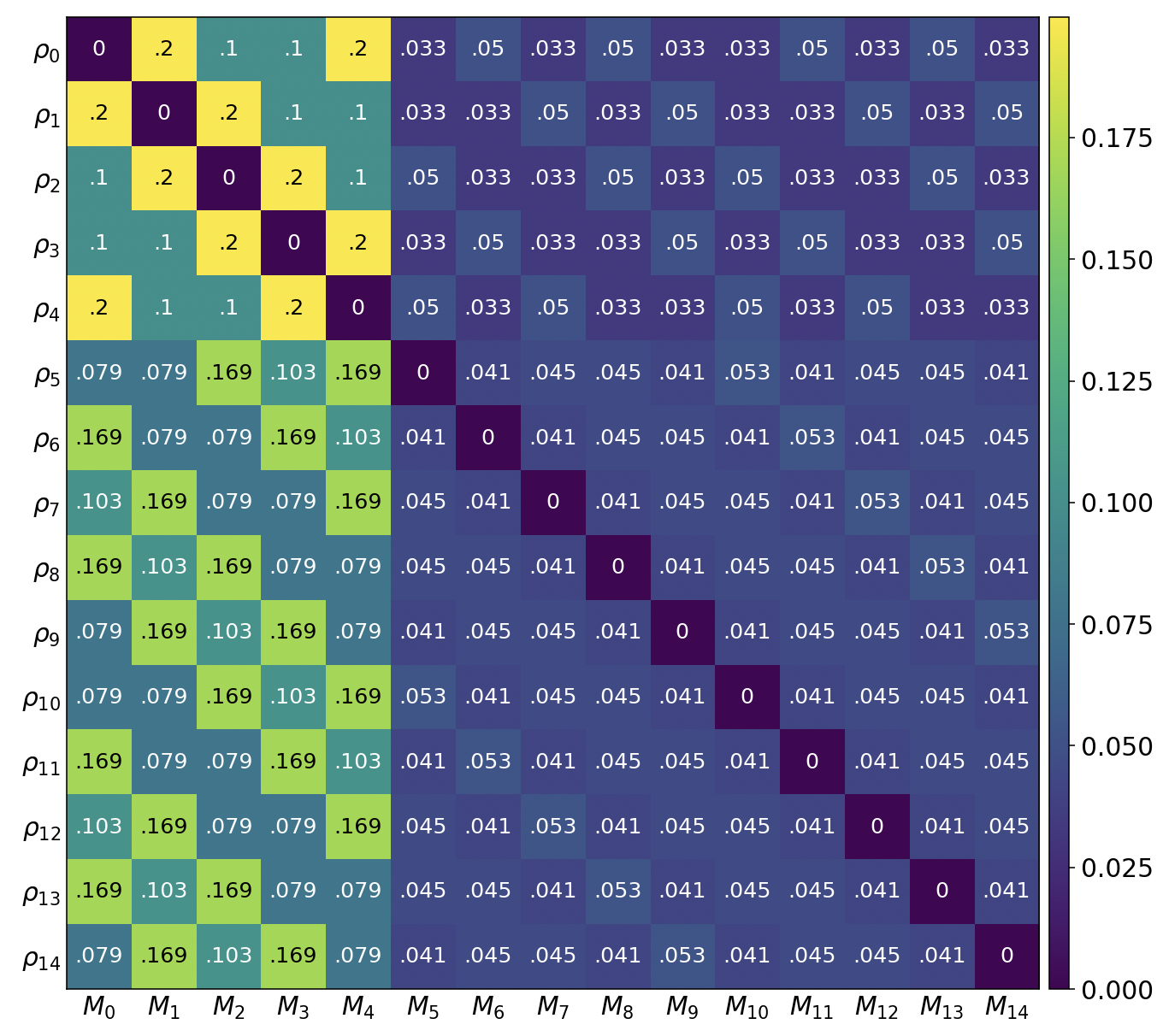}
  {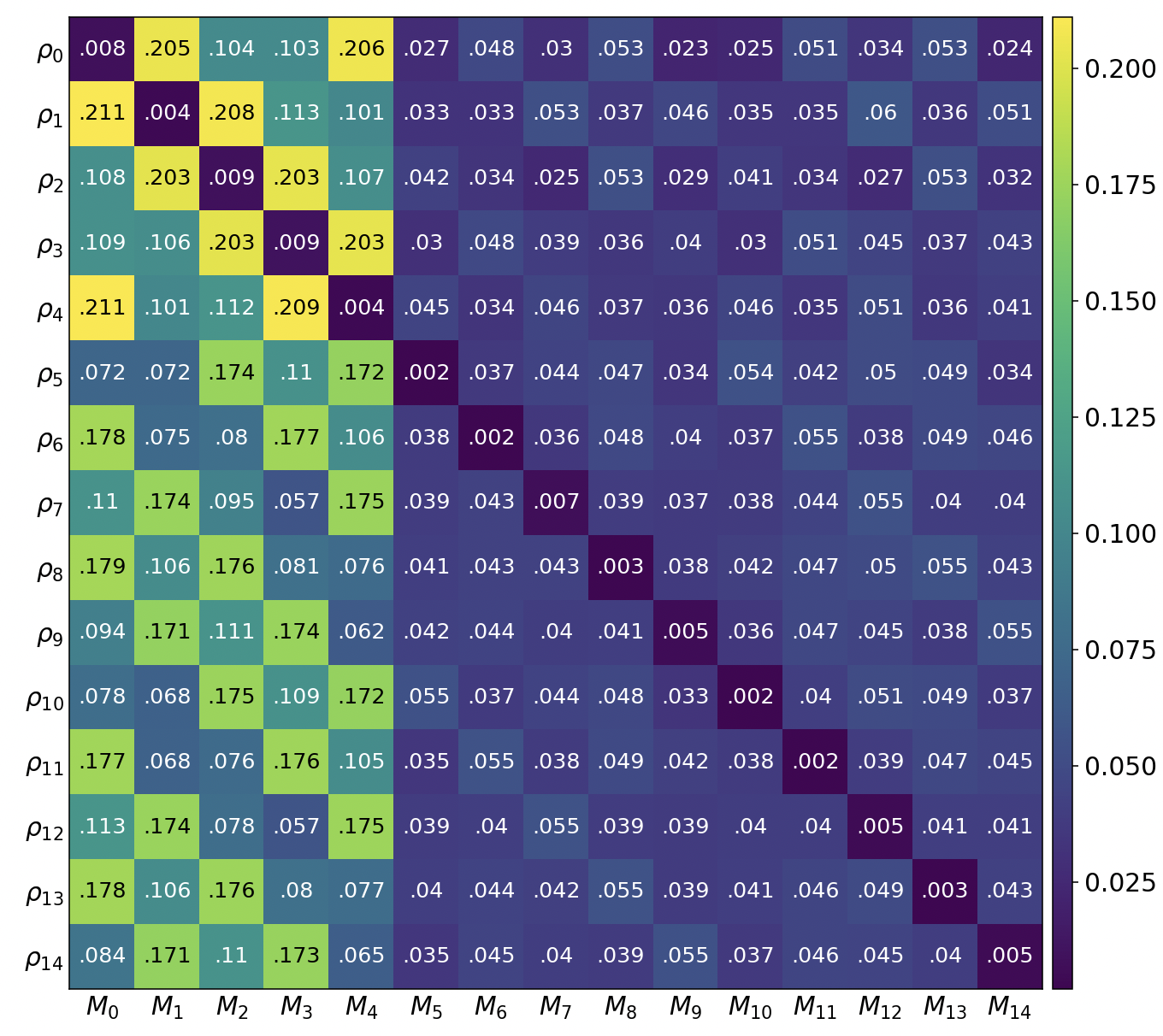}
  {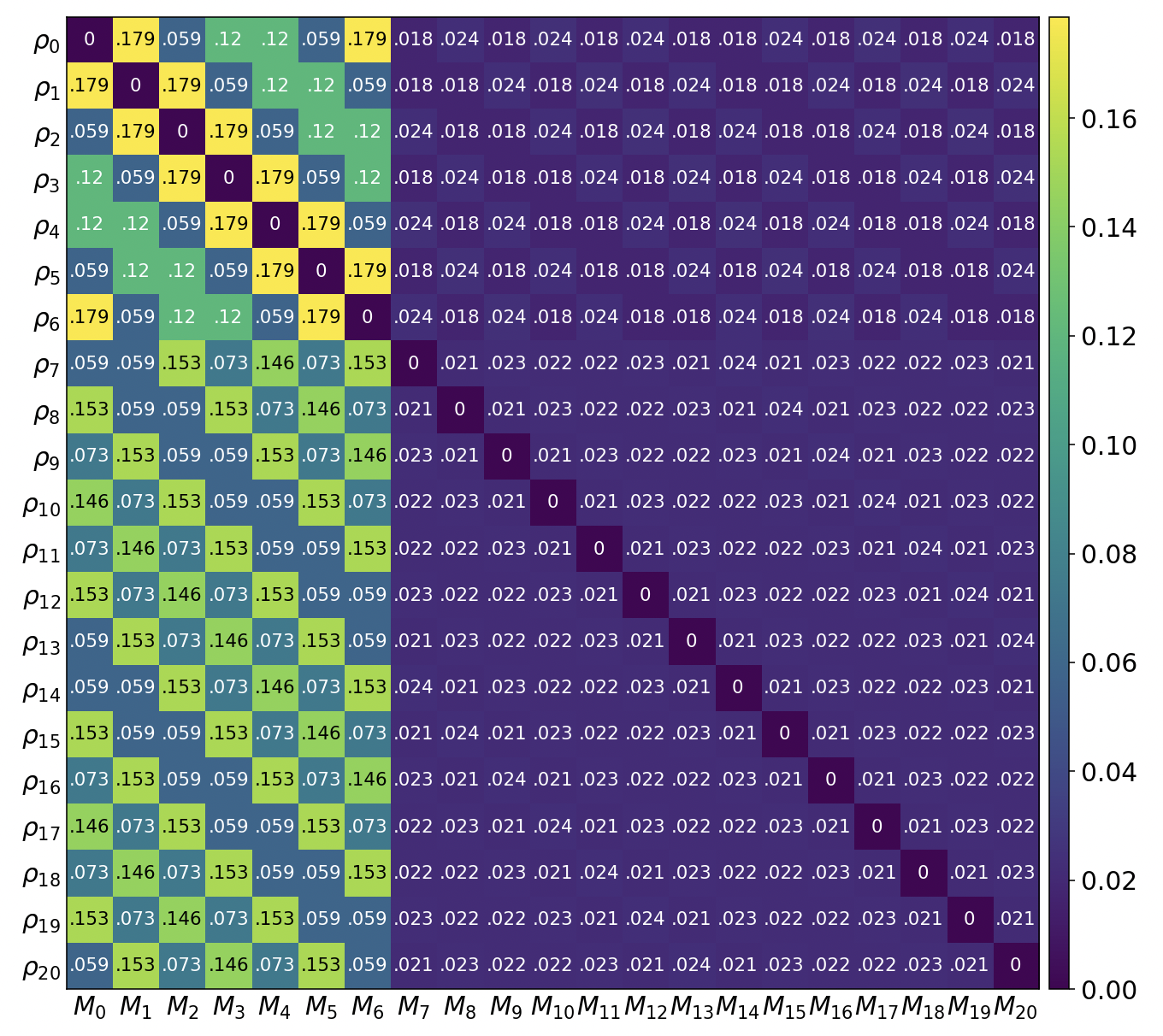}
  {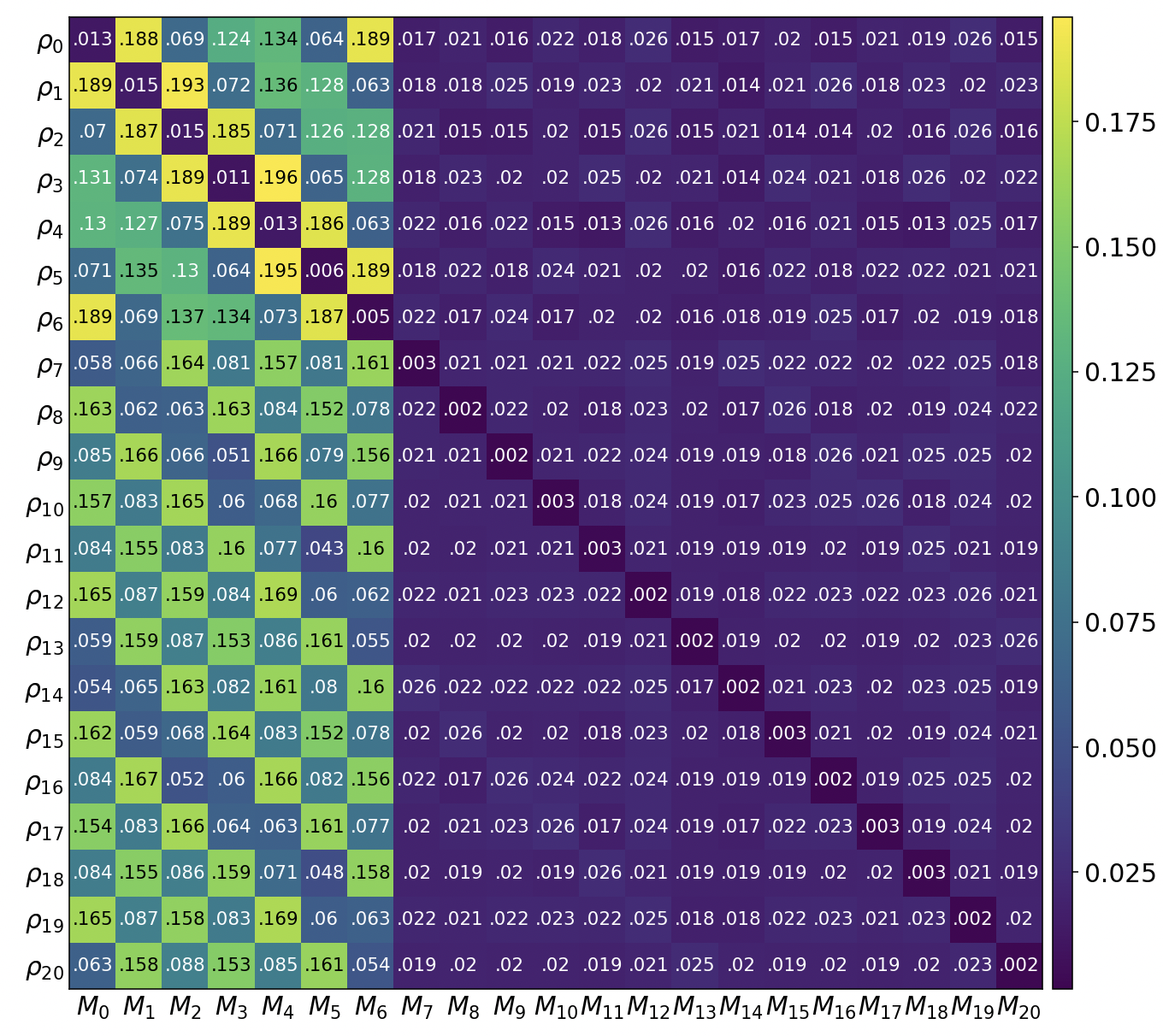}
  {\textbf{Experimental results taken for the incoherent protocol.} 
   Theoretical $P$-matrices with zeros on the diagonal (left) and their experimentally reconstructed counterparts (right) for $d=3,5,7$ (top–bottom).
   The OAM ($\ell$) $\leftrightarrow$ qudit mappings (order of $\ket{0},\ket{1},\ldots,\ket{6}$) chosen to minimize $\|E\|_2$ are: 
   $P_{3}\!:\!(\ell=+0,+1,-1)$; 
   $P_{5}\!:\!(\ell=+1,-2,+2,+3,-3)$; 
   $P_{7}\!:\!(\ell=+0,-2,+1,+2,-1,+3,-3)$.}%
  {fig:incoherent-noperms}
  {{}}

\subsubsection{\label{sec:Ap3} Maximal\vspace{-1cm}}

\ThreePmatrixMainFigure
  {qcom-outputs-M3/Theory/d3}
  {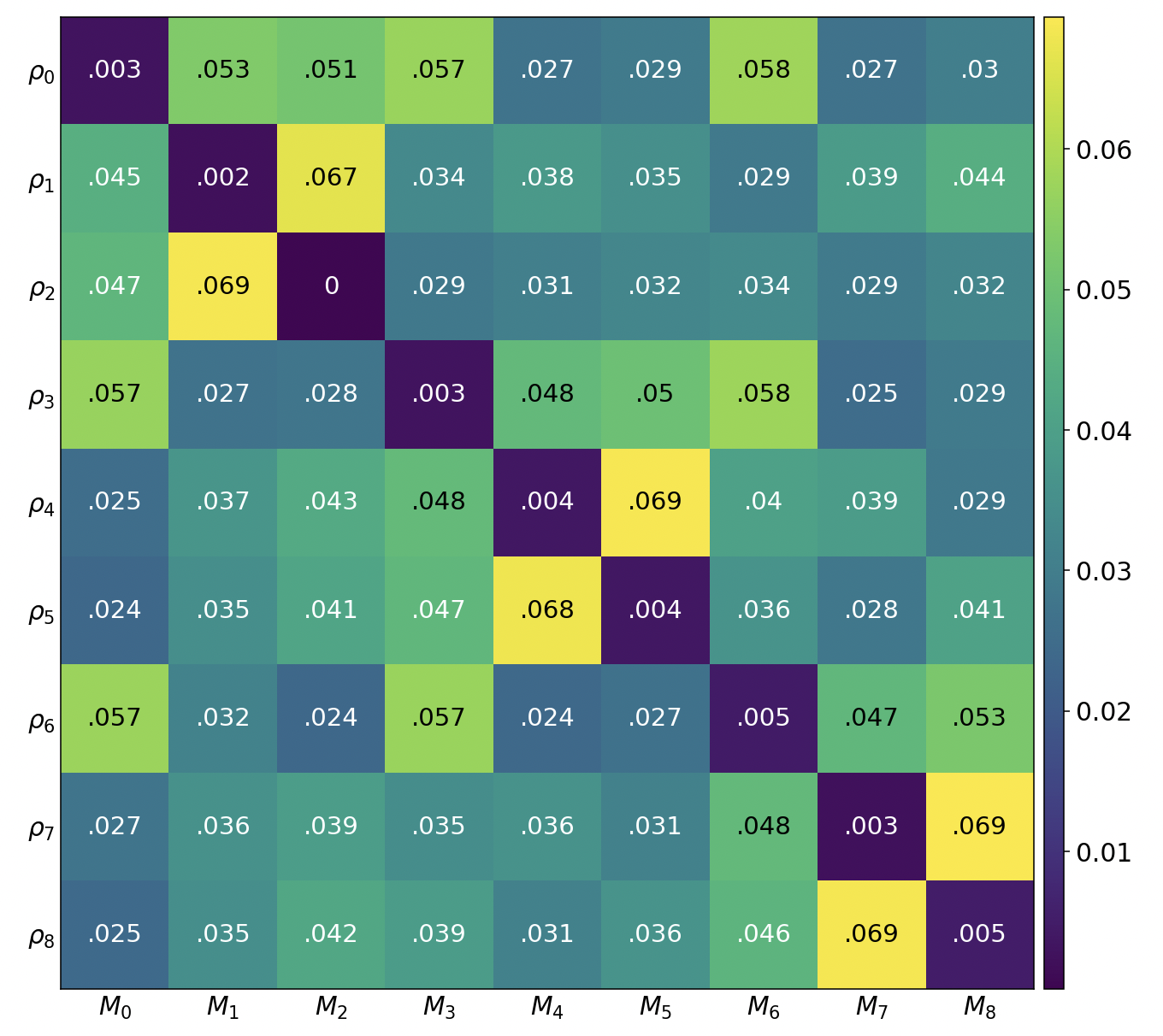}
  {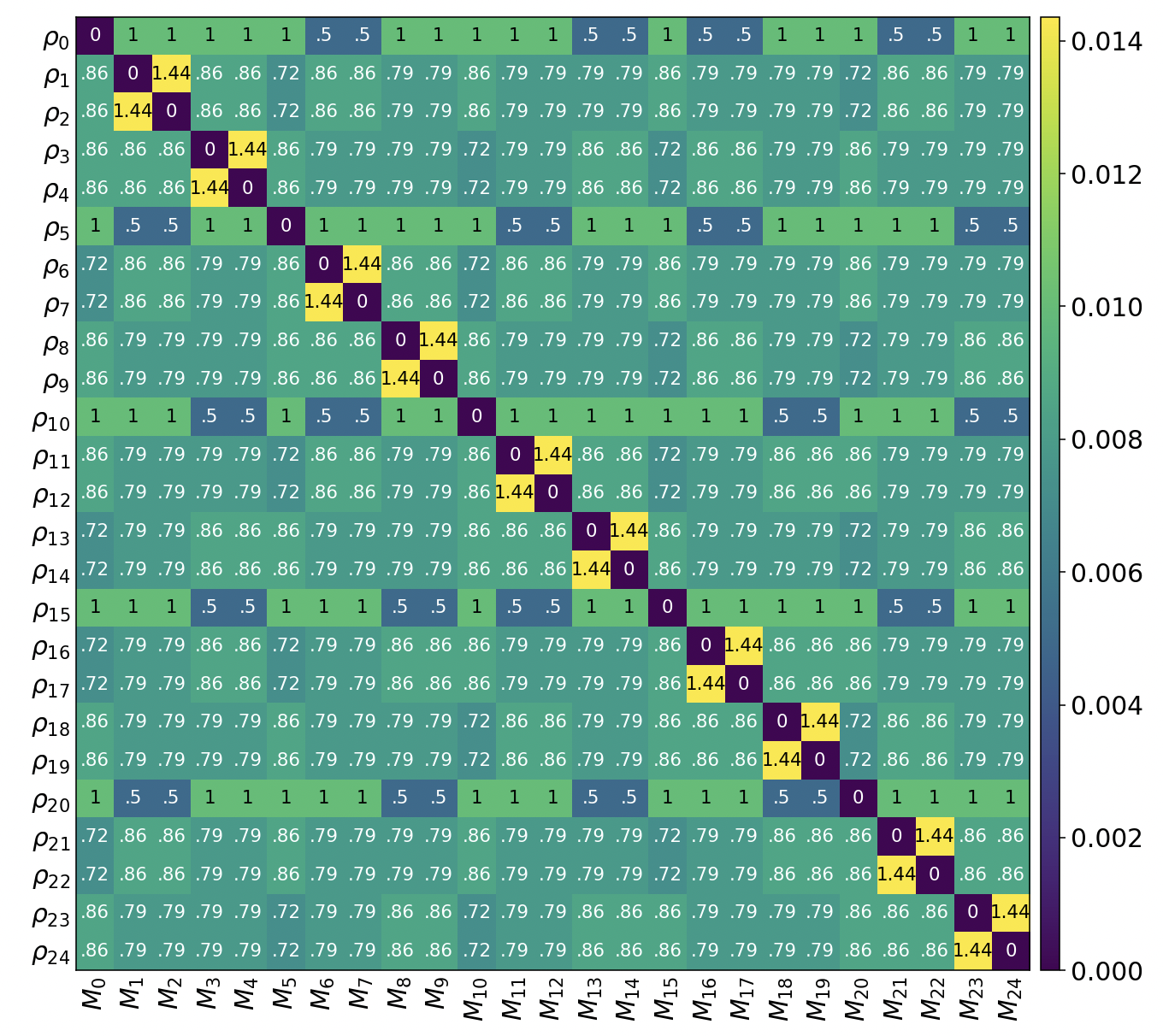}
  {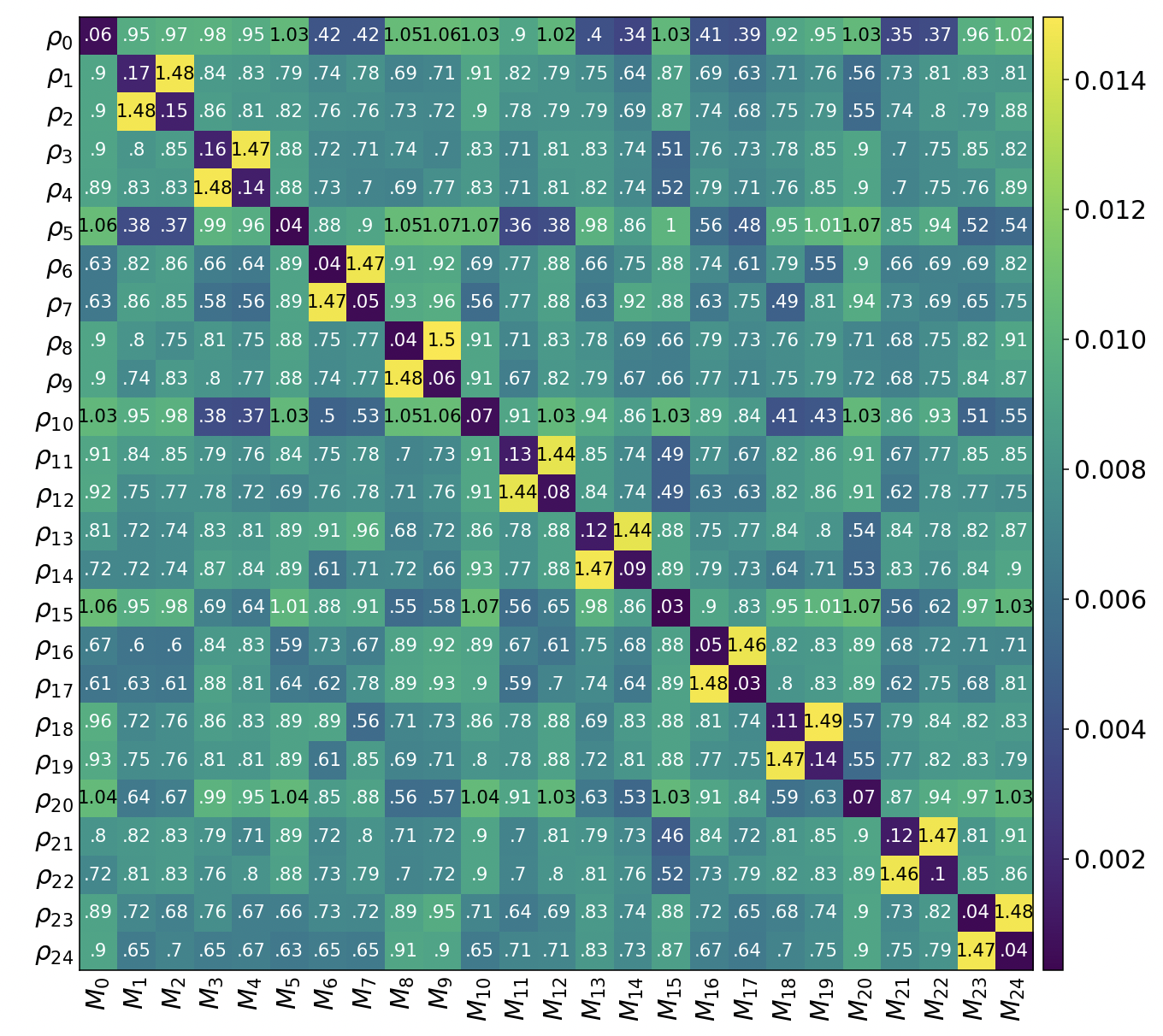}
  {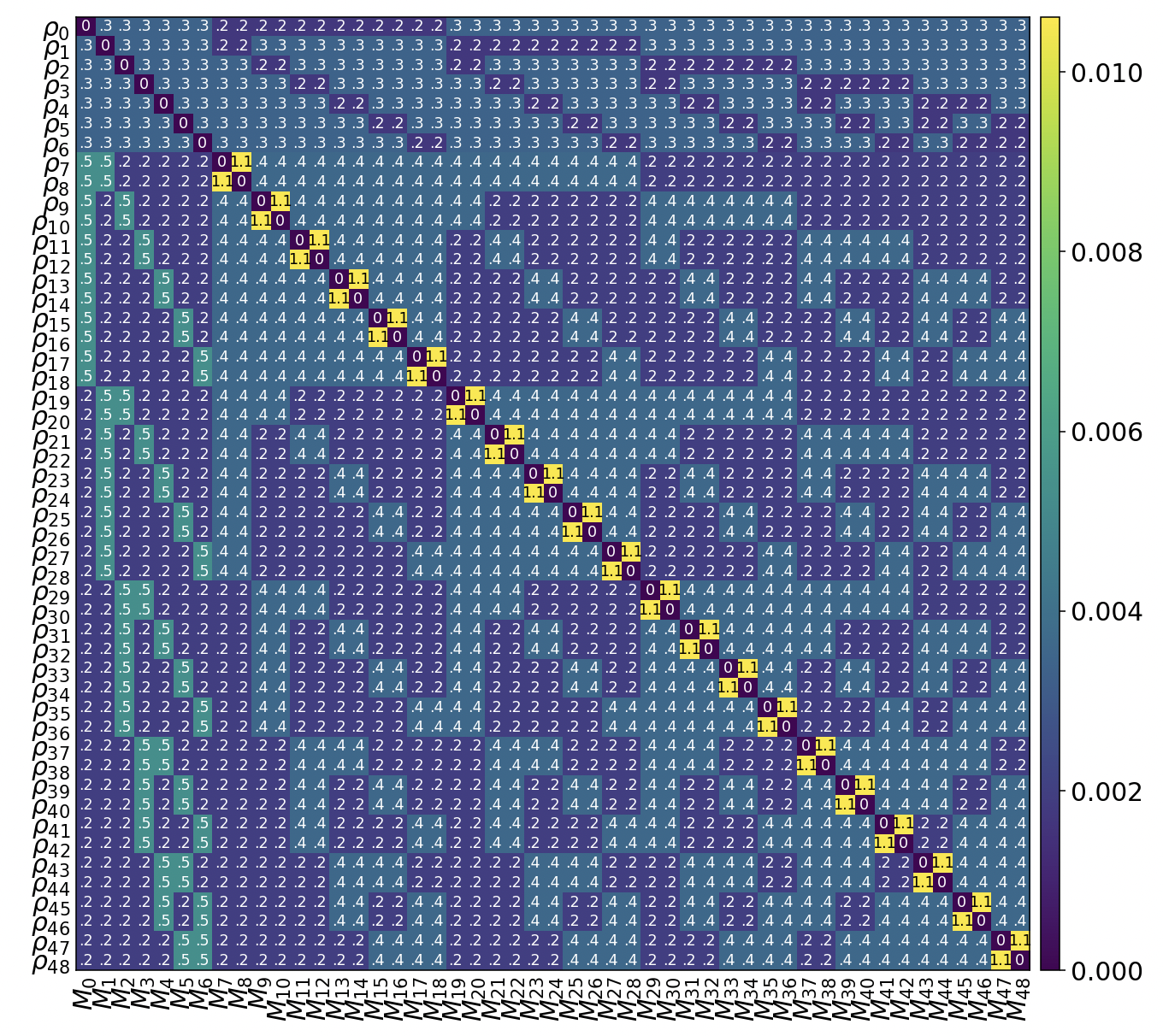}
  {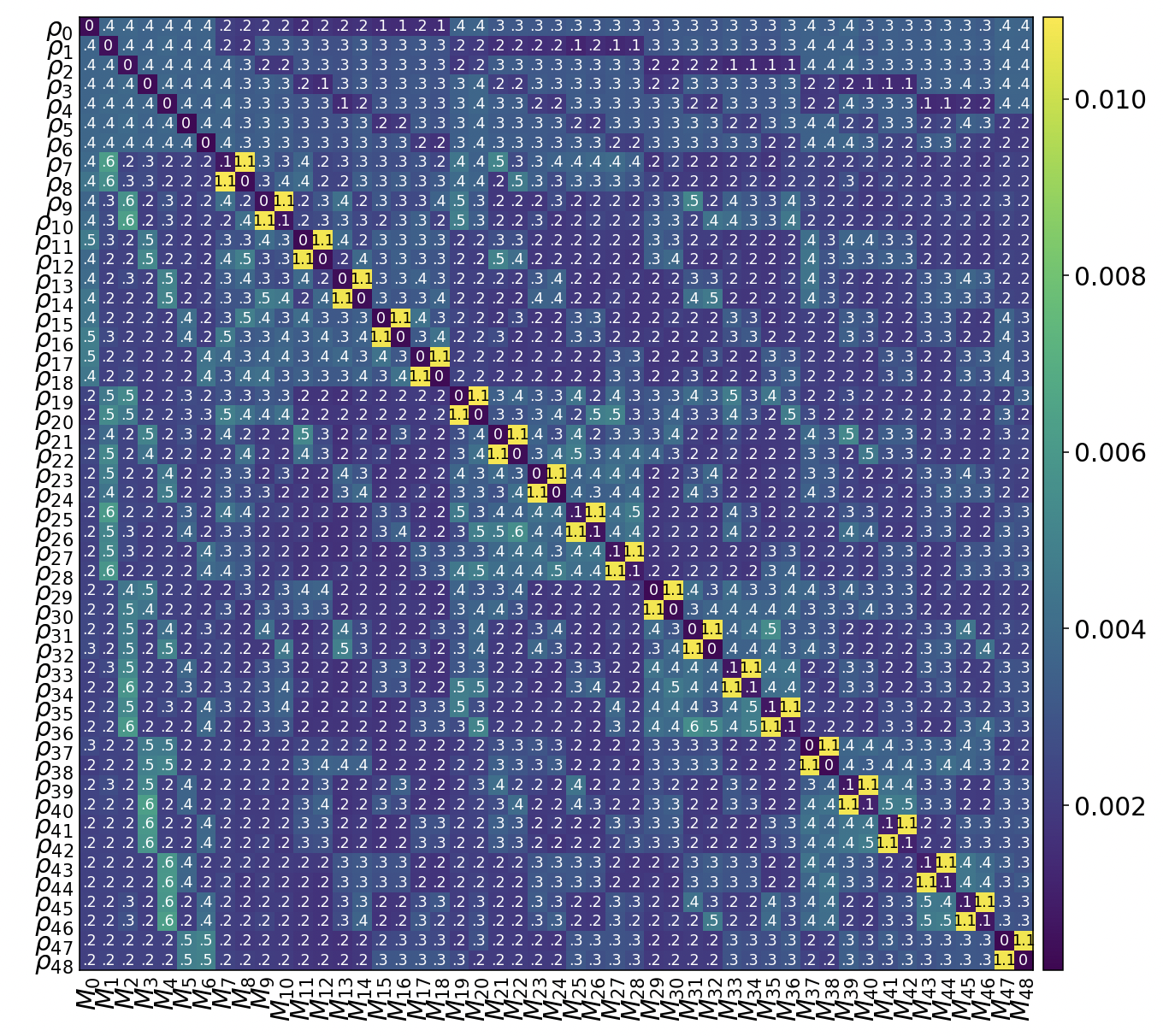}
  {\textbf{Experimental results taken for the maximal protocol.} 
   Theoretical $P$-matrices with zeros on the diagonal (left) and their experimentally reconstructed counterparts (right) for $d=3,5,7$ (top–bottom).
   The OAM ($\ell$) $\leftrightarrow$ qudit mappings (order of $\ket{0},\ket{1},\ldots,\ket{6}$) chosen to minimize $\|E\|_2$ are: 
   $P_{3}\!:\!(\ell=+0,+1,-1)$; 
   $P_{5}\!:\!(\ell=-1,+2,-2,+3,-3)$; 
   $P_{7}\!:\!(\ell=+0,+1,-1,+2,-2,+3,-3)$.\\ Note: For $d=5$ and $d=7$, the cell values are shown multiplied by 100, while the color bars show the true values.}%
  {fig:maximal-noperms}
  {{}}

\subsection{\label{sec:ap-Setup} About the experimental setup}

\subsubsection{\label{sec:s-details} Details}

A Toptica CTL 780 laser delivering 75 mW at 775.0 nm was focused to a 500 m spot at the entrance face of a 5 mm-long periodically poled KTiOPO$_4$ (PPKTP) crystal, phase-matched for type-0 spontaneous parametric down-conversion. Residual pump light was removed with an 800 nm long-pass filter, after which the down-converted photons were spectrally cleaned by a 3 nm FWHM band-pass filter centred at 1550 nm. The signal beam was expanded to 6 mm diameter and addressed by reflective SLMs for preparation and measurement.

\begin{figure}[H]
\centering
\includegraphics[width=0.9\linewidth]{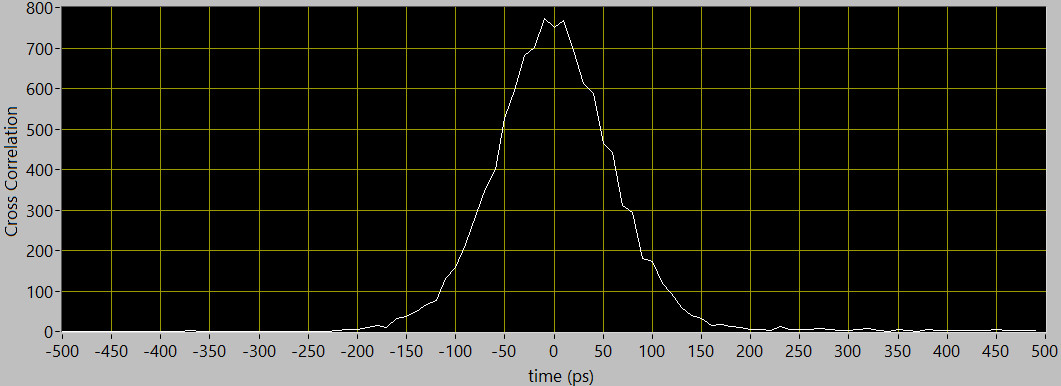}
\caption{\textbf{Cross-correlation plot.} Coincidences were recorded with a time resolution of 10~ps over a 1000~ps span. For a delay of 51825~ps, we observe a centered Gaussian-shaped peak and a reasonably good coincidence window length of 300~ps.}
\label{fig:path_delay}
\end{figure}

Detection was performed with fibre-coupled superconducting nanowire single-photon detectors (SNSPs) operated at $\sim$90~\% detection efficiency, and dark counts of less than 500~Hz. A Swabian time-tagger module registered coincidences within a 300~ps coincidence window; the free space optics path added a fixed 51825~ps delay see \cref{fig:path_delay}, which was subtracted in post-processing. Data were acquired in $0.5$~s sampling intervals and averaged over $50$ independent runs, yielding a single-photon heralding efficiency of approximately 27~\% (defined as the conditional detection probability of the idler given a signal click).

\subsubsection{\label{sec:s-align} Alignment}
To align the setup prior to single‑photon measurements, we followed two main steps:

\begin{enumerate}
\item \textbf{Back‑propagation and spiral‑bandwidth metric.}
We performed a standard back‑propagation using a $1550\text{ nm}$, $50\text{ mW}$ laser. Specifically, a Gaussian beam was injected backward through the fiber normally connected to APD$_A$ (see \cref{fig:setup-simple}), and the coupled power was measured at the output fiber (normally connected to APD$_B$) with a power meter (Thorlabs PM101). After maximizing the coupling, we recorded a ``spiral bandwidth'' scan (overlaps of integer OAM modes for $\ell=-3,-2,\ldots,3$). As an alignment metric, we used the average crosstalk of the central element ($\ell=0$) with its four nearest neighbors (up, down, left, right), see \cref{fig:spiral-b}. In the classical back‑propagation test this value was $0.4\%$. Once satisfactory, the setup was restored to its normal configuration and the same spiral‑bandwidth scan was performed with single photons from the SPDC source; in this work, the corresponding (quantum) crosstalk was $3.3\%$.

\begin{figure}[H]
\centering
\includegraphics[width=0.375\linewidth]{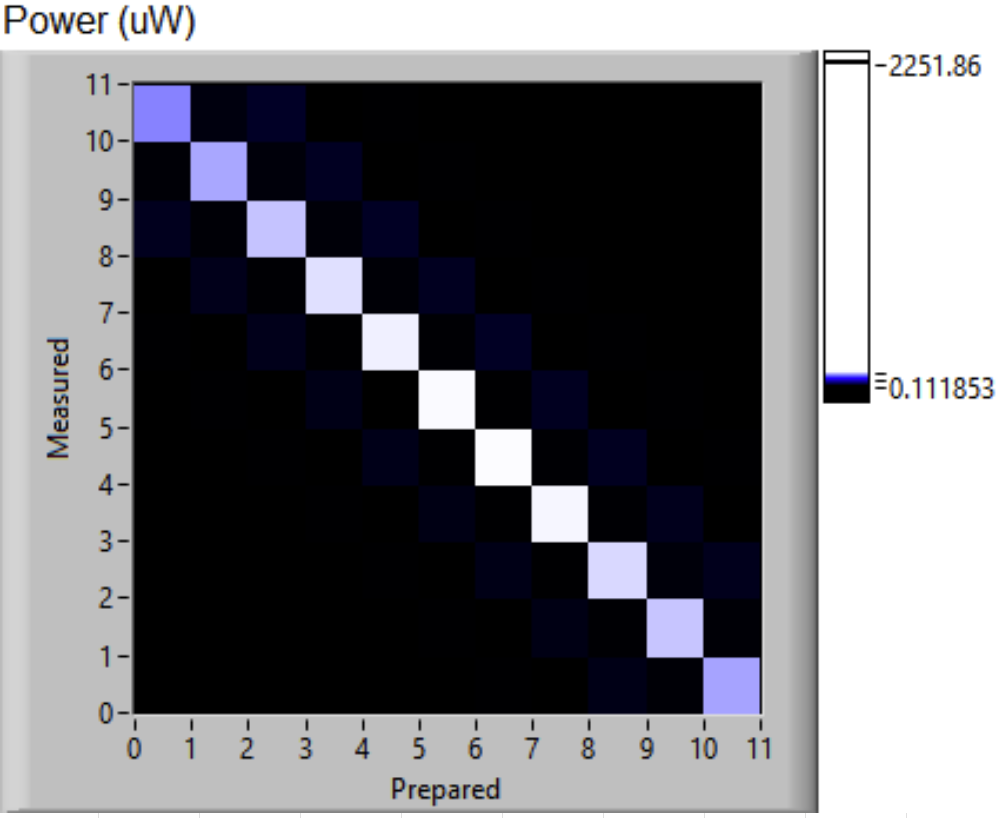}
\hspace{0.2cm}
\includegraphics[width=0.3525\linewidth]{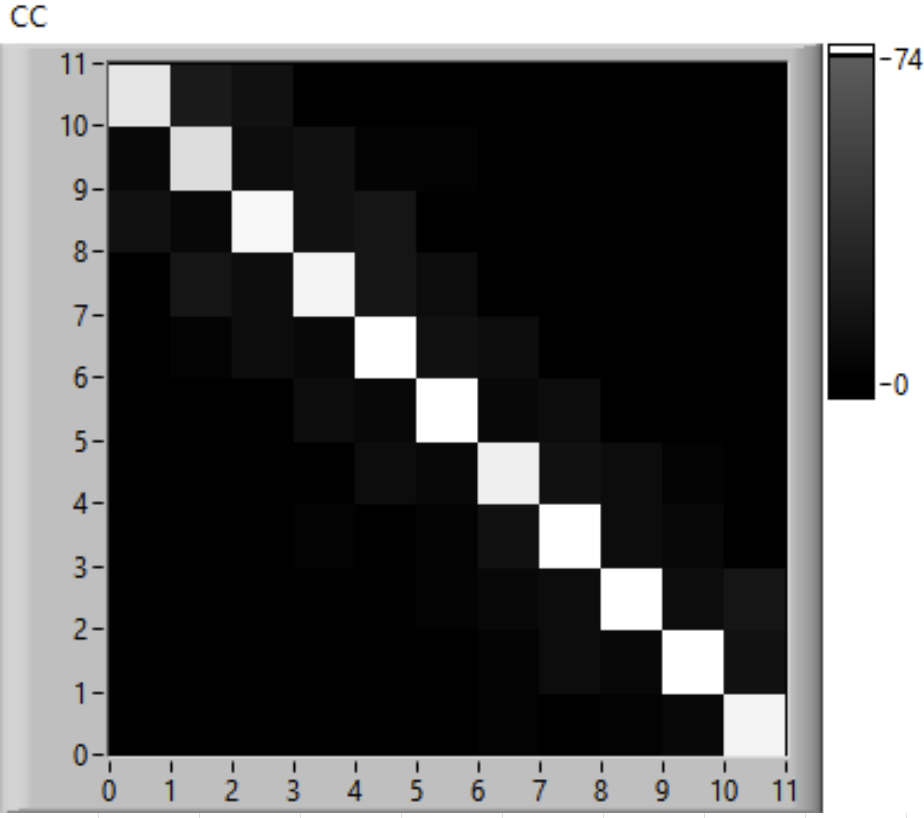}
\caption{\textbf{Spiral bandwidths used for alignment.} Left: classical back‑propagation with a $1550\text{ nm}$ laser; right: single‑photon scan with the SPDC source. The average crosstalk of the central element ($\ell=0$) with its four nearest neighbors is $0.4\%$ (left) and $3.3\%$ (right). Note that the LabVIEW labeling maps $0,1,\ldots,11 \rightarrow \ell=-5,-4,\ldots,5$.}
\label{fig:spiral-b}
\end{figure}

\item \textbf{Mask scaling for unbalanced superpositions.}
In the LabVIEW interface we explicitly compensated, at the mask‑design stage, for the different effective beam diameters that arise in \emph{unbalanced} superpositions such as $\big(\ket{\ell=3}+\ket{\ell=1}\big)/\sqrt{2}$, since the two constituent OAM modes have different radial profiles. Accordingly, the SLM masks for unbalanced superpositions were radially rescaled relative to the theoretical beam size expected at the SLM. This correction is important because the phase singularities that generate the desired OAM superposition must lie within the beam footprint: if the actual beam is smaller than assumed, it may miss the top and bottom singularities and effectively acquire only a single‑charge phase (e.g., $\ell=1$), reducing the superposition quality. For \emph{balanced} superpositions such as $\big(\ket{\ell=3}+\ket{\ell=-3}\big)/\sqrt{2}$ this issue is mitigated, as the required masks do not vary radially.

\begin{figure}[H]
\centering
\includegraphics[width=0.4\linewidth]{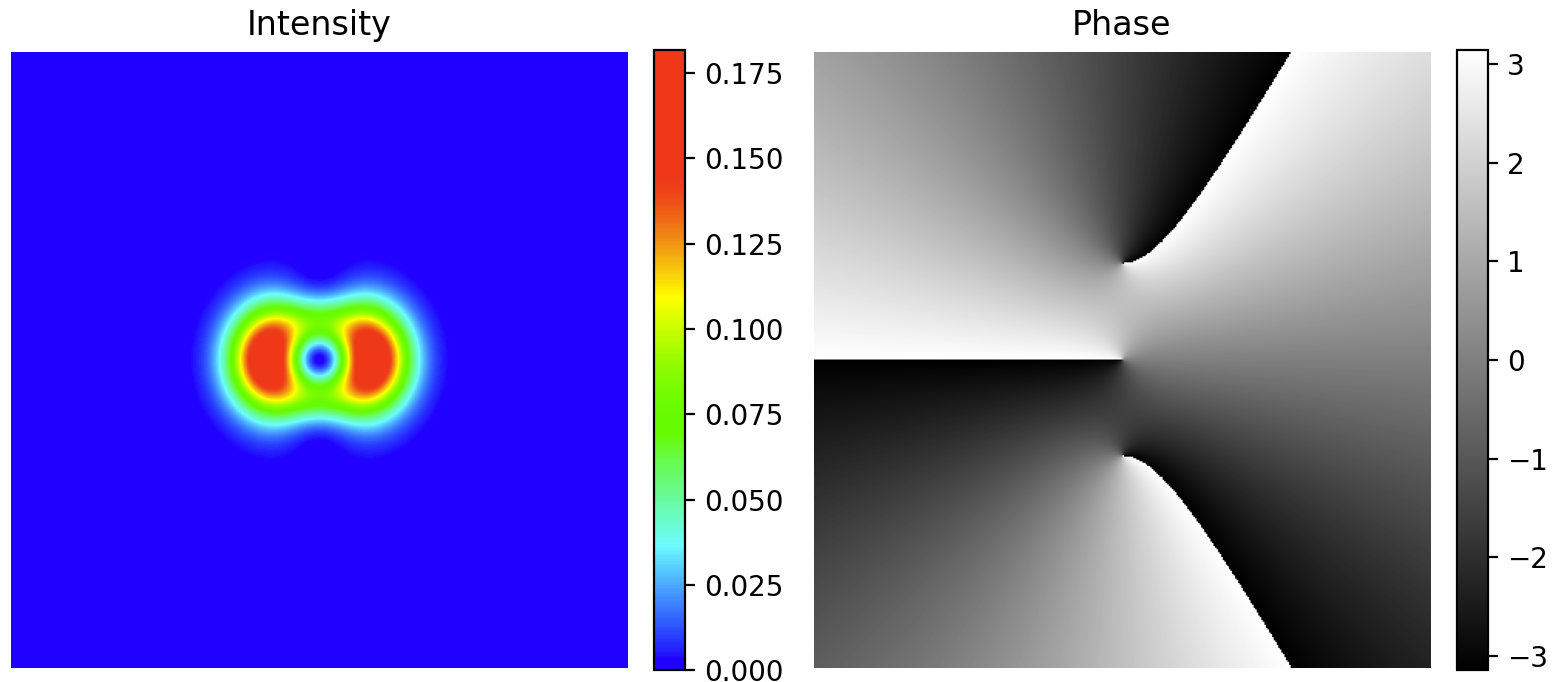}
\hspace{0.2cm}
\includegraphics[width=0.4\linewidth]{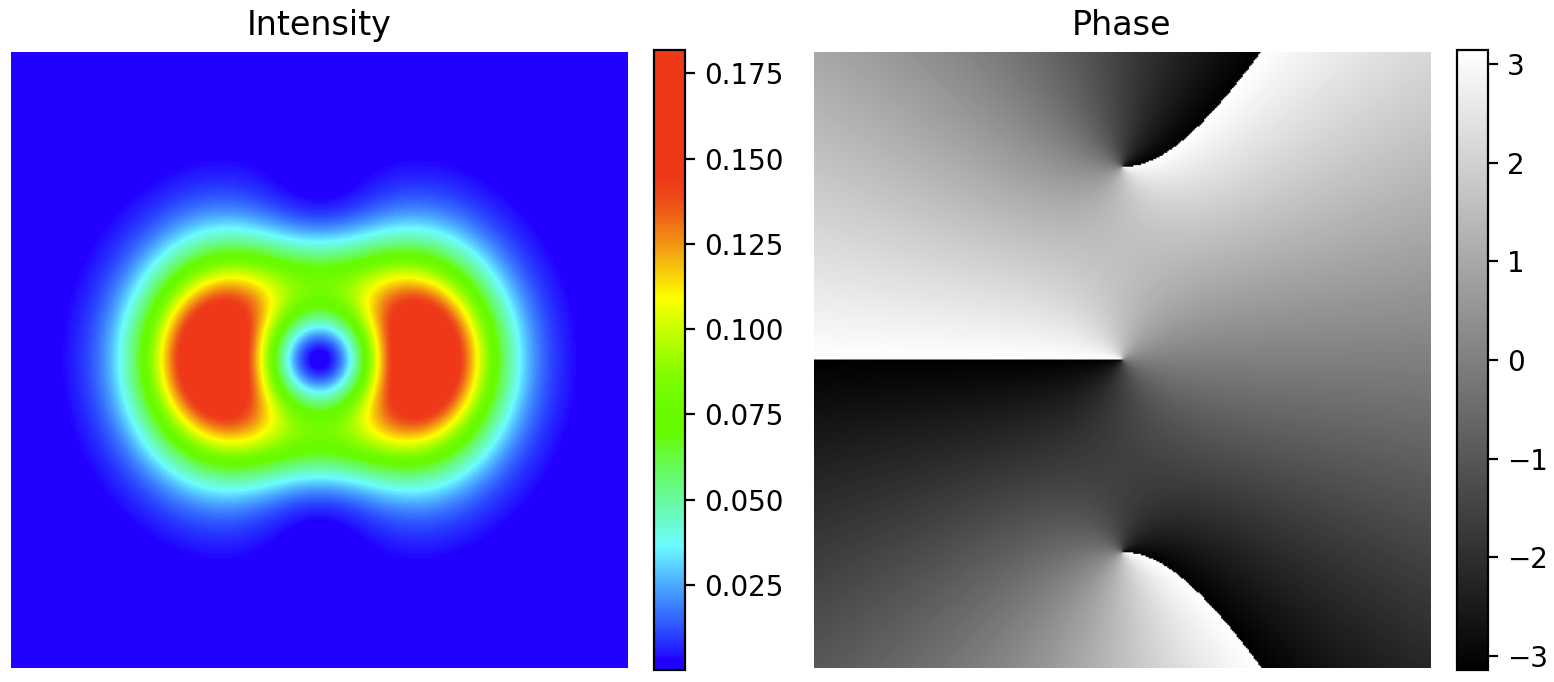}
\caption{\textbf{Intensity and phase for $\big(\ket{\ell=3}+\ket{\ell=1}\big)/\sqrt{2}$ .} The phase plot shows the SLM mask (omitting the blazed grating typically applied to select the less noisy first diffracted order, $m=1$). The intensity plots illustrate the generated superposition profile for a Gaussian input beam.}
\label{fig:intensity-phase}
\end{figure}

\end{enumerate}

\subsubsection{\label{sec:s-perms} Permutations}

Since Alice and Bob have the freedom to decide which element from the qudit labels $(q)$ corresponds to each element of the OAM labels $(\ell)$, the reconstructed $P$-matrix could be passed through a symmetry-based averaging procedure designed to dilute configuration-dependent systematic errors, such as SLM pixel discretization, small but fixed beam misalignments, or slightly different but fixed beam diameters.\\

The key idea is to exploit all \emph{physically allowed} permutations of the already chosen qudit basis. In our case, these are generated by cyclic permutations of an initial labeling and their mirror (reversed) versions. Concretely, consider as an example a $d=7$ qudit encoded as: $$
\ket{q=0,1,2,...,6} \;\longleftrightarrow\;
\ket{\ell = 0,\,-1,\,+1,\,-3,\,+2,\,-2,\,+3}.$$
Once this is fixed, Alice and Bob have agreed also on the finite set of permutations to consider, since all the possible permutations that will not change the POVMs from  \cref{fig:PVMd=5} are listed in \cref{tab:all-permutations}. Each row of these tables specifies a relabeling $\pi$ of the logical basis.\\

\begin{table}[H]
    \centering
    \caption{Cyclic and mirror permutations of the initial OAM labeling example
    $(0,\,-1,\,+1,\,-3,\,+2,\,-2,\,+3)$.
    For the cyclic block, each row $\pi^{(k)}$ corresponds to a shift by $k$ positions of the original sequence.
    For the mirror block, each row $\tilde{\pi}^{(k)}$ is obtained by a cyclic shift by $k$ positions of the mirrored sequence.
    Columns are labeled by the logical index $q$, and entries give the OAM charge $\ell$ assigned to that label under the permutation.}
    \label{tab:all-permutations}
    \setlength{\tabcolsep}{6pt}
    \renewcommand{\arraystretch}{1.2}
    \begin{tabular}{|c c|ccccccc|}
        \hline
        & & \multicolumn{7}{c|}{\textbf{Logical basis} $\ket{q}$} \\
        & & 0 & 1 & 2 & 3 & 4 & 5 & 6\\
        \hline
        \multirow{4}{*}{\rotatebox[origin=c]{90}{\textbf{Cyclic} $\ket{\ell}$}}
        & $\pi^{(0)}$ & $0$  & $-1$ & $+1$ & $-3$ & $+2$ & $-2$ & $+3$ \\
        & $\pi^{(1)}$ & $-1$ & $+1$ & $-3$ & $+2$ & $-2$ & $+3$ & $0$  \\
        & $\vdots$ & \multicolumn{7}{c|}{$\vdots$}\\
        & $\pi^{(6)}$ & $+3$ & $0$  & $-1$ & $+1$ & $-3$ & $+2$ & $-2$ \\
        \hline
        \multirow{4}{*}{\rotatebox[origin=c]{90}{\textbf{Mirror} $\ket{\ell}$}}
        & $\tilde{\pi}^{(0)}$ & $+3$ & $-2$ & $+2$ & $-3$ & $+1$ & $-1$ & $0$  \\
        & $\tilde{\pi}^{(1)}$ & $-2$ & $+2$ & $-3$ & $+1$ & $-1$ & $0$  & $+3$ \\
        & $\vdots$ & \multicolumn{7}{c|}{$\vdots$}\\
        & $\tilde{\pi}^{(6)}$ & $0$  & $+3$ & $-2$ & $+2$ & $-3$ & $+1$ & $-1$ \\
        \hline
    \end{tabular}
\end{table}

Before data acquisition, Alice and Bob agree on a fixed sequence of permutations, specifying for each run which row of \cref{tab:all-permutations} will be used. This sequence is chosen so that each permutation appears the same number of times over the full data set. Thus, every time a given logical state $\rho$ is prepared and measured, it is implemented in one of the encoding choices from the table, and by the end of the experiment all labelings have been used equally often. In this way, fixed configuration-dependent imperfections are effectively \textit{scrambled} across the different qudit basis mappings, reducing their systematic impact on the final reconstructed probability matrix $P$.\\

Furthermore, when combining the reconstructions obtained from all permutations, it is important to note that these are not statistically independent realizations, but merely re-indexings of the same underlying $49\times 49$ dataset. For each physically allowed permutation $\pi$ of the logical basis (generated as described above), we obtain a corresponding reconstructed probability matrix $P^{(\pi)}$ and an associated per-cell standard deviation matrix $\sigma^{(\pi)}$. We then define the permutation-averaged probability matrix as
$
P = m^{-1}\sum_{\pi} P^{(\pi)},
$
where $m$ is the total number of cyclic and mirror permutations used. However, for the associated uncertainties we avoid the usual $m^{-1/2}$ reduction appropriate for independent samples. Instead, we use a correlation-agnostic expression:
$$
\sigma_{\mathrm{ag}}(i,j) = \sqrt{\sum_{\pi} \bigl[\sigma^{(\pi)}(i,j)\bigr]^2},
$$
\emph{without} dividing by $m$, precisely because each $P^{(\pi)}$ is built from the same raw data, only relabeled. In this way, the permutation averaging can dilute configuration-dependent systematic effects in $P$, while \textbf{the reported statistical errors are not artificially diminished} by treating permutation-related reconstructions as independent measurements.

In our case, permutations can be used to boost the certification for $d=5$, as shown in \cref{tab:robust-cert-n15}. By averaging over the physically allowed values, permutations effectively dilute the fixed experimental error, yielding an experimental $P$ matrix that is closer to the theoretical one. This results in a smaller value of $\|E\|_2$ and therefore allows for a higher certification.

\begin{table}[h]
  \centering
  \caption{Comparison for the certification case $n=15$ ($d_Q = 5$) from \cref{tab:robust-cert}. Using permutations yields a significant improvement, increasing the certified value from $d^{\mathrm{exp}}_C=8$ to the theoretical limit $2d_Q=10$.}
  \label{tab:robust-cert-n15}
  \setlength{\tabcolsep}{3pt}
  \renewcommand{\arraystretch}{1.4}

  \resizebox{0.6\textwidth}{!}{%
  \begin{tabular}{|c|c c c|c c c|}
    \hline
    Permutations? & $n$ & $d_Q$ & $d^{\mathrm{exp}}_C$ & $\sigma_{d^{\mathrm{exp}}_C}(P)$ & $\|E\|_2$ & $\sigma_{d^{\mathrm{exp}}_C+1}(P)$ \\
    \hline
    No  & 15 & 5 & 8  &
    $1.24\,\text{\footnotesize$\times 10^{-1}$}$ &
    $(5.87\pm0.13)\,\text{\footnotesize$\times 10^{-2}$}$ &
    $4.00\,\text{\footnotesize$\times 10^{-2}$}$ \\
    Yes & 15 & 5 & 10 &
    $4.00\,\text{\footnotesize$\times 10^{-2}$}$ &
    $(2.49\pm0.16)\,\text{\footnotesize$\times 10^{-2}$}$ &
    $7\,\text{\footnotesize$\times 10^{-17}$}$ \\
    \hline
  \end{tabular}
  }
\end{table}

We now present the corresponding plots for all protocols when permutations are included. Significant improvement in uniformity is achieved.

\ThreePmatrixMainFigure
  {qcom-outputs-M3/Theory/d3}
  {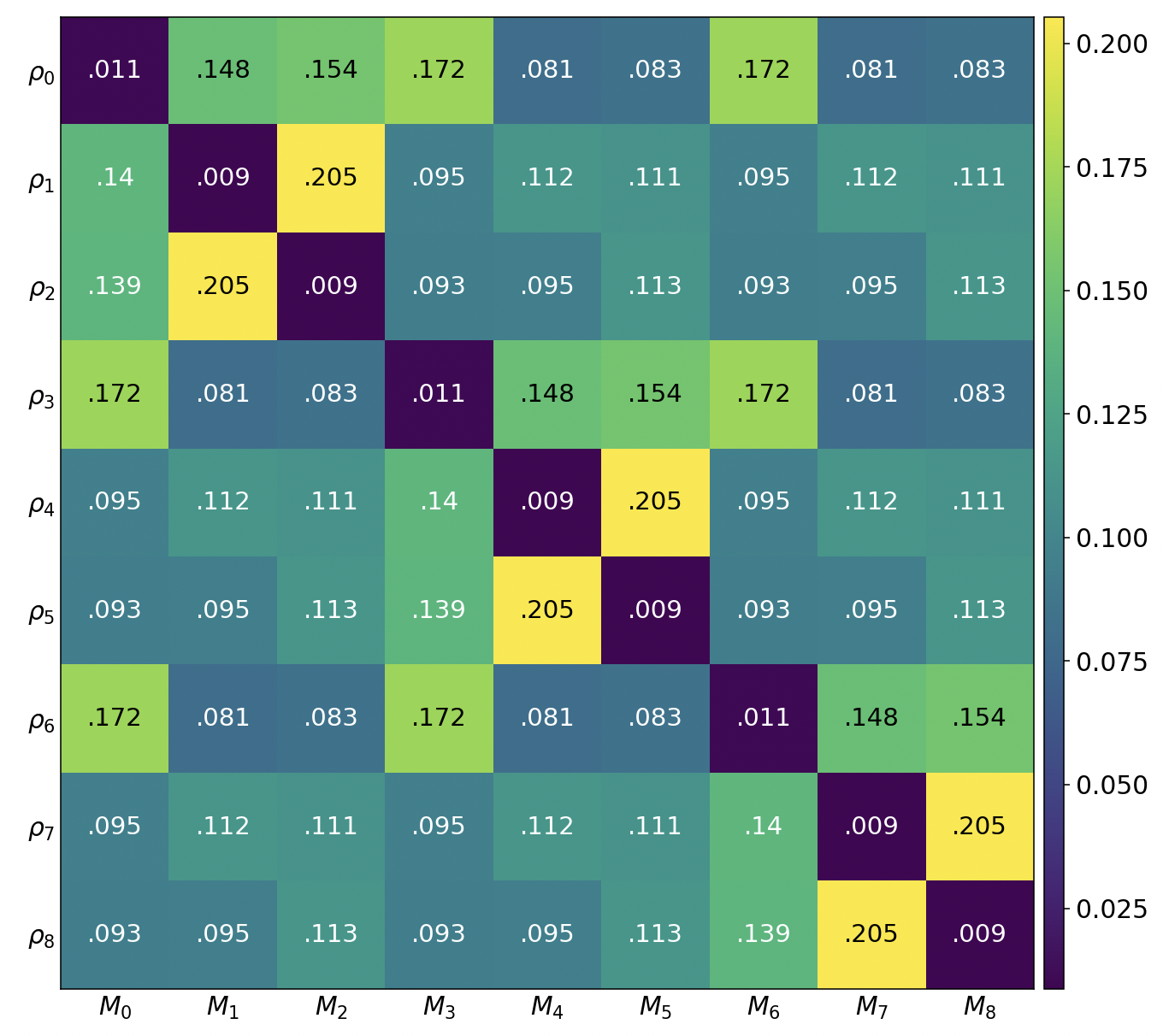}
  {qcom-outputs-M3/Theory/convex-d5}
  {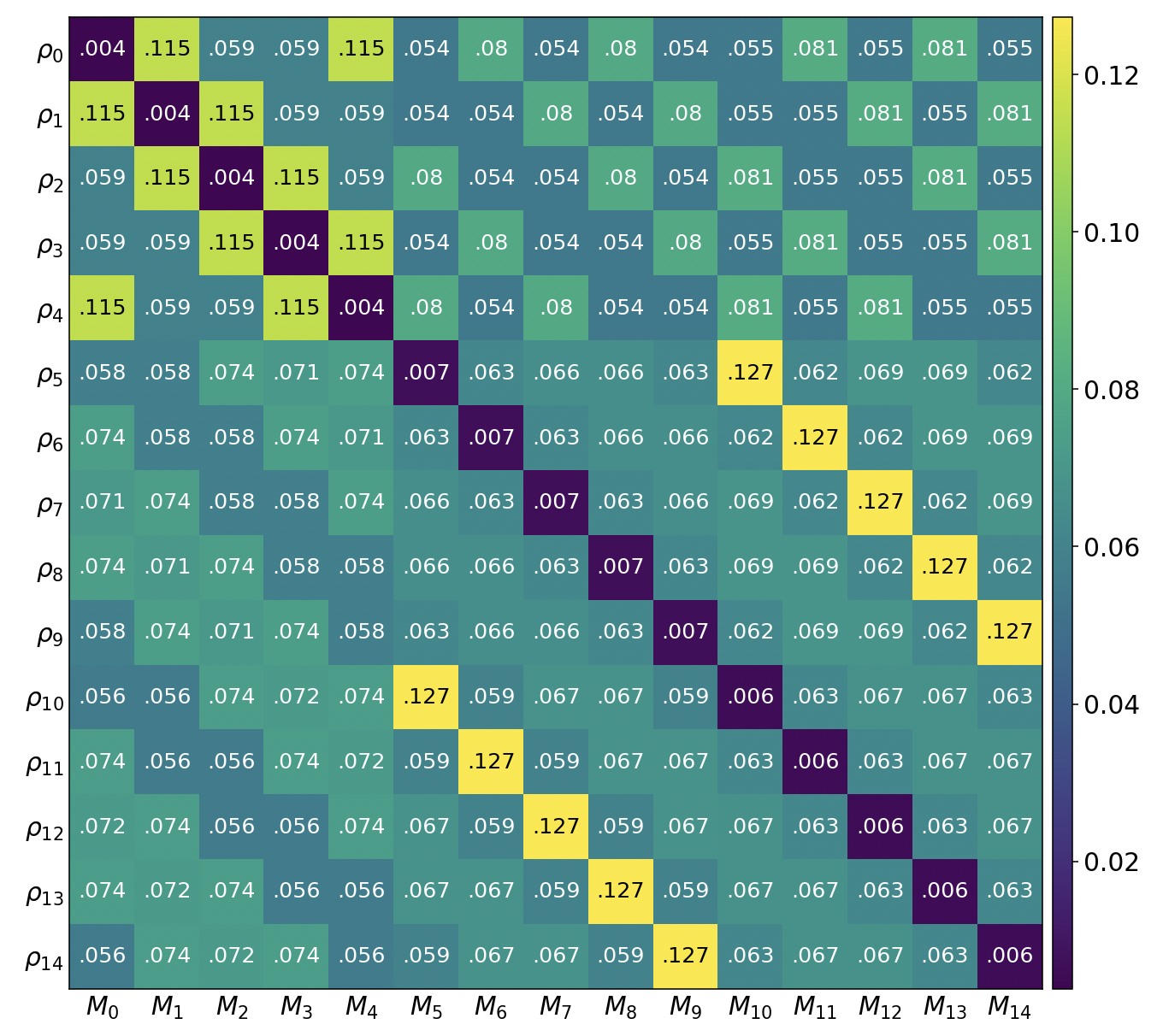}
  {qcom-outputs-M3/Theory/convex-d7}
  {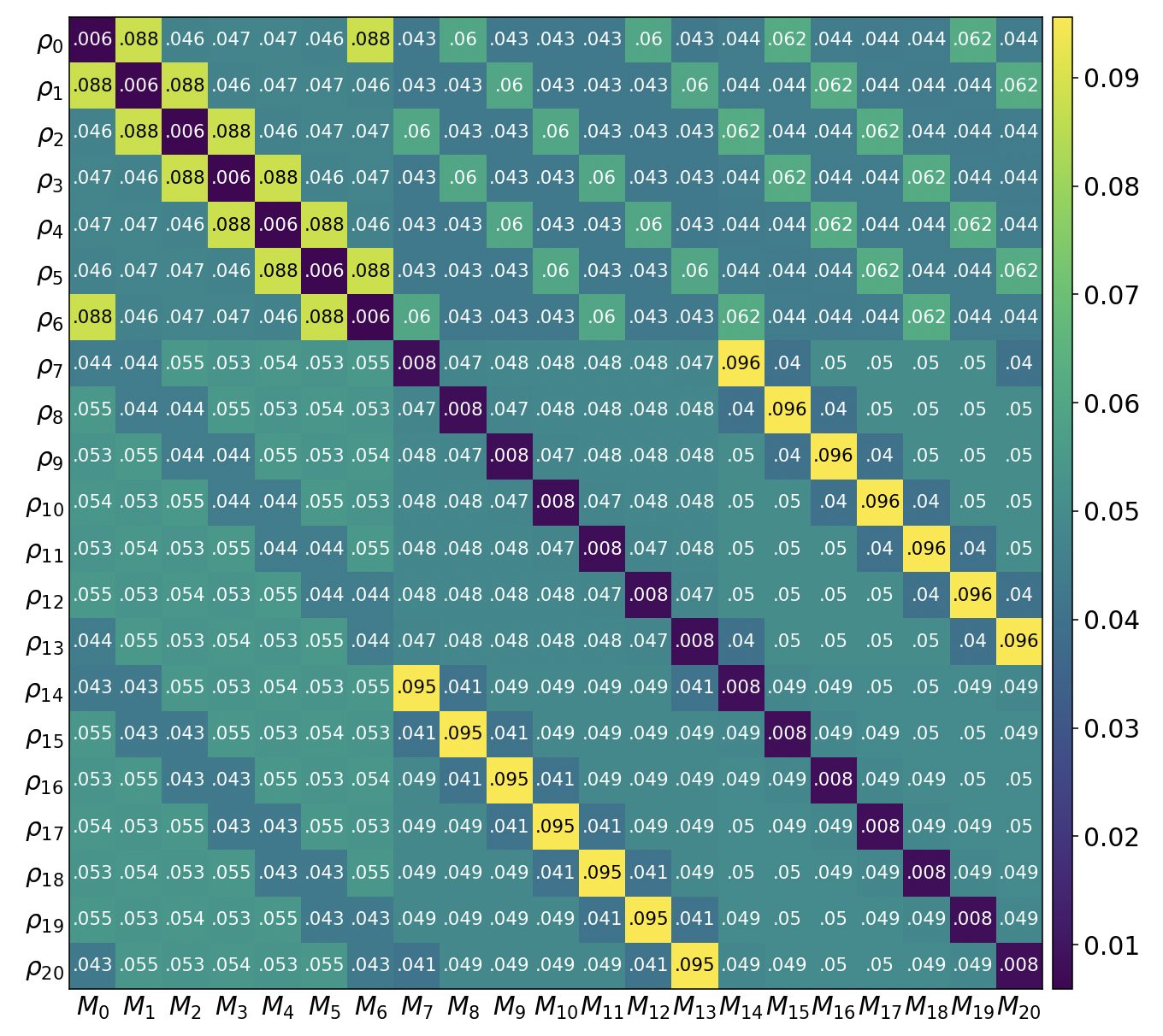}
  {\textbf{Experimental results taken for the convex protocol, with permutations.} 
   Theoretical $P$-matrices with zeros on the diagonal (left) and their experimentally reconstructed counterparts (right) for $d=3,5,7$ (top–bottom).
   The OAM ($\ell$) $\leftrightarrow$ qudit mappings (order of $\ket{0},\ket{1},\ldots,\ket{6}$) chosen to minimize $\|E\|_2$ are: 
   $P_{3}\!:\!(\ell=+0,+1,-1)$; 
   $P_{5}\!:\!(\ell=+1,+2,-2,+3,-3)$; 
   $P_{7}\!:\!(\ell=+0,+1,+2,-3,+3,-2,-1)$.}%
  {fig:convex-siperms}
  {{ (with permutations)}}

\ThreePmatrixMainFigure
  {qcom-outputs-M3/Theory/d3}
  {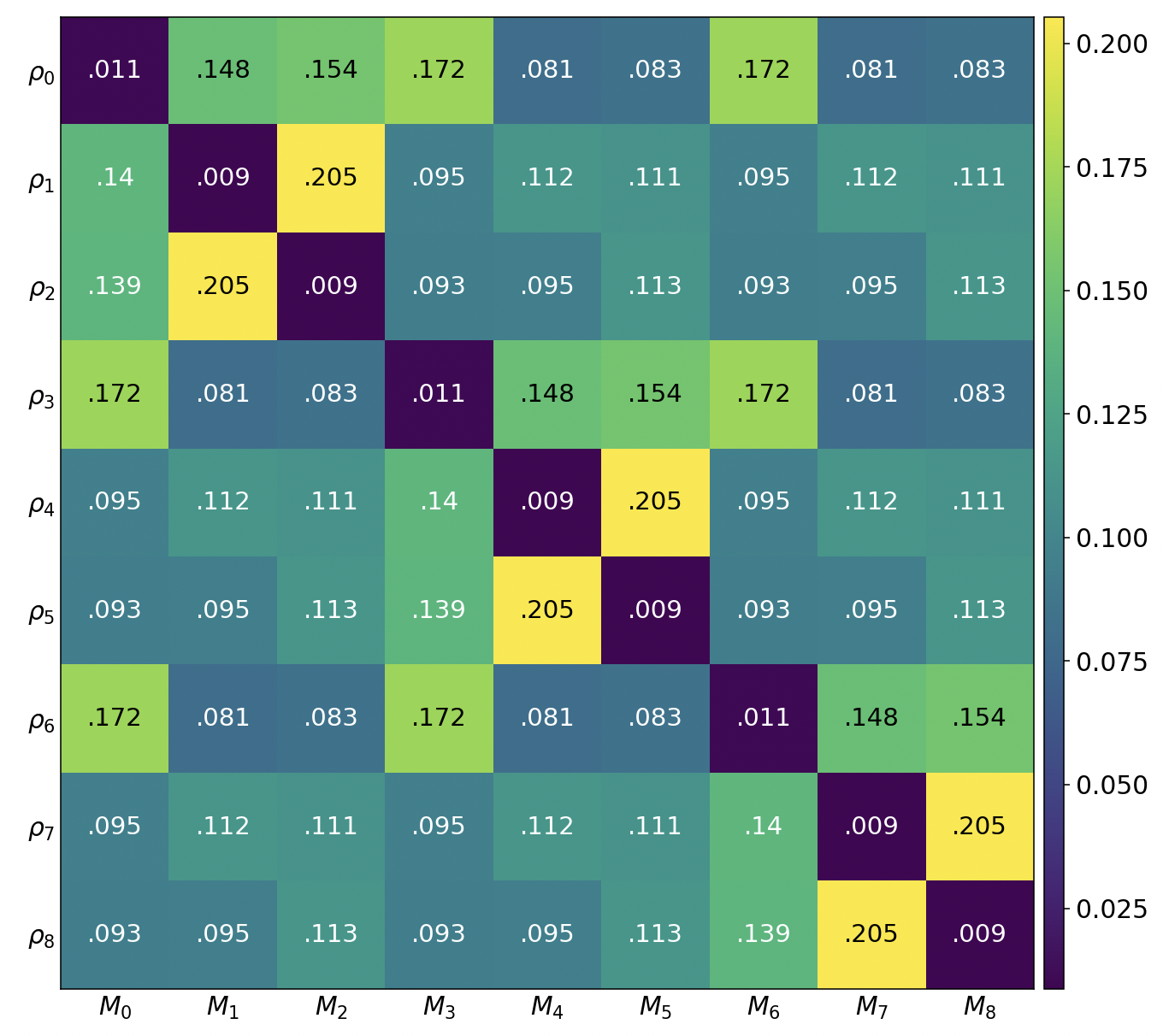}
  {qcom-outputs-M3/Theory/coherent-d5}
  {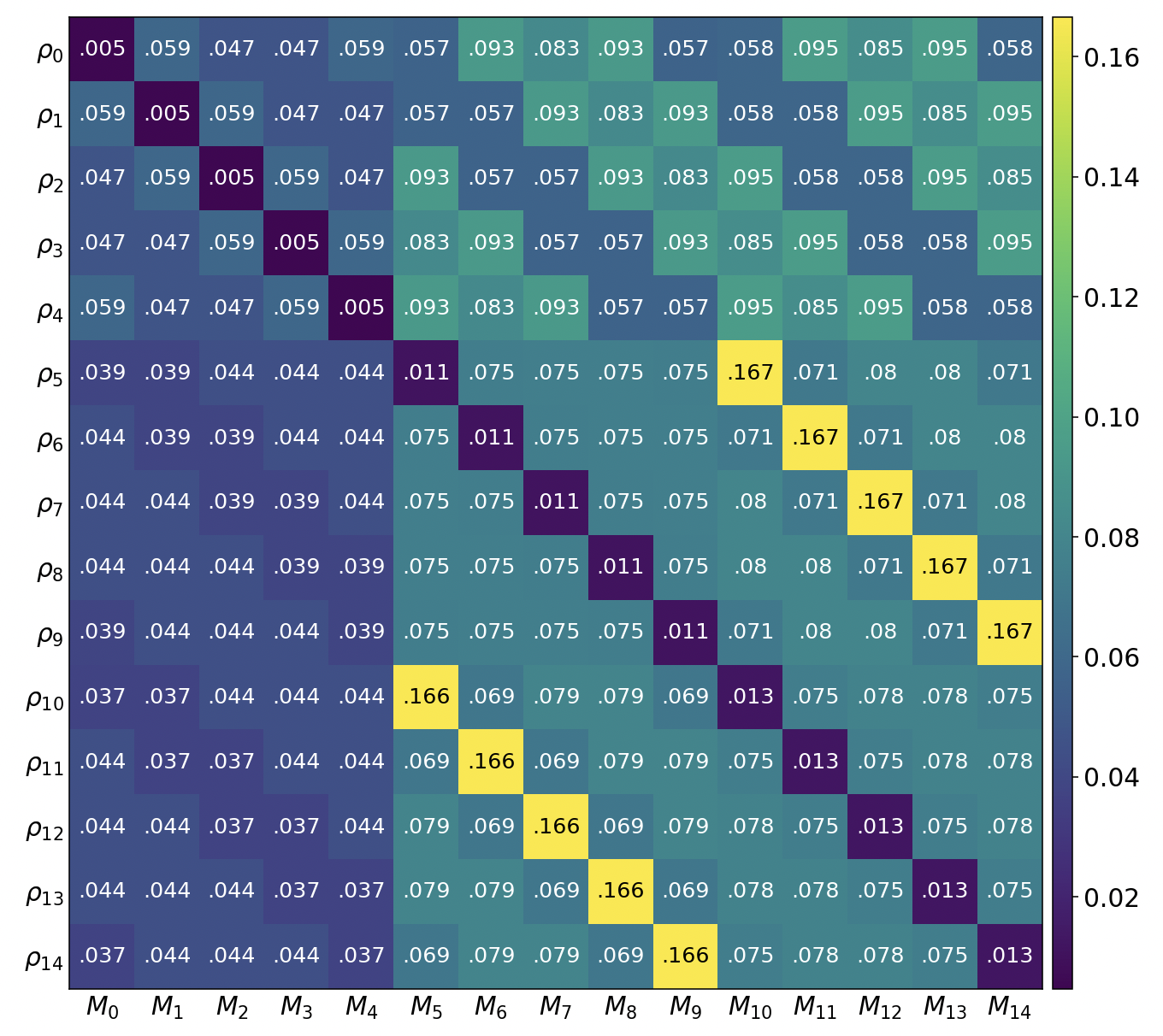}
  {qcom-outputs-M3/Theory/coherent-d7}
  {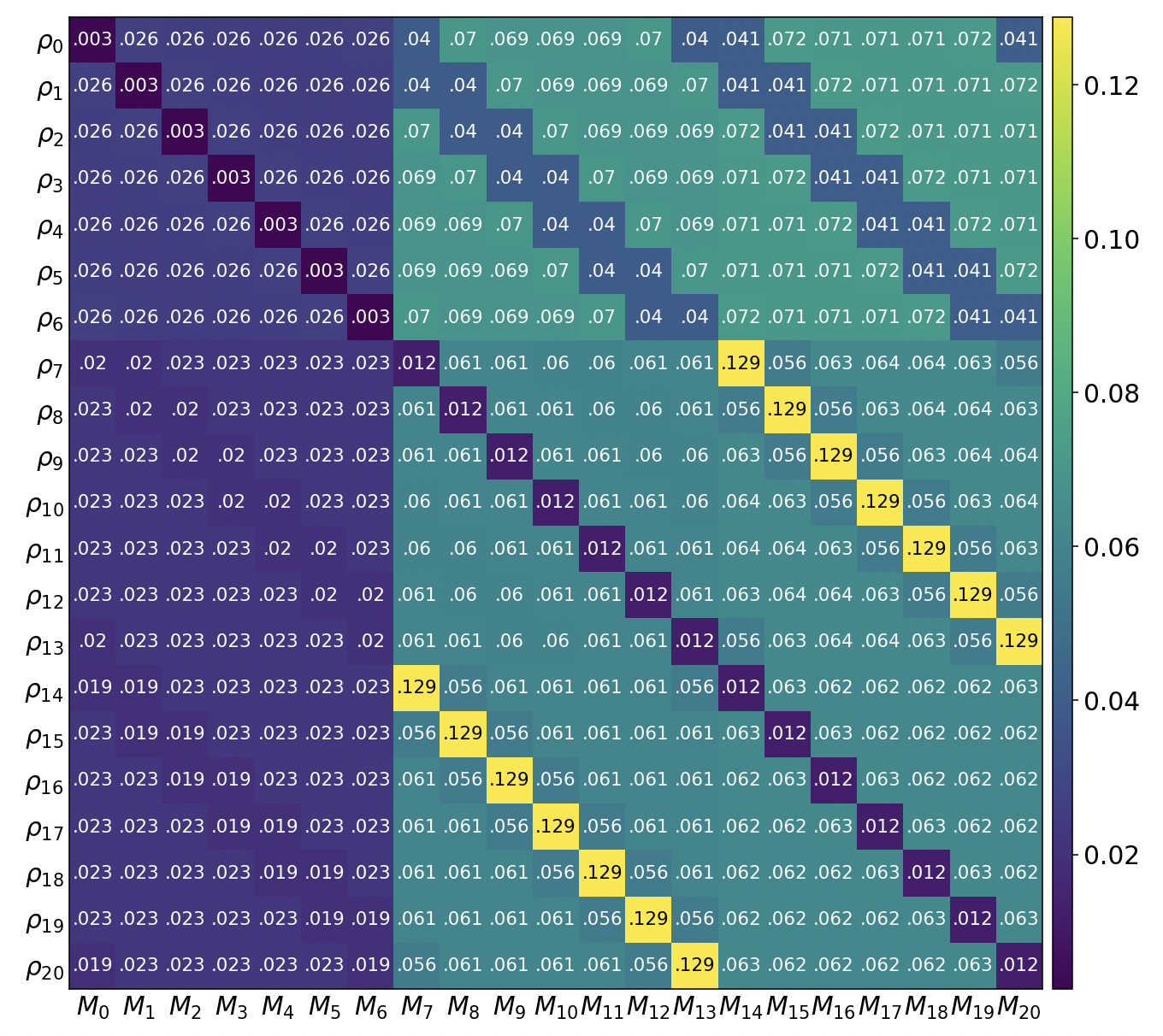}
  {\textbf{Experimental results taken for the coherent protocol, with permutations.} 
   Theoretical $P$-matrices with zeros on the diagonal (left) and their experimentally reconstructed counterparts (right) for $d=3,5,7$ (top–bottom).
   The OAM ($\ell$) $\leftrightarrow$ qudit mappings (order of $\ket{0},\ket{1},\ldots,\ket{6}$) chosen to minimize $\|E\|_2$ are: 
   $P_{3}\!:\!(\ell=+0,+1,-1)$; 
   $P_{5}\!:\!(\ell=+0,+1,-2,-3,-1)$; 
   $P_{7}\!:\!(\ell=+0,+1,+2,-3,-2,+3,-1)$.}%
  {fig:coherent-siperms}
  {{(with permutations)}}

\ThreePmatrixMainFigure
  {qcom-outputs-M3/Theory/d3}
  {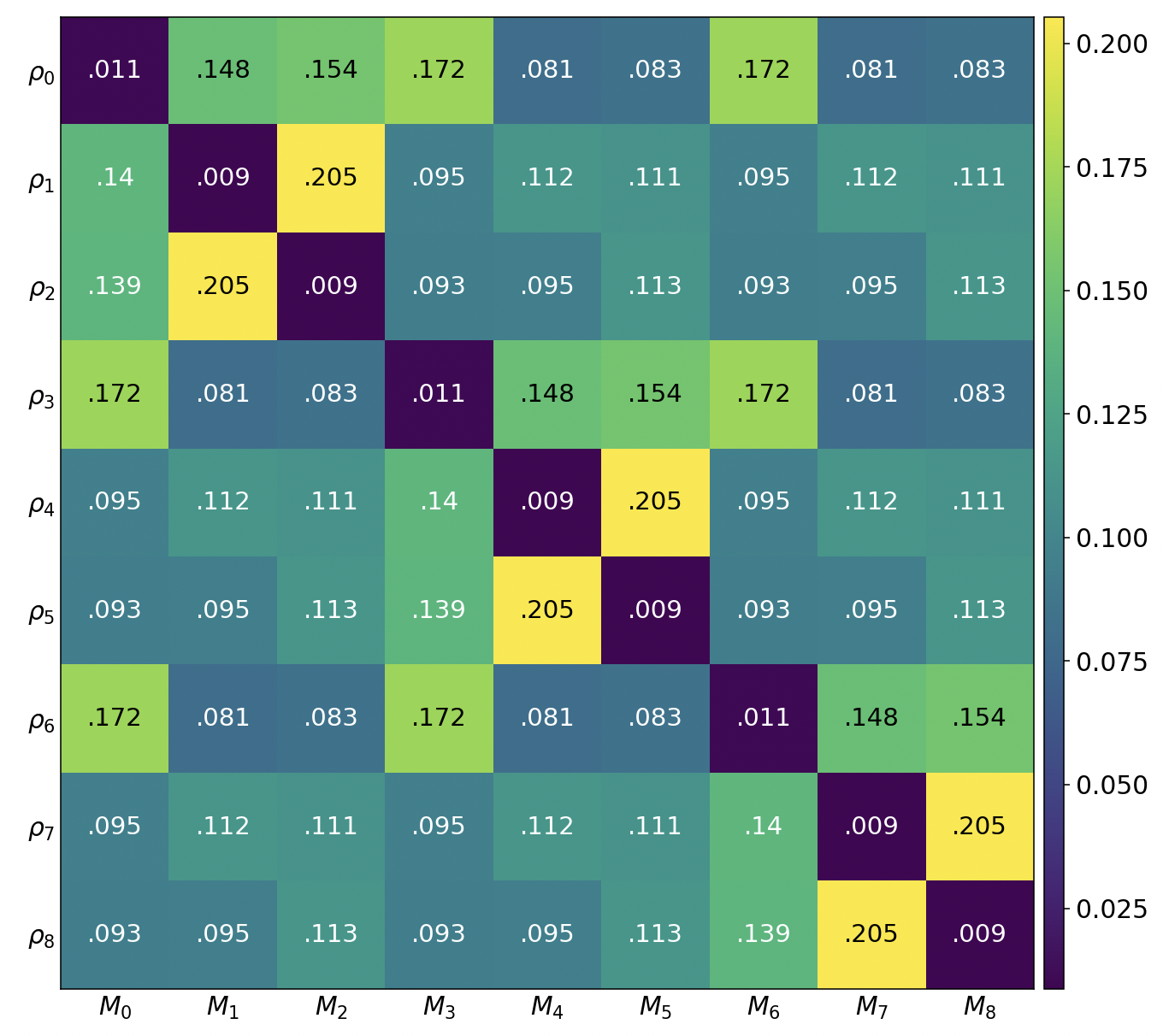}
  {qcom-outputs-M3/Theory/incoherent-d5}
  {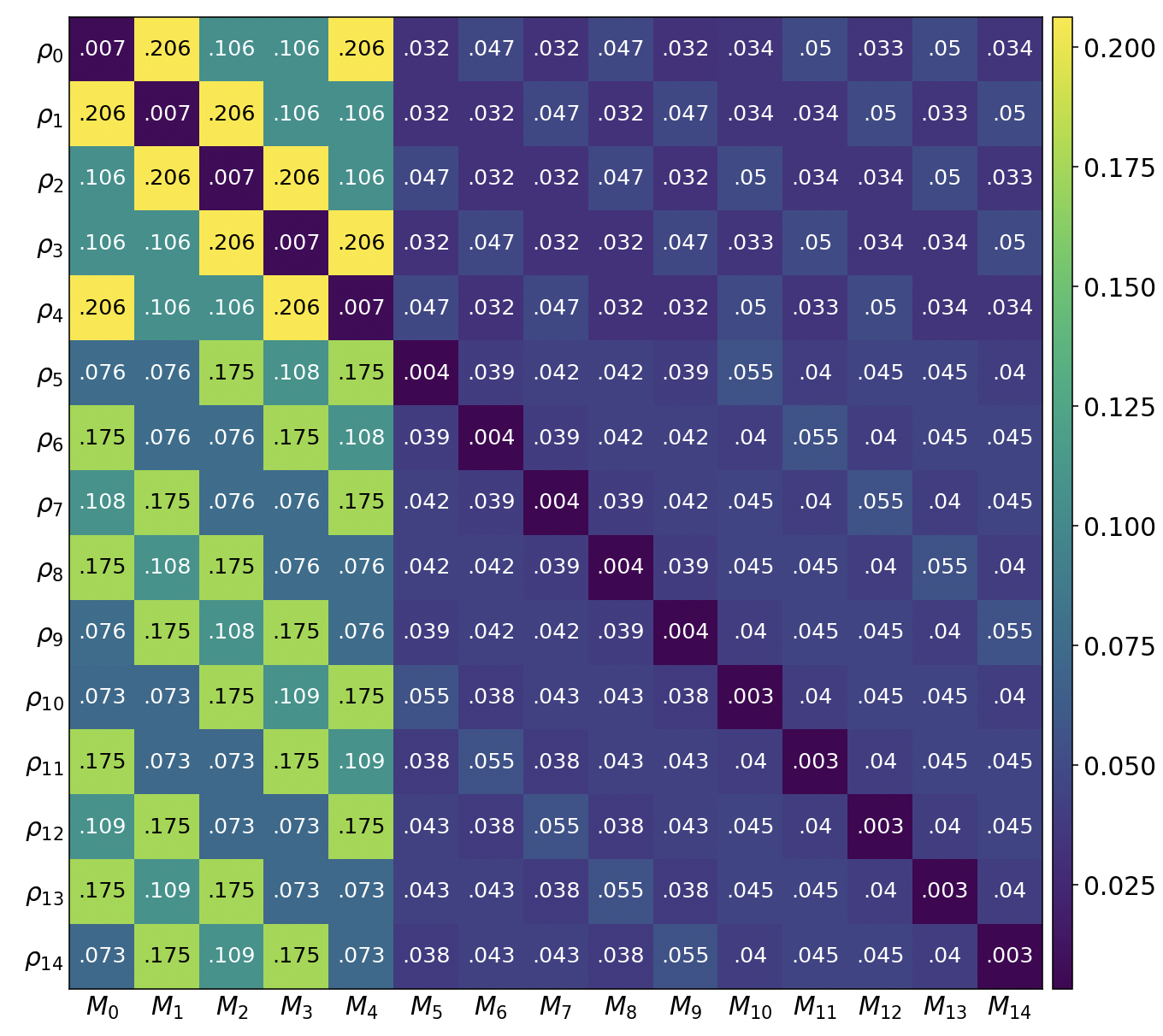}
  {qcom-outputs-M3/Theory/incoherent-d7}
  {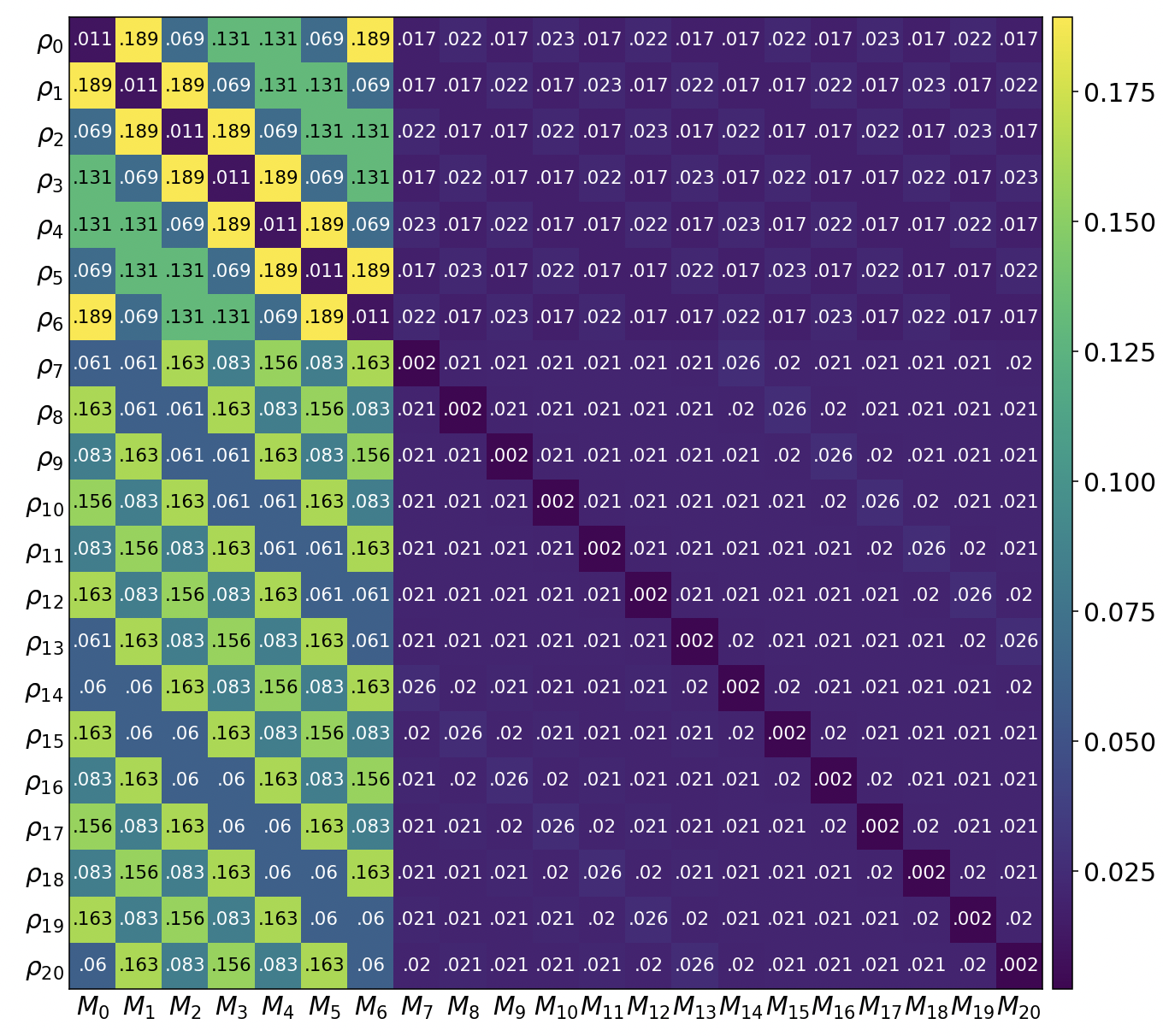}
  {\textbf{Experimental results taken for the incoherent protocol, with permutations.} 
   Theoretical $P$-matrices with zeros on the diagonal (left) and their experimentally reconstructed counterparts (right) for $d=3,5,7$ (top–bottom).
   The OAM ($\ell$) $\leftrightarrow$ qudit mappings (order of $\ket{0},\ket{1},\ldots,\ket{6}$) chosen to minimize $\|E\|_2$ are: 
   $P_{3}\!:\!(\ell=+0,+1,-1)$; 
   $P_{5}\!:\!(\ell=+1,-2,+2,+3,-3)$; 
   $P_{7}\!:\!(\ell=+0,-2,+1,+2,-1,+3,-3)$.}%
  {fig:incoherent-siperms}
  {{ (with permutations)}}

\ThreePmatrixMainFigure
  {qcom-outputs-M3/Theory/d3}
  {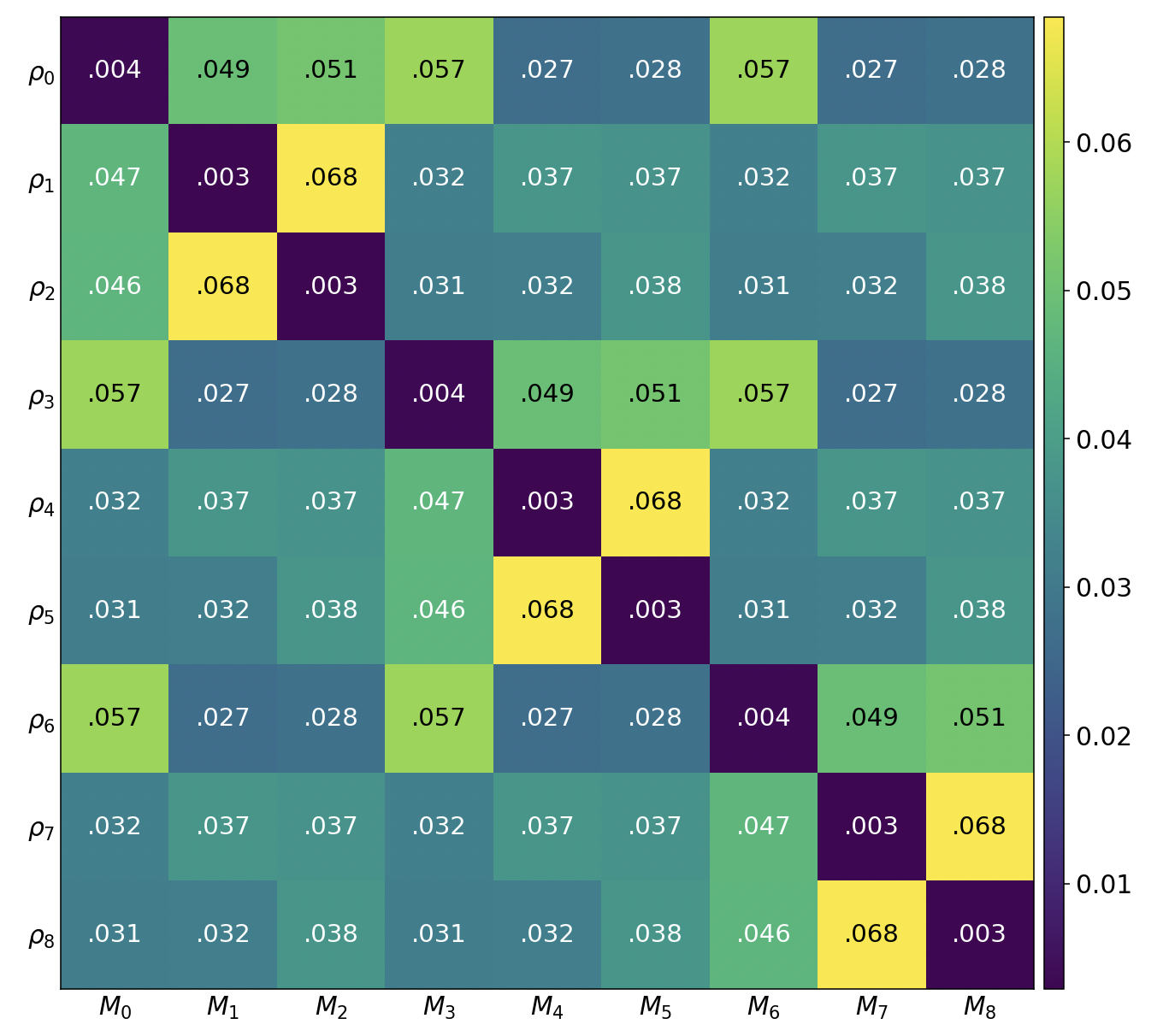}
  {qcom-outputs-M3/Theory/maximal-d5}
  {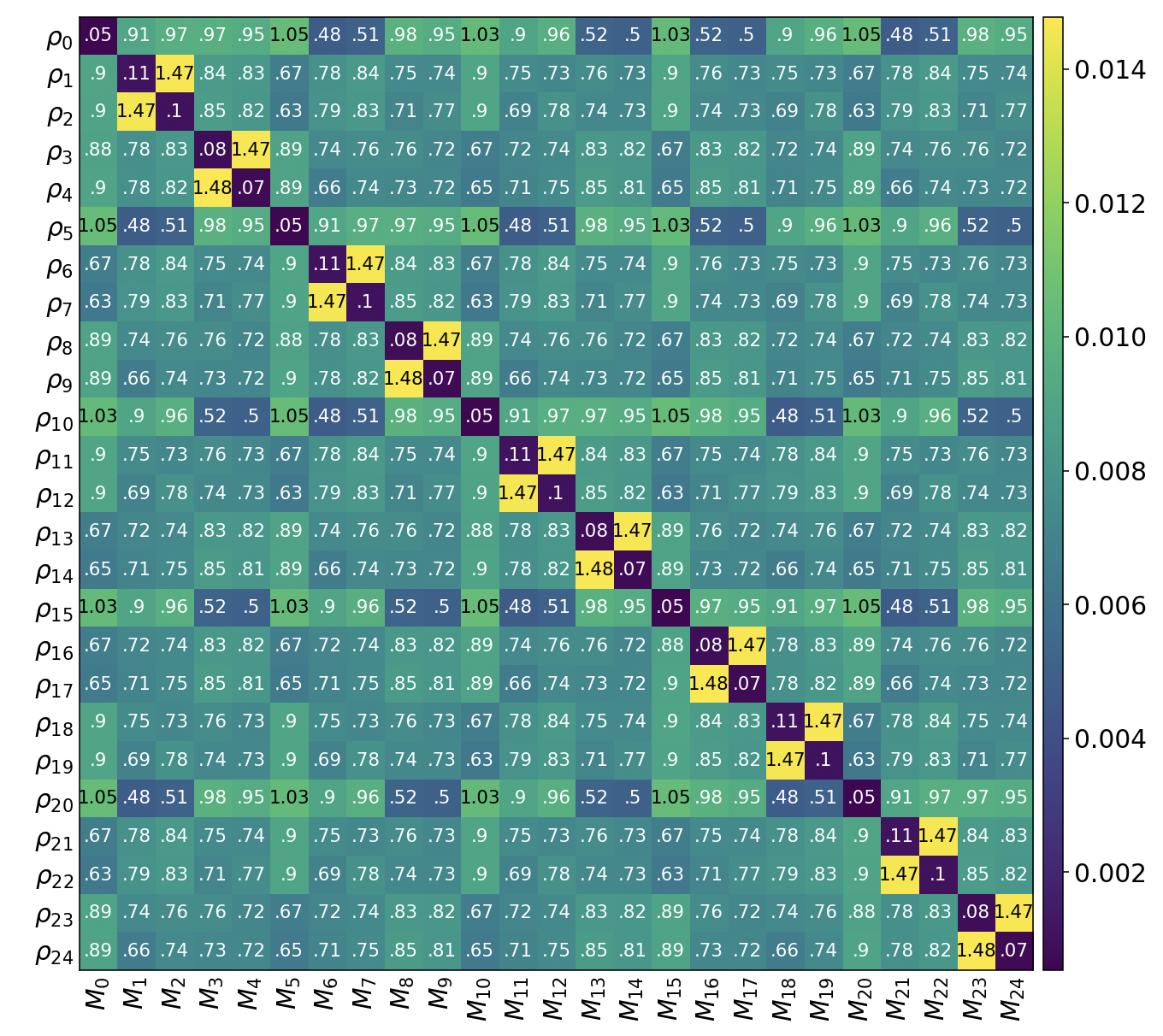}
  {qcom-outputs-M3/Theory/maximal-d7}
  {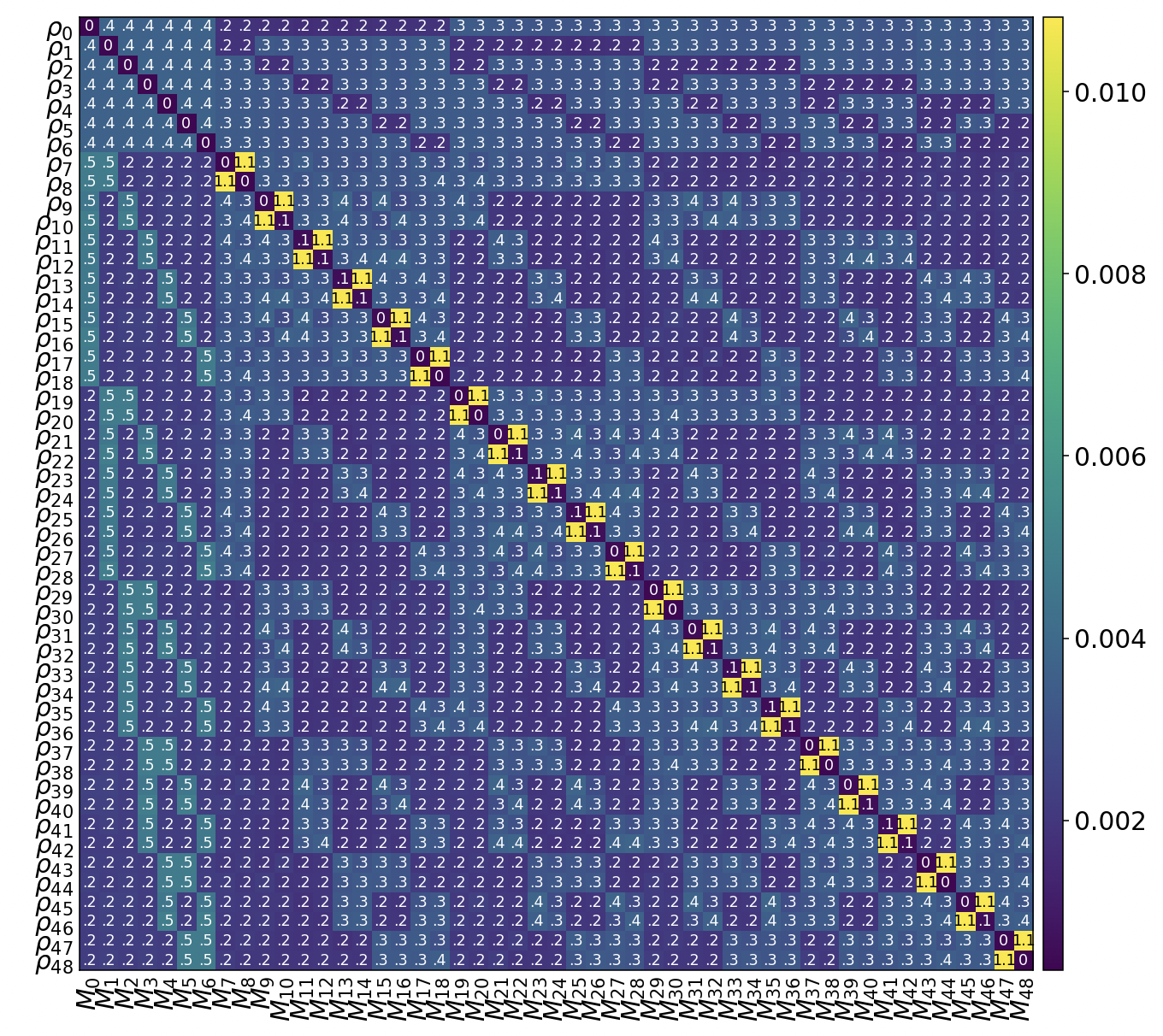}
  {\textbf{Experimental results taken for the maximal protocol, with permutations.} 
   Theoretical $P$-matrices with zeros on the diagonal (left) and their experimentally reconstructed counterparts (right) for $d=3,5,7$ (top–bottom).
   The OAM ($\ell$) $\leftrightarrow$ qudit mappings (order of $\ket{0},\ket{1},\ldots,\ket{6}$) chosen to minimize $\|E\|_2$ are: 
   $P_{3}\!:\!(\ell=+0,+1,-1)$; 
   $P_{5}\!:\!(\ell=-1,+2,-2,+3,-3)$; 
   $P_{7}\!:\!(\ell=+0,+1,-1,+2,-2,+3,-3)$.\\ Note: For $d=5$ and $d=7$, the cell values are shown multiplied by 100, while the color bars show the true values.}%
  {fig:maximal-siperms}
  {{ (with permutations)}}

\ThreePmatrixMainFigure
  {qcom-outputs-M3/Theory/d3}
  {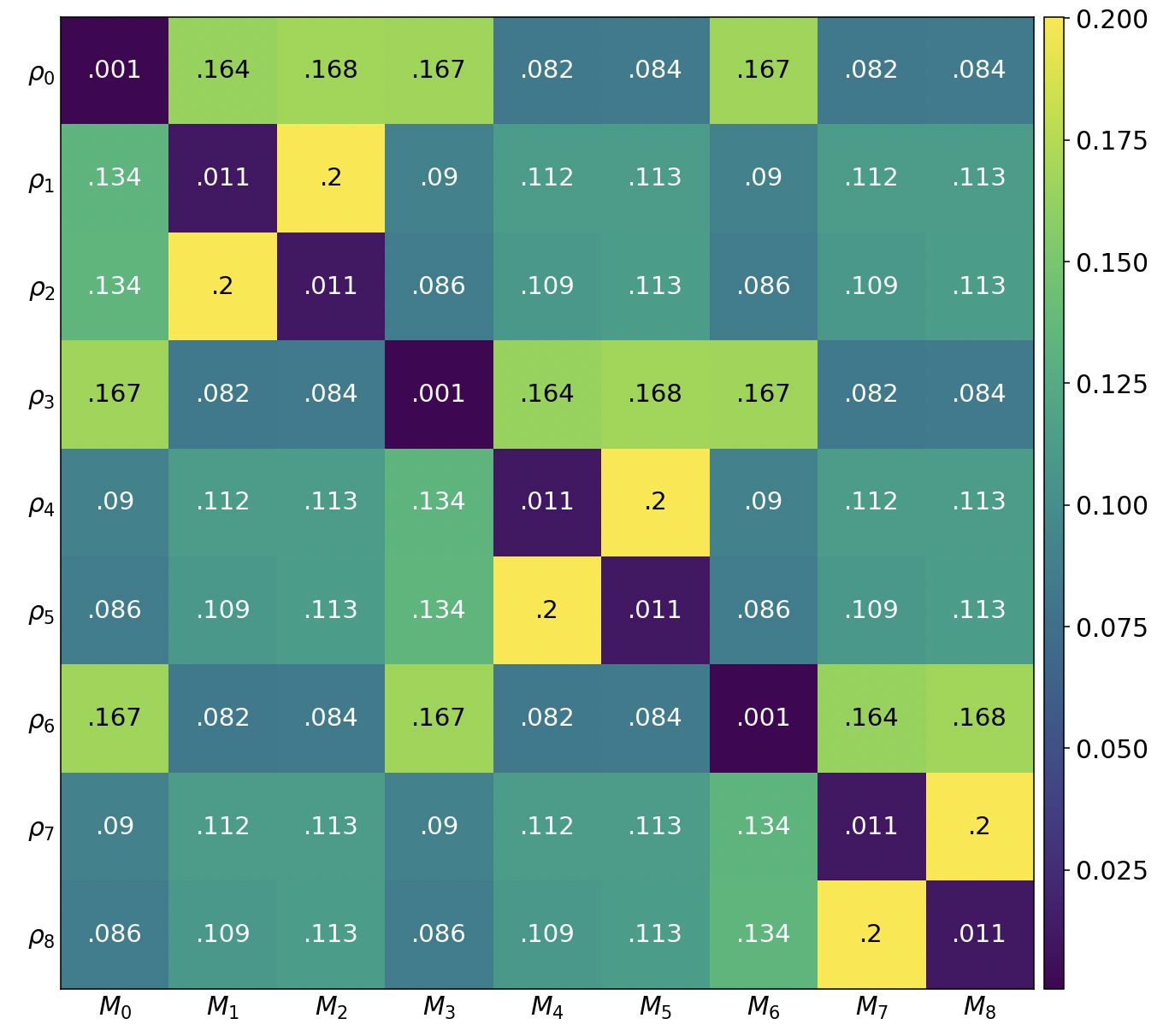}
  {qcom-outputs-M3/Theory/convex-d5}
  {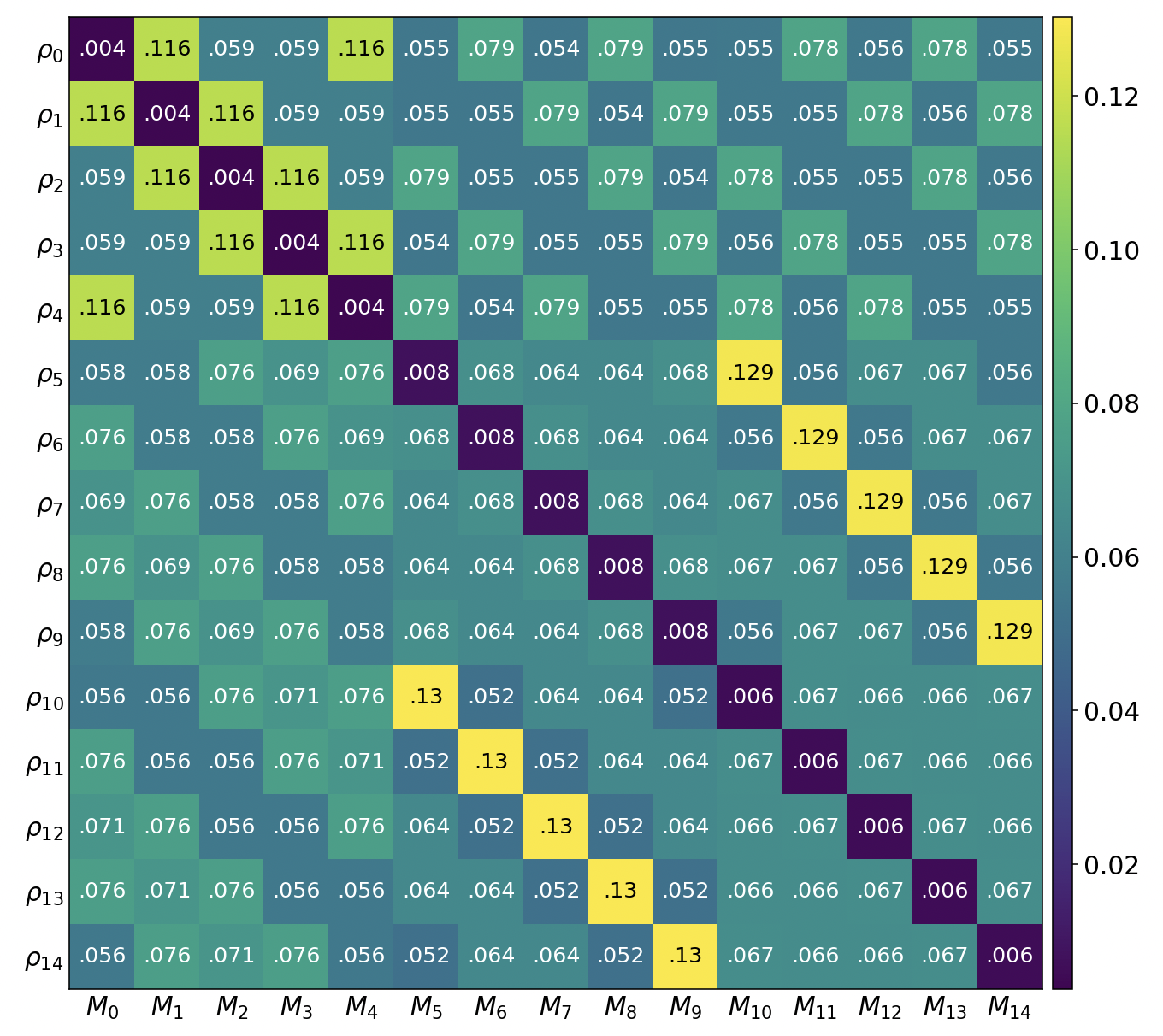}
  {qcom-outputs-M3/Theory/convex-d7}
  {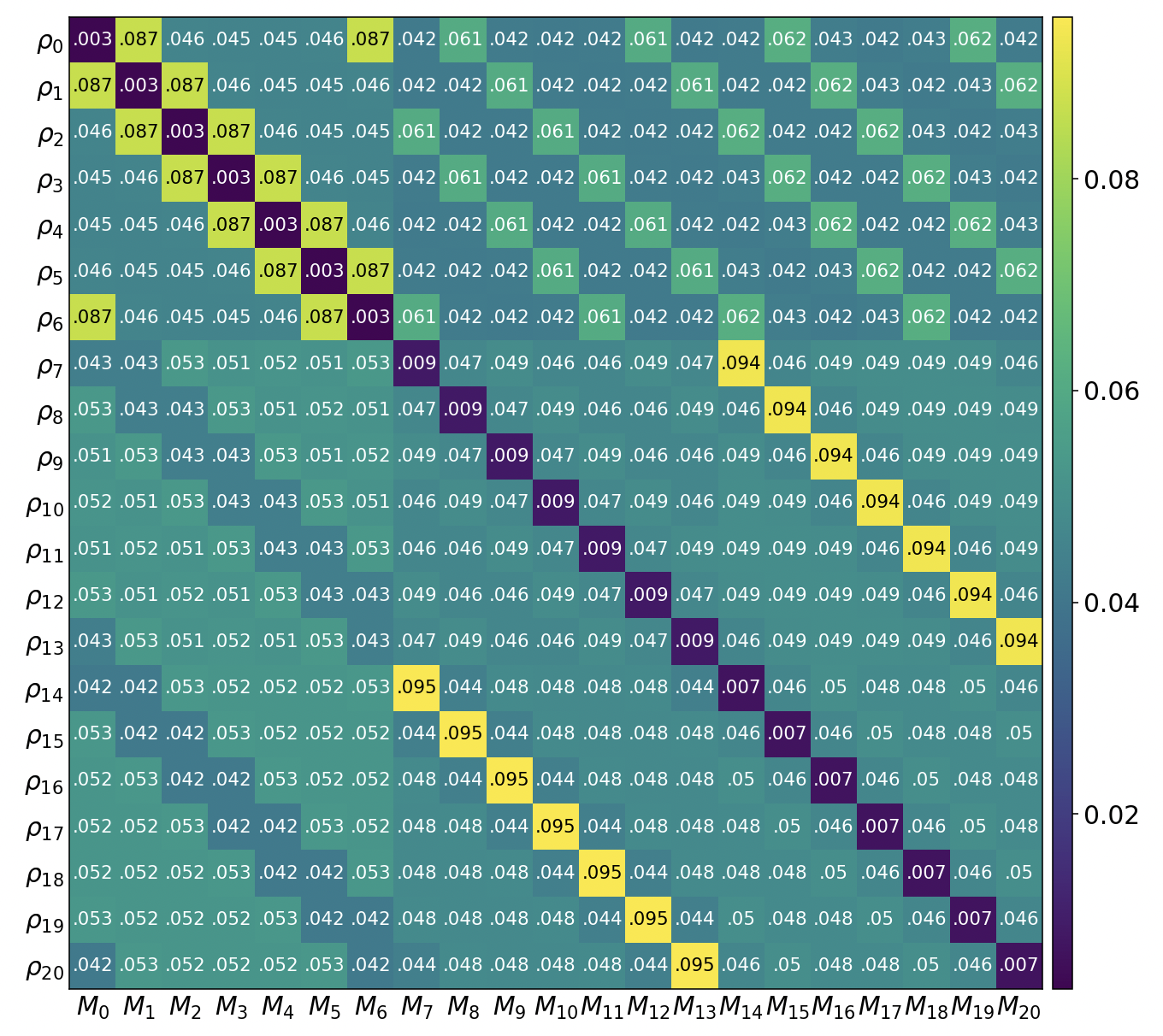}
  {\textbf{Experimental results taken for the convex protocol with an intense beam, applying permutations.} 
   Theoretical $P$-matrices with zeros on the diagonal (left) and their experimentally reconstructed counterparts (right) for $d=3,5,7$ (top–bottom).
   The OAM ($\ell$) $\leftrightarrow$ qudit mappings chosen to minimize $\|E\|_2$ are: 
   $P_{3}\!:\!(\ell=-3,+0,+1)$; 
   $P_{5}\!:\!(\ell=-2,+2,+3,-3,+0)$; 
   $P_{7}\!:\!(\ell=+0,+1,-3,-2,+2,+3,-1)$.\\}%
  {fig:classical-siperms}
  {{ (with permutations)}}

\subsubsection{\label{sec:s-future} Possible direct implementation}
In this section, a ``triangular interferometer'' built from OAM tools that can provide a practical way to perform general three-element POVMs on superpositions of at most two topological charges, without resorting to indirect tomographic reconstruction is presented.\\ 

A PPBS–waveplate block which serves Bob to send the desired superpositions is replaced by a Mach–Zehnder interferometer: the first $50{:}50$ beam-splitter creates two paths, SLMs in the arms apply the required $SU(2)$ rotations, and the second beam-splitter—together with a Dove prism that fixes the $\ell$-dependent phase and a piezo mirror that tunes the common delay—acts as an \emph{OAM-dependent partial splitter} that steers controllable fractions of the $|\ell=0\rangle$ and $|\ell=1\rangle$ amplitudes into distinct output ports.

After the first BS the photon is in $(\ket{a}+i\ket{b})/\sqrt2$.  
SLM$_a$ imprints the state 
\(
\ket{\phi_A}=(\ket{0}+\ket{1})/\sqrt2
\)
with diffraction efficiency $\sqrt{0.667}$, while SLM$_b$ imprints  
\(
\ket{\phi_B}=(\ket{2}+\ket{-1})/\sqrt2
\)
with efficiency $\sqrt{0.333}$.  
If $\ket{\phi_A}$ and $\ket{\phi_B}$ are orthogonal (as here) the relative phase $\delta$ controlled by the piezo does not change the output statistics; if they overlap, $\delta$ is scanned rapidly so that the coherence averages to zero.  
Blocking one output port of the second BS leaves a single beam carrying the mixed state  
\[
\rho \;=\; 0.667\,\ket{\phi_A}\!\bra{\phi_A}
          \;+\; 0.333\,\ket{\phi_B}\!\bra{\phi_B},
\]
which Alice sends to Bob through one spatial channel.

Bob uses an identical Mach–Zehnder.  
A Dove prism in one arm fixes the mode-dependent phase $e^{i2\ell\theta}$; the piezo sets a tunable common phase~$\delta$.  
Together they route adjustable fractions of the logical basis $\{\ket{0},\ket{1}\}$ into the two arms before the second BS.  
SLM$_1$ and SLM$_2$ apply the required single-qubit rotations $U\in SU(2)$.  
Choosing $(\theta,\delta)$ so that $\ket{\phi_A}$ interferes constructively in output port~$c$ and destructively in port~$d$ implements the projector  
\(
\Pi=\ket{\phi_A}\!\bra{\phi_A}
\)
at port~$c$.  
For the incoming state $\rho$ the click probability is therefore  
\[
P_c \;=\; \operatorname{Tr}(\rho\,\Pi) \;=\; 0.667 .
\]
Reprogramming the two SLMs and readjusting $(\theta,\delta)$ aligns any other superposition of two OAM values with port~$c$; blocking one of the four nominal outputs again leaves three detectors whose statistics realise the desired three-element POVM.

\subsubsection{\label{sec:class-d3} Implementation with laser beam}

For the laser-beam implementation of the setup, a 1550 nm laser with an output power of 50mW replaced the single-photon source. Since coincidence detection was no longer required, the second detection arm was removed, and the two APDs were substituted by a single power meter. This device measured the intensity coupled into the single-mode fiber for each prepared and measured OAM superposition.

\ThreePmatrixMainFigure
  {qcom-outputs-M3/Theory/d3}
  {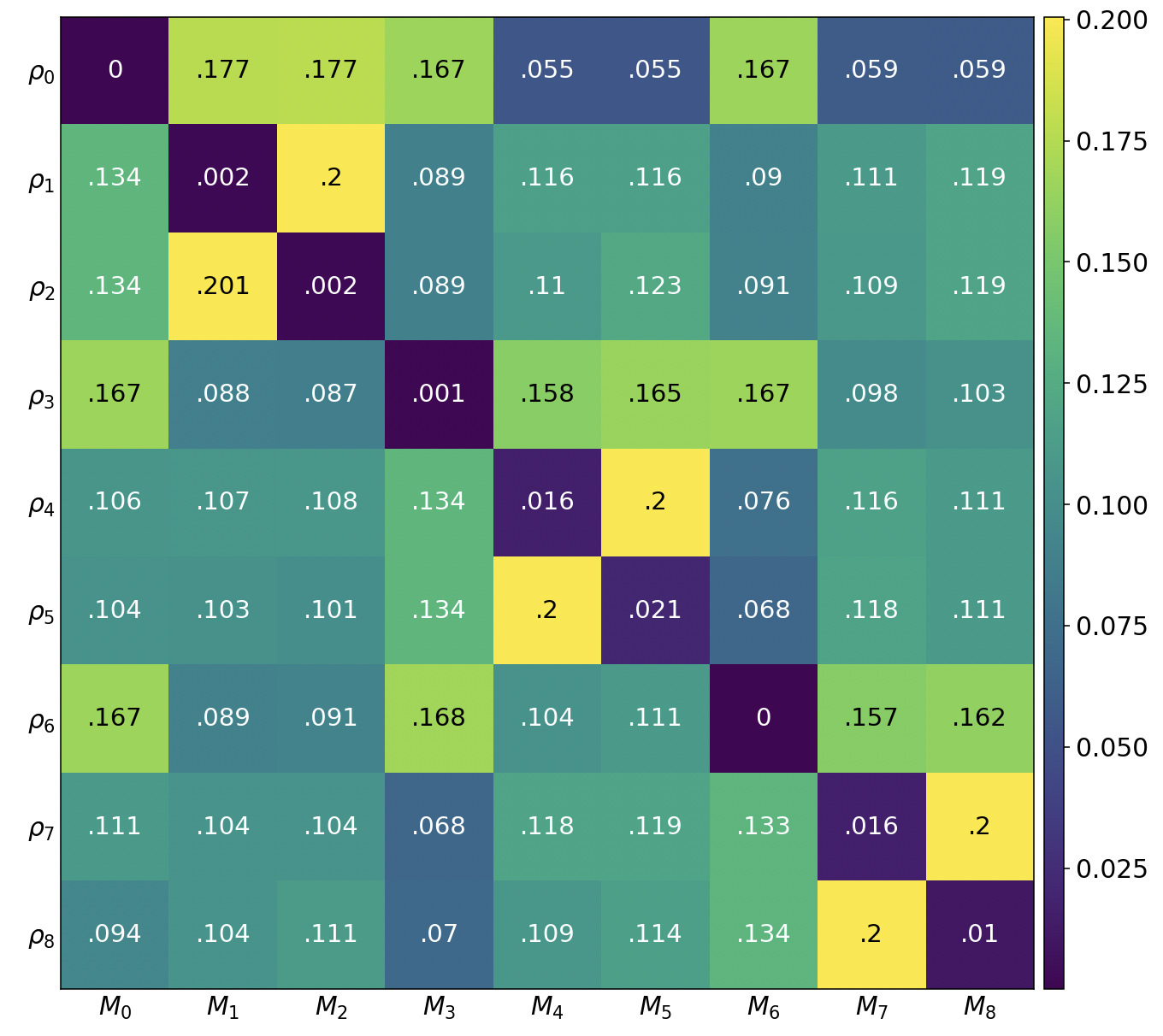}
  {qcom-outputs-M3/Theory/convex-d5}
  {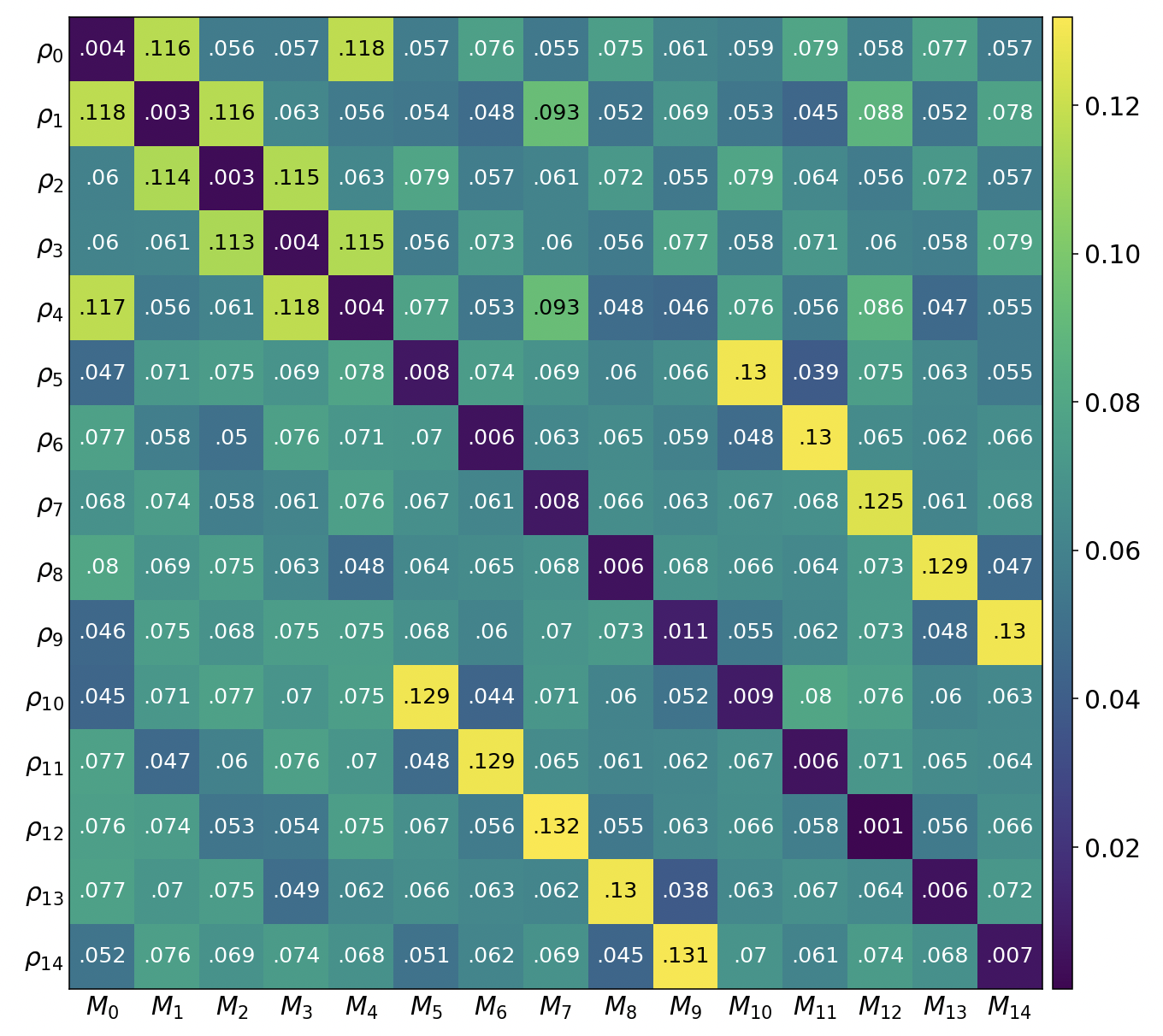}
  {qcom-outputs-M3/Theory/convex-d7}
  {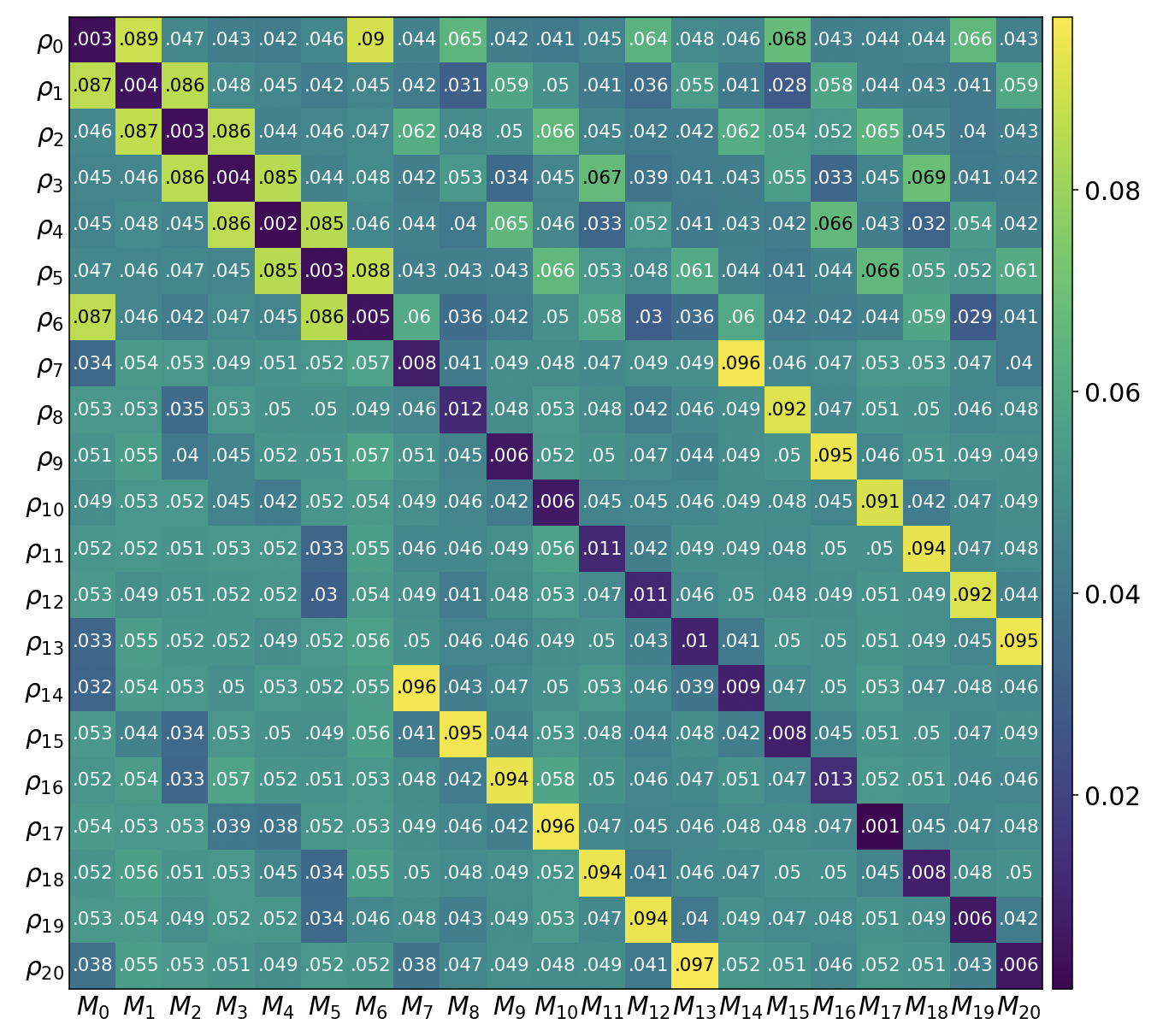}
  {\textbf{Experimental results taken for the convex protocol with an intense beam.} 
  Theoretical $P$-matrices with zeros on the diagonal (left) and their experimentally reconstructed counterparts (right) for $d=3,5,7$ (top–bottom).
   The OAM ($\ell$) $\leftrightarrow$ qudit mappings (order of $\ket{0},\ket{1},\ldots,\ket{6}$) chosen to minimize $\|E\|_2$ are: 
   $P_{3}\!:\!(\ell=-3,+0,+1)$; 
   $P_{5}\!:\!(\ell=-2,+2,+3,-3,+0)$; 
   $P_{7}\!:\!(\ell=+0,+1,-3,-2,+2,+3,-1)$.\\}%
  {fig:classical-noperms}
  {{}}

\subsection{\label{sec:PrepMeas} State preparations and probability matrices}
\subsubsection{Dimension 3}
Here we will write explicitly the measurements, to built the intuition behind Eqs.~\eqref{eq:measdiag},~\eqref{eq:meas+} and~\eqref{eq:meas-}. Bob chooses at random to measure on one of the bases
\begin{align}
    B_1 &= \Bigg \{\ket{M_0}=\ket{0}, \quad \ket{M_1}=\frac{\ket{1}+\ket{2}}{\sqrt{2}},\quad  \ket{M_2}=\frac{\ket{1}-\ket{2}}{\sqrt{2}}\Bigg \}\\
    B_2 &= \Bigg \{\ket{M_3}=\ket{1}, \quad \ket{M_4}=\frac{\ket{0}+\ket{2}}{\sqrt{2}},\quad  \ket{M_5}=\frac{\ket{0}-\ket{2}}{\sqrt{2}}\Bigg \}\\
    B_3 &= \Bigg \{\ket{M_6}=\ket{2},\quad  \ket{M_7}=\frac{\ket{0}+\ket{1}}{\sqrt{2}}, \quad \ket{M_8}=\frac{\ket{0}-\ket{1}}{\sqrt{2}}\Bigg \}
\end{align}
with equal probability. Alice prepares the states
\begin{align}
    \rho_0&=0.5\dyad{1}+0.5\dyad{2},
    \quad \rho_1=0.4\dyad{0}+0.3\dyad{1-2},\quad
    \rho_2=0.4\dyad{0}+0.3\dyad{1+2},\\
    \rho_3 &=0.5\dyad{0}+0.5\dyad{2}, \quad \rho_4=0.4\dyad{1}+0.3\dyad{0-2},\quad  \rho_5=0.4\dyad{1}+0.3\dyad{0+2}\\
    \rho_6 &=0.5\dyad{0}+0.5\dyad{1}, \quad \rho_7=0.4\dyad{2}+0.3\dyad{0-1},\quad  \rho_8=0.4\dyad{2}+0.3\dyad{0+1}\,.
\end{align}
The conditional probabilities $P_{xb}:=p(b|x)=\tr(\rho_xM_b)$ define a matrix $P=(P_{xb})$ with entries $P_{xb}$. Each row is the probability distribution of obtaining on outcome $b$, given that the state prepared was $\rho_x$ (the matrix is thus row-stochastic, meaning that rows sum to 1). The explicit matrix to be obtained with the protocol above reads:
\begin{equation}
P = \begin{pmatrix}
0.     & 0.167  & 0.167  & 0.167  & 0.083  & 0.083  & 0.167  & 0.083  & 0.083\\
 0.133  & 0.     & 0.2    & 0.1    & 0.117  & 0.117  & 0.1    & 0.117  & 0.117\\
  0.133  & 0.2    & 0.     & 0.1    & 0.117  & 0.117  & 0.1    & 0.117  & 0.117\\
  0.167  & 0.083  & 0.083  & 0.     & 0.167  & 0.167  & 0.167  & 0.083  & 0.083\\
  0.1    & 0.117  & 0.117  & 0.133  & 0.     & 0.2    & 0.1    & 0.117  & 0.117\\
  0.1    & 0.117  & 0.117  & 0.133  & 0.2    & 0.     & 0.1    & 0.117  & 0.117\\
  0.167  & 0.083  & 0.083  & 0.167  & 0.083  & 0.083  & 0.     & 0.167  & 0.167\\
  0.1    & 0.117  & 0.117  & 0.1    & 0.117  & 0.117  & 0.133  & 0.     & 0.2  \\
  0.1    & 0.117  & 0.117  & 0.1    & 0.117  & 0.117  & 0.133  & 0.2   & 0.
\end{pmatrix}
\end{equation}
As desired, it is a $3d=9$-dimensional matrix with rank $2d=6$ (which implies that at least $6$ digits are needed for classical implementation), and can be implemented with a quantum system of dimension $d=3$.

\subsubsection{Dimension 5}
{\bf First approach:} Computational basis measurements repeat $d-2$ times, coherent superposition measurements do not repeat.
\begin{align}
    B_1 &= \Bigg \{\ket{M_0}=\ket{0}, \,\,\, \ket{M_1}=\ket{1},\,\,\, \ket{M_2}=\ket{2},\,\,\,\ket{M_8}=\frac{\ket{3}+\ket{4}}{\sqrt{2}},\,\,\, \ket{M_{13}}=\frac{\ket{3}-\ket{4}}{\sqrt{2}}\Bigg \}\\
    B_2 &= \Bigg \{\ket{M_1}=\ket{1}, \,\,\, \ket{M_2}=\ket{2}, \,\,\, \ket{M_3}=\ket{3},\,\,\,\ket{M_9}=\frac{\ket{4}+\ket{0}}{\sqrt{2}},\,\,\,\ket{M_{14}}=\frac{\ket{4}-\ket{0}}{\sqrt{2}} \Bigg \}\\
    B_3 &= \Bigg \{\ket{M_2}=\ket{2}, \,\,\, \ket{M_3}=\ket{3}, \,\,\, \ket{M_4}=\ket{4},\,\,\, \ket{M_{5}}=\frac{\ket{0}+\ket{1}}{\sqrt{2}} \,\,\, \ket{M_{10}}=\frac{\ket{0}-\ket{1}}{\sqrt{2}} \Bigg \}\\
    B_4 &= \Bigg \{\ket{M_3}=\ket{3}, \,\,\, \ket{M_4}=\ket{4}, \,\,\, \ket{M_0}=\ket{0},\,\,\, \ket{M_{6}}=\frac{\ket{1}+\ket{2}}{\sqrt{2}} \,\,\, \ket{M_{11}}=\frac{\ket{1}-\ket{2}}{\sqrt{2}} \Bigg \}\\
    B_5 &= \Bigg \{\ket{M_4}=\ket{4}, \,\,\, \ket{M_0}=\ket{0}, \,\,\, \ket{M_1}=\ket{1},\,\,\, \ket{M_{7}}=\frac{\ket{2}+\ket{3}}{\sqrt{2}} \,\,\, \ket{M_{12}}=\frac{\ket{2}-\ket{3}}{\sqrt{2}} \Bigg \}
\end{align}

{\bf Second approach:} Computational measurements do not repeat, coherent superposition measurements repeat twice.
\begin{align}
    B_1 &= \Bigg \{\ket{M_0}=\ket{0}, \,\,\, \ket{M_6}=\frac{\ket{1}+\ket{2}}{\sqrt{2}},\,\,\, \ket{M_{11}}=\frac{\ket{1}-\ket{2}}{\sqrt{2}},\,\,\,\ket{M_8}=\frac{\ket{3}+\ket{4}}{\sqrt{2}},\,\,\, \ket{M_{13}}=\frac{\ket{3}-\ket{4}}{\sqrt{2}}\Bigg \}\\
    B_2 &= \Bigg \{\ket{M_1}=\ket{1}, \,\,\, \ket{M_7}=\frac{\ket{2}+\ket{3}}{\sqrt{2}}, \,\,\, \ket{M_{12}}=\frac{\ket{2}-\ket{3}}{\sqrt{2}},\,\,\,\ket{M_9}=\frac{\ket{4}+\ket{0}}{\sqrt{2}},\,\,\,\ket{M_{14}}=\frac{\ket{4}-\ket{0}}{\sqrt{2}} \Bigg \}\\
    B_3 &= \Bigg \{\ket{M_2}=\ket{2}, \,\,\, \ket{M_8}=\frac{\ket{3}+\ket{4}}{\sqrt{2}}, \,\,\, \ket{M_{13}}=\frac{\ket{3}-\ket{4}}{\sqrt{2}},\,\,\, \ket{M_{5}}=\frac{\ket{0}+\ket{1}}{\sqrt{2}} \,\,\, \ket{M_{10}}=\frac{\ket{0}-\ket{1}}{\sqrt{2}} \Bigg \}\\
    B_4 &= \Bigg \{\ket{M_3}=\ket{3}, \,\,\, \ket{M_9}=\frac{\ket{4}+\ket{0}}{\sqrt{2}}, \,\,\, \ket{M_{14}}=\frac{\ket{4}-\ket{0}}{\sqrt{2}},\,\,\, \ket{M_{6}}=\frac{\ket{1}+\ket{2}}{\sqrt{2}} \,\,\, \ket{M_{11}}=\frac{\ket{1}-\ket{2}}{\sqrt{2}} \Bigg \}\\
    B_5 &= \Bigg \{\ket{M_4}=\ket{4}, \,\,\, \ket{M_5}=\frac{\ket{0}+\ket{1}}{\sqrt{2}}, \,\,\, \ket{M_{10}}=\frac{\ket{0}-\ket{1}}{\sqrt{2}},\,\,\, \ket{M_{7}}=\frac{\ket{2}+\ket{3}}{\sqrt{2}} \,\,\, \ket{M_{12}}=\frac{\ket{2}-\ket{3}}{\sqrt{2}} \Bigg \}
\end{align}

{\bf Combination of both:} Taking $1/3\times$ the probabilities obtained with the first measurement choices (in first approach), plus $2/3\times$ the probabilities obtained in the second measurement choices. The states we use in both measurements are the following:

\[ \rho_{0} = 0.167\ket{2}\bra{2} + 0.167\ket{3}\bra{3} + 0.333\ket{1}\bra{1} + 0.333\ket{4}\bra{4} \]
\[ \rho_{1} = 0.167\ket{3}\bra{3} + 0.167\ket{4}\bra{4} + 0.333\ket{0}\bra{0} + 0.333\ket{2}\bra{2} \]
\[ \rho_{2} = 0.167\ket{0}\bra{0} + 0.167\ket{4}\bra{4} + 0.333\ket{3}\bra{3} + 0.333\ket{1}\bra{1} \]
\[ \rho_{3} = 0.167\ket{1}\bra{1} + 0.167\ket{0}\bra{0} + 0.333\ket{2}\bra{2} + 0.333\ket{4}\bra{4} \]
\[ \rho_{4} = 0.167\ket{1}\bra{1} + 0.167\ket{2}\bra{2} + 0.333\ket{3}\bra{3} + 0.333\ket{0}\bra{0} \]
\[ \rho_{5} = 0.203\ket{3}\bra{3} + 0.212\ket{2}\bra{2} + 0.212\ket{4}\bra{4} + 0.373\ket{0-1}\bra{0-1} \]
\[ \rho_{6} = 0.203\ket{4}\bra{4} + 0.212\ket{0}\bra{0} + 0.212\ket{3}\bra{3} + 0.373\ket{1-2}\bra{1-2} \]
\[ \rho_{7} = 0.203\ket{0}\bra{0} + 0.212\ket{1}\bra{1} + 0.212\ket{4}\bra{4} + 0.373\ket{2-3}\bra{2-3} \]
\[ \rho_{8} = 0.203\ket{1}\bra{1} + 0.212\ket{2}\bra{2} + 0.212\ket{0}\bra{0} + 0.373\ket{3-4}\bra{3-4} \]
\[ \rho_{9} = 0.203\ket{2}\bra{2} + 0.212\ket{1}\bra{1} + 0.212\ket{3}\bra{3} + 0.373\ket{0-4}\bra{0-4} \]
\[ \rho_{10} = 0.203\ket{3}\bra{3} + 0.212\ket{2}\bra{2} + 0.212\ket{4}\bra{4} + 0.373\ket{0+1}\bra{0+1} \]
\[ \rho_{11} = 0.203\ket{4}\bra{4} + 0.212\ket{0}\bra{0} + 0.212\ket{3}\bra{3} + 0.373\ket{1+2}\bra{1+2} \]
\[ \rho_{12} = 0.203\ket{0}\bra{0} + 0.212\ket{1}\bra{1} + 0.212\ket{4}\bra{4} + 0.373\ket{2+3}\bra{2+3} \]
\[ \rho_{13} = 0.203\ket{1}\bra{1} + 0.212\ket{2}\bra{2} + 0.212\ket{0}\bra{0} + 0.373\ket{3+4}\bra{3+4} \]
\[ \rho_{14} = 0.203\ket{2}\bra{2} + 0.212\ket{1}\bra{1} + 0.212\ket{3}\bra{3} + 0.373\ket{0+4}\bra{0+4} \]

The final $P$ matrix is
\[ P=
\left(\begin{array}{ccccccccccccccc}
0.000 & 0.111 & 0.056 & 0.056 & 0.111 & 0.056 & 0.083 & 0.056 & 0.083 & 0.056 & 0.056 & 0.083 & 0.056 & 0.083 & 0.056 \\
0.111 & 0.000 & 0.111 & 0.056 & 0.056 & 0.056 & 0.056 & 0.083 & 0.056 & 0.083 & 0.056 & 0.056 & 0.083 & 0.056 & 0.083 \\
0.056 & 0.111 & 0.000 & 0.111 & 0.056 & 0.083 & 0.056 & 0.056 & 0.083 & 0.056 & 0.083 & 0.056 & 0.056 & 0.083 & 0.056 \\
0.056 & 0.056 & 0.111 & 0.000 & 0.111 & 0.056 & 0.083 & 0.056 & 0.056 & 0.083 & 0.056 & 0.083 & 0.056 & 0.056 & 0.083 \\
0.111 & 0.056 & 0.056 & 0.111 & 0.000 & 0.083 & 0.056 & 0.083 & 0.056 & 0.056 & 0.083 & 0.056 & 0.083 & 0.056 & 0.056 \\
0.062 & 0.062 & 0.071 & 0.068 & 0.071 & 0.000 & 0.066 & 0.069 & 0.069 & 0.066 & 0.124 & 0.066 & 0.069 & 0.069 & 0.066 \\
0.071 & 0.062 & 0.062 & 0.071 & 0.068 & 0.066 & 0.000 & 0.066 & 0.069 & 0.069 & 0.066 & 0.124 & 0.066 & 0.069 & 0.069 \\
0.068 & 0.071 & 0.062 & 0.062 & 0.071 & 0.069 & 0.066 & 0.000 & 0.066 & 0.069 & 0.069 & 0.066 & 0.124 & 0.066 & 0.069 \\
0.071 & 0.068 & 0.071 & 0.062 & 0.062 & 0.069 & 0.069 & 0.066 & 0.000 & 0.066 & 0.069 & 0.069 & 0.066 & 0.124 & 0.066 \\
0.062 & 0.071 & 0.068 & 0.071 & 0.062 & 0.066 & 0.069 & 0.069 & 0.066 & 0.000 & 0.066 & 0.069 & 0.069 & 0.066 & 0.124 \\
0.062 & 0.062 & 0.071 & 0.068 & 0.071 & 0.124 & 0.066 & 0.069 & 0.069 & 0.066 & 0.000 & 0.066 & 0.069 & 0.069 & 0.066 \\
0.071 & 0.062 & 0.062 & 0.071 & 0.068 & 0.066 & 0.124 & 0.066 & 0.069 & 0.069 & 0.066 & 0.000 & 0.066 & 0.069 & 0.069 \\
0.068 & 0.071 & 0.062 & 0.062 & 0.071 & 0.069 & 0.066 & 0.124 & 0.066 & 0.069 & 0.069 & 0.066 & 0.000 & 0.066 & 0.069 \\
0.071 & 0.068 & 0.071 & 0.062 & 0.062 & 0.069 & 0.069 & 0.066 & 0.124 & 0.066 & 0.069 & 0.069 & 0.066 & 0.000 & 0.066 \\
0.062 & 0.071 & 0.068 & 0.071 & 0.062 & 0.066 & 0.069 & 0.069 & 0.066 & 0.124 & 0.066 & 0.069 & 0.069 & 0.066 & 0.000 \\
\end{array}\right)
\]

\subsubsection{Dimension 7}
The measurements are performed similarly as for dimensions $3$ and $5$, merging two approaches with different coherences. The states are given as follows:

\[ \rho_{0} = 0.125\ket{2}\bra{2} + 0.125\ket{3}\bra{3} + 0.125\ket{4}\bra{4} + 0.125\ket{5}\bra{5} + 0.250\ket{1}\bra{1} + 0.250\ket{6}\bra{6} \]
\[ \rho_{1} = 0.125\ket{3}\bra{3} + 0.125\ket{4}\bra{4} + 0.125\ket{5}\bra{5} + 0.125\ket{6}\bra{6} + 0.250\ket{0}\bra{0} + 0.250\ket{2}\bra{2} \]
\[ \rho_{2} = 0.125\ket{0}\bra{0} + 0.125\ket{4}\bra{4} + 0.125\ket{5}\bra{5} + 0.125\ket{6}\bra{6} + 0.250\ket{3}\bra{3} + 0.250\ket{1}\bra{1} \]
\[ \rho_{3} = 0.125\ket{1}\bra{1} + 0.125\ket{0}\bra{0} + 0.125\ket{5}\bra{5} + 0.125\ket{6}\bra{6} + 0.250\ket{2}\bra{2} + 0.250\ket{4}\bra{4} \]
\[ \rho_{4} = 0.125\ket{1}\bra{1} + 0.125\ket{2}\bra{2} + 0.125\ket{0}\bra{0} + 0.125\ket{6}\bra{6} + 0.250\ket{5}\bra{5} + 0.250\ket{3}\bra{3} \]
\[ \rho_{5} = 0.125\ket{1}\bra{1} + 0.125\ket{2}\bra{2} + 0.125\ket{3}\bra{3} + 0.125\ket{0}\bra{0} + 0.250\ket{4}\bra{4} + 0.250\ket{6}\bra{6} \]
\[ \rho_{6} = 0.125\ket{1}\bra{1} + 0.125\ket{2}\bra{2} + 0.125\ket{3}\bra{3} + 0.125\ket{4}\bra{4} + 0.250\ket{5}\bra{5} + 0.250\ket{0}\bra{0} \]
\[ \rho_{7} = 0.144\ket{3}\bra{3} + 0.144\ket{5}\bra{5} + 0.145\ket{4}\bra{4} + 0.148\ket{2}\bra{2} + 0.148\ket{6}\bra{6} + 0.271\ket{0-1}\bra{0-1} \]
\[ \rho_{8} = 0.144\ket{4}\bra{4} + 0.144\ket{6}\bra{6} + 0.145\ket{5}\bra{5} + 0.148\ket{3}\bra{3} + 0.148\ket{0}\bra{0} + 0.271\ket{1-2}\bra{1-2} \]
\[ \rho_{9} = 0.144\ket{0}\bra{0} + 0.144\ket{5}\bra{5} + 0.145\ket{6}\bra{6} + 0.148\ket{4}\bra{4} + 0.148\ket{1}\bra{1} + 0.271\ket{2-3}\bra{2-3} \]
\[ \rho_{10} = 0.144\ket{1}\bra{1} + 0.144\ket{6}\bra{6} + 0.145\ket{0}\bra{0} + 0.148\ket{5}\bra{5} + 0.148\ket{2}\bra{2} + 0.271\ket{3-4}\bra{3-4} \]
\[ \rho_{11} = 0.144\ket{2}\bra{2} + 0.144\ket{0}\bra{0} + 0.145\ket{1}\bra{1} + 0.148\ket{3}\bra{3} + 0.148\ket{6}\bra{6} + 0.271\ket{4-5}\bra{4-5} \]
\[ \rho_{12} = 0.144\ket{1}\bra{1} + 0.144\ket{3}\bra{3} + 0.145\ket{2}\bra{2} + 0.148\ket{4}\bra{4} + 0.148\ket{0}\bra{0} + 0.271\ket{5-6}\bra{5-6} \]
\[ \rho_{13} = 0.144\ket{4}\bra{4} + 0.144\ket{2}\bra{2} + 0.145\ket{3}\bra{3} + 0.148\ket{1}\bra{1} + 0.148\ket{5}\bra{5} + 0.271\ket{0-6}\bra{0-6} \]
\[ \rho_{14} = 0.144\ket{3}\bra{3} + 0.144\ket{5}\bra{5} + 0.145\ket{4}\bra{4} + 0.148\ket{2}\bra{2} + 0.148\ket{6}\bra{6} + 0.271\ket{0+1}\bra{0+1} \]
\[ \rho_{15} = 0.144\ket{4}\bra{4} + 0.144\ket{6}\bra{6} + 0.145\ket{5}\bra{5} + 0.148\ket{3}\bra{3} + 0.148\ket{0}\bra{0} + 0.271\ket{1+2}\bra{1+2} \]
\[ \rho_{16} = 0.144\ket{0}\bra{0} + 0.144\ket{5}\bra{5} + 0.145\ket{6}\bra{6} + 0.148\ket{4}\bra{4} + 0.148\ket{1}\bra{1} + 0.271\ket{2+3}\bra{2+3} \]
\[ \rho_{17} = 0.144\ket{1}\bra{1} + 0.144\ket{6}\bra{6} + 0.145\ket{0}\bra{0} + 0.148\ket{5}\bra{5} + 0.148\ket{2}\bra{2} + 0.271\ket{3+4}\bra{3+4} \]
\[ \rho_{18} = 0.144\ket{2}\bra{2} + 0.144\ket{0}\bra{0} + 0.145\ket{1}\bra{1} + 0.148\ket{3}\bra{3} + 0.148\ket{6}\bra{6} + 0.271\ket{4+5}\bra{4+5} \]
\[ \rho_{19} = 0.144\ket{1}\bra{1} + 0.144\ket{3}\bra{3} + 0.145\ket{2}\bra{2} + 0.148\ket{4}\bra{4} + 0.148\ket{0}\bra{0} + 0.271\ket{5+6}\bra{5+6} \]
\[ \rho_{20} = 0.144\ket{4}\bra{4} + 0.144\ket{2}\bra{2} + 0.145\ket{3}\bra{3} + 0.148\ket{1}\bra{1} + 0.148\ket{5}\bra{5} + 0.271\ket{0+6}\bra{0+6} \]

The final matrix is

{\footnotesize 
\begin{equation}
P=
\left(\begin{array}{ccccccccccccccccccccc}
0.00 & 0.08 & 0.04 & 0.04 & 0.04 & 0.04 & 0.08 & 0.04 & 0.06 & 0.04 & 0.04 & 0.04 & 0.06 & 0.04 & 0.04 & 0.06 & 0.04 & 0.04 & 0.04 & 0.06 & 0.04 \\
0.08 & 0.00 & 0.08 & 0.04 & 0.04 & 0.04 & 0.04 & 0.04 & 0.04 & 0.06 & 0.04 & 0.04 & 0.04 & 0.06 & 0.04 & 0.04 & 0.06 & 0.04 & 0.04 & 0.04 & 0.06 \\
0.04 & 0.08 & 0.00 & 0.08 & 0.04 & 0.04 & 0.04 & 0.06 & 0.04 & 0.04 & 0.06 & 0.04 & 0.04 & 0.04 & 0.06 & 0.04 & 0.04 & 0.06 & 0.04 & 0.04 & 0.04 \\
0.04 & 0.04 & 0.08 & 0.00 & 0.08 & 0.04 & 0.04 & 0.04 & 0.06 & 0.04 & 0.04 & 0.06 & 0.04 & 0.04 & 0.04 & 0.06 & 0.04 & 0.04 & 0.06 & 0.04 & 0.04 \\
0.04 & 0.04 & 0.04 & 0.08 & 0.00 & 0.08 & 0.04 & 0.04 & 0.04 & 0.06 & 0.04 & 0.04 & 0.06 & 0.04 & 0.04 & 0.04 & 0.06 & 0.04 & 0.04 & 0.06 & 0.04 \\
0.04 & 0.04 & 0.04 & 0.04 & 0.08 & 0.00 & 0.08 & 0.04 & 0.04 & 0.04 & 0.06 & 0.04 & 0.04 & 0.06 & 0.04 & 0.04 & 0.04 & 0.06 & 0.04 & 0.04 & 0.06 \\
0.08 & 0.04 & 0.04 & 0.04 & 0.04 & 0.08 & 0.00 & 0.06 & 0.04 & 0.04 & 0.04 & 0.06 & 0.04 & 0.04 & 0.06 & 0.04 & 0.04 & 0.04 & 0.06 & 0.04 & 0.04 \\
0.05 & 0.05 & 0.05 & 0.05 & 0.05 & 0.05 & 0.05 & 0.00 & 0.05 & 0.05 & 0.05 & 0.05 & 0.05 & 0.05 & 0.09 & 0.05 & 0.05 & 0.05 & 0.05 & 0.05 & 0.05 \\
0.05 & 0.05 & 0.05 & 0.05 & 0.05 & 0.05 & 0.05 & 0.05 & 0.00 & 0.05 & 0.05 & 0.05 & 0.05 & 0.05 & 0.05 & 0.09 & 0.05 & 0.05 & 0.05 & 0.05 & 0.05 \\
0.05 & 0.05 & 0.05 & 0.05 & 0.05 & 0.05 & 0.05 & 0.05 & 0.05 & 0.00 & 0.05 & 0.05 & 0.05 & 0.05 & 0.05 & 0.05 & 0.09 & 0.05 & 0.05 & 0.05 & 0.05 \\
0.05 & 0.05 & 0.05 & 0.05 & 0.05 & 0.05 & 0.05 & 0.05 & 0.05 & 0.05 & 0.00 & 0.05 & 0.05 & 0.05 & 0.05 & 0.05 & 0.05 & 0.09 & 0.05 & 0.05 & 0.05 \\
0.05 & 0.05 & 0.05 & 0.05 & 0.05 & 0.05 & 0.05 & 0.05 & 0.05 & 0.05 & 0.05 & 0.00 & 0.05 & 0.05 & 0.05 & 0.05 & 0.05 & 0.05 & 0.09 & 0.05 & 0.05 \\
0.05 & 0.05 & 0.05 & 0.05 & 0.05 & 0.05 & 0.05 & 0.05 & 0.05 & 0.05 & 0.05 & 0.05 & 0.00 & 0.05 & 0.05 & 0.05 & 0.05 & 0.05 & 0.05 & 0.09 & 0.05 \\
0.05 & 0.05 & 0.05 & 0.05 & 0.05 & 0.05 & 0.05 & 0.09 & 0.05 & 0.05 & 0.05 & 0.05 & 0.05 & 0.00 & 0.05 & 0.05 & 0.05 & 0.05 & 0.05 & 0.05 & 0.09 \\
0.05 & 0.05 & 0.05 & 0.05 & 0.05 & 0.05 & 0.05 & 0.05 & 0.09 & 0.05 & 0.05 & 0.05 & 0.05 & 0.05 & 0.00 & 0.05 & 0.05 & 0.05 & 0.05 & 0.05 & 0.05 \\
0.05 & 0.05 & 0.05 & 0.05 & 0.05 & 0.05 & 0.05 & 0.05 & 0.05 & 0.09 & 0.05 & 0.05 & 0.05 & 0.05 & 0.05 & 0.00 & 0.05 & 0.05 & 0.05 & 0.05 & 0.05 \\
0.05 & 0.05 & 0.05 & 0.05 & 0.05 & 0.05 & 0.05 & 0.05 & 0.05 & 0.05 & 0.09 & 0.05 & 0.05 & 0.05 & 0.05 & 0.05 & 0.00 & 0.05 & 0.05 & 0.05 & 0.05 \\
0.05 & 0.05 & 0.05 & 0.05 & 0.05 & 0.05 & 0.05 & 0.05 & 0.05 & 0.05 & 0.05 & 0.09 & 0.05 & 0.05 & 0.05 & 0.05 & 0.05 & 0.00 & 0.05 & 0.05 & 0.05 \\
0.05 & 0.05 & 0.05 & 0.05 & 0.05 & 0.05 & 0.05 & 0.05 & 0.05 & 0.05 & 0.05 & 0.05 & 0.09 & 0.05 & 0.05 & 0.05 & 0.05 & 0.05 & 0.00 & 0.05 & 0.05 \\
0.05 & 0.05 & 0.05 & 0.05 & 0.05 & 0.05 & 0.05 & 0.05 & 0.05 & 0.05 & 0.05 & 0.05 & 0.05 & 0.09 & 0.05 & 0.05 & 0.05 & 0.05 & 0.05 & 0.00 & 0.05 \\
0.05 & 0.05 & 0.05 & 0.05 & 0.05 & 0.05 & 0.05 & 0.05 & 0.05 & 0.05 & 0.05 & 0.05 & 0.05 & 0.09 & 0.05 & 0.05 & 0.05 & 0.05 & 0.05 & 0.05 & 0.00
\end{array}\right)
\end{equation}
}

\end{document}